\documentclass[
	a4paper,
	fontsize=10pt,
	twoside=true,
    numbers=noenddot,
    parskip=half-
]{kaobook}

\usepackage[UKenglish]{babel}
\usepackage[autostyle]{csquotes}

\usepackage[framed=true]{kaotheorems}

\usepackage{kaobiblio}
\AtEveryBibitem{
    \clearfield{urlyear}
    \clearfield{urlmonth}
}
\graphicspath{{images/}} % Paths in which to look for images

\usepackage{icomma}
\usepackage{subcaption}
\usepackage[dvipsnames]{xcolor}
\usepackage{tikz}
\usepackage{pgfplots}
\usepackage{tikz-3dplot}
\usetikzlibrary{matrix,shapes,decorations.pathreplacing,fit,backgrounds,calc,babel, positioning}
\usepgfplotslibrary{groupplots, colormaps}
\pgfplotsset{compat=1.18}
\tdplotsetmaincoords{70}{115}

\pgfplotsset{every major tick/.append style={thick}}

\makeatletter
\pgfplotsset{
    groupplot xlabel/.initial={},
    every groupplot x label/.style={
        at={($({\pgfplots@group@name\space c1r\pgfplots@group@rows.west}|-{\pgfplots@group@name\space c1r\pgfplots@group@rows.outer south})!0.5!({\pgfplots@group@name\space c\pgfplots@group@columns r\pgfplots@group@rows.east}|-{\pgfplots@group@name\space c\pgfplots@group@columns r\pgfplots@group@rows.outer south})$)},
        anchor=north,
    },
    groupplot ylabel/.initial={},
    every groupplot y label/.style={
            rotate=90,
        at={($({\pgfplots@group@name\space c1r1.north}-|{\pgfplots@group@name\space c1r1.outer
west})!0.5!({\pgfplots@group@name\space c1r\pgfplots@group@rows.south}-|{\pgfplots@group@name\space c1r\pgfplots@group@rows.outer west})$)},
        anchor=south
    },
    execute at end groupplot/.code={%
      \node [/pgfplots/every groupplot x label]
{\pgfkeysvalueof{/pgfplots/groupplot xlabel}};  
      \node [/pgfplots/every groupplot y label] 
{\pgfkeysvalueof{/pgfplots/groupplot ylabel}};  
    }
}

\def\endpgfplots@environment@groupplot{%
    \endpgfplots@environment@opt%
    \pgfkeys{/pgfplots/execute at end groupplot}%
    \endgroup%
}
\makeatother

\makeatletter
\pgfmathdeclarefunction{distance}{2}{%
\begingroup%
\pgfextractx{\pgf@xa}{\pgfpointanchor{#1}{center}}%
\pgfextracty{\pgf@ya}{\pgfpointanchor{#1}{center}}%
\pgfextractx{\pgf@xb}{\pgfpointanchor{#2}{center}}%
\pgfextracty{\pgf@yb}{\pgfpointanchor{#2}{center}}%
\pgfmathparse{sqrt((\pgf@xa-\pgf@xb)*(\pgf@xa-\pgf@xb)+(\pgf@ya-\pgf@yb)*(\pgf@ya-\pgf@yb))}%
\pgfmathsmuggle\pgfmathresult\endgroup%
}%
\makeatother

\pgfplotsset{
    colormap/twilight/.style={%
        /pgfplots/colormap={twilight}{%
            rgb=(0.885750, 0.850009, 0.887974)
            rgb=(0.794543, 0.822451, 0.843922)
            rgb=(0.638026, 0.743314, 0.791151)
            rgb=(0.502885, 0.648972, 0.765214)
            rgb=(0.415722, 0.544949, 0.748092)
            rgb=(0.378414, 0.430809, 0.722599)
            rgb=(0.369433, 0.306929, 0.672051)
            rgb=(0.355374, 0.183804, 0.580932)
            rgb=(0.299425, 0.089682, 0.420216)
            rgb=(0.214828, 0.065940, 0.261917)
            rgb=(0.219305, 0.067466, 0.224510)
            rgb=(0.335334, 0.083532, 0.276325)
            rgb=(0.473841, 0.124264, 0.312255)
            rgb=(0.588949, 0.199970, 0.312889)
            rgb=(0.679075, 0.307695, 0.316655)
            rgb=(0.742599, 0.429943, 0.353527)
            rgb=(0.781742, 0.559751, 0.446247)
            rgb=(0.812699, 0.687130, 0.600052)
            rgb=(0.857757, 0.798020, 0.778296)
            rgb=(0.885712, 0.850022, 0.885725)
        },
      },
    }

\pgfplotsset{
    colormap/seismic/.style={%
        /pgfplots/colormap={seismic}{%
            rgb=(0.000000, 0.000000, 0.300000)
            rgb=(0.000000, 0.000000, 0.442745)
            rgb=(0.000000, 0.000000, 0.585490)
            rgb=(0.000000, 0.000000, 0.739216)
            rgb=(0.000000, 0.000000, 0.881961)
            rgb=(0.050980, 0.050980, 1.000000)
            rgb=(0.254902, 0.254902, 1.000000)
            rgb=(0.474510, 0.474510, 1.000000)
            rgb=(0.678431, 0.678431, 1.000000)
            rgb=(0.898039, 0.898039, 1.000000)
            rgb=(1.000000, 0.898039, 0.898039)
            rgb=(1.000000, 0.678431, 0.678431)
            rgb=(1.000000, 0.474510, 0.474510)
            rgb=(1.000000, 0.254902, 0.254902)
            rgb=(1.000000, 0.050980, 0.050980)
            rgb=(0.915686, 0.000000, 0.000000)
            rgb=(0.813725, 0.000000, 0.000000)
            rgb=(0.703922, 0.000000, 0.000000)
            rgb=(0.601961, 0.000000, 0.000000)
            rgb=(0.500000, 0.000000, 0.000000)
            },
        },
    }

\usepackage{mathtools}
\usepackage{amsthm}
\usepackage{bm}
\usepackage{array}

\usepackage{kaorefs}

\makeatletter
\providecommand*{\toclevel@mtocsection}{3}%
\makeatother

\SetKwComment{Comment}{/* }{ */}

\newcolumntype{L}[1]{>{\raggedright\let\newline\\\arraybackslash\hspace{0pt}}p{#1}}
\newcolumntype{C}[1]{>{\centering\let\newline\\\arraybackslash\hspace{0pt}}p{#1}}
\newcolumntype{R}[1]{>{\raggedleft\let\newline\\\arraybackslash\hspace{0pt}}p{#1}}
\newcolumntype{J}[1]{>{\let\newline\\\arraybackslash\hspace{0pt}}p{#1}}

\renewcommand{\Re}{\mathrm{Re}}
\renewcommand{\Im}{\mathrm{Im}}
\renewcommand{\vec}{\bm}

\newcommand{\imag}{i}
\newcommand{\eul}{e}

\newcommand{\bra}[1]{\left\langle #1 \right|}
\newcommand{\ket}[1]{\left| #1 \right\rangle}
\newcommand{\braket}[2]{\left\langle #1 | #2 \right\rangle}

\newcommand{\derivative}[2]{\frac{\mathrm{d} #1}{\mathrm{d} #2}}
\newcommand{\derivativen}[3]{\frac{\mathrm{d}^{#3} #1}{\mathrm{d} #2^{#3}}}
\NewDocumentCommand{\der}{m m g}{\IfNoValueTF{#3}{\derivative{#1}{#2}}{\derivativen{#1}{#2}{#3}}}

\newcommand{\pderivative}[2]{\frac{\partial #1}{\partial #2}}
\newcommand{\pderivativen}[3]{\frac{\partial^{#3} #1}{\partial #2^{#3}}}
\NewDocumentCommand{\pder}{m m g}{\IfNoValueTF{#3}{\pderivative{#1}{#2}}{\pderivativen{#1}{#2}{#3}}}

\newcommand{\dd}{\mathrm{d}}

\newcommand{\dia}{\mathrm{dia}}
\newcommand{\adia}{\mathrm{adia}}
\newcommand{\el}{\mathrm{el}}
\newcommand{\nuc}{\mathrm{nuc}}

\newcommand{\sgn}{\mathrm{sgn}}

\newcommand{\tr}[1]{\mathrm{tr}\left[ #1 \right]}
\newcommand{\Tr}[1]{\mathrm{Tr}\left[ #1 \right]}

\newcommand{\avg}[1]{\left\langle #1 \right\rangle}

\newcommand{\lmin}{\mathrm{min}}
\newcommand{\lmax}{\mathrm{max}}

\newcommand{\MASH}{\mathrm{MASH}}
\newcommand{\MISH}{\mathrm{msMASH}}

\newcommand{\hnorm}{\hat{H}_{\mathrm{norm}}}

\makeatletter
\@body@par
\makeatother

\begin{document}

% info

\title{An Analysis of the Mapping Approach to Surface Hopping}
\author{Jan Vavrin}
\date{\today}

\frontmatter

% -------------------------------
% TITLE PAGE
% -------------------------------
\makeatletter
\begin{titlepage}
    
    \begin{center}
        \vspace*{2cm}
            
        \Huge
        \textbf{An Analysis of the Mapping\\Approach to Surface Hopping}
            
        \vspace{2.5cm}

        \includegraphics[width=0.275\textwidth]{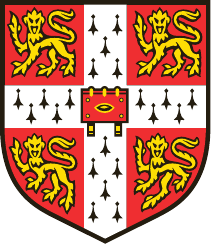}      

        \vspace{1.5cm}

        \Large
        \textbf{Jan Vavřín\\Trinity College}

        \vspace{1.5cm}
            
        \Large
        Department of Chemistry\\
        University of Cambridge

        \vspace{1.5cm}

        \Large
        \textit{Part III project\\2024\hspace{1pt}--25}

        \vfill

    \end{center}
\end{titlepage}
\makeatother

% -------------------------------
% DECLARATION
% -------------------------------
\cleardoubleoddpage
\thispagestyle{plain}

\vspace*{1cm}

\begin{center}
    \Large
    \textbf{Declaration}
\end{center}
% \addcontentsline{toc}{chapter}{Declaration}

\leftskip=2.25cm\rightskip=2.25cm
This dissertation is submitted in partial fulfilment of the requirements for Part~III Chemistry. It describes work carried out in the Department of Chemistry in~the Michaelmas Term 2024 and the Lent Term 2025. Unless otherwise indicated, the~research described is my own and not the product of collaboration.\\ \\
\null \hfill \textit{Jan Vavřín}\\
\null \hfill \textit{April 2025}

\leftskip=0pt\rightskip=0pt
\vfill
\pagebreak

% -------------------------------
% ACKNOWLEDGEMENTS
% -------------------------------
\cleardoubleoddpage 
\thispagestyle{plain}

\vspace*{1cm}

\begin{center}
    \Large
    \textbf{Acknowledgements}
\end{center}
\addcontentsline{toc}{chapter}{Acknowledgements}

\leftskip=2.25cm\rightskip=2.25cm

I am very grateful to Prof Stuart Althorpe not just for his guidance and support over the course of my project, but also for his patience with my last-minute writing of this thesis.

I've learned an enormous number of things this year -- I'll just highlight Chebyshev propagation as one of them, because it is one of the cleverest methods I have ever seen. If you've never heard about it, just skip the main text and go directly to Appendix F; it's the best thing you can do.

I would like to thank other members of the Althorpe group, especially Tom Drayton and Cole Hunt, who have provided a lot of scientific and non-scientific entertainment in the office. I couldn't have a\texttt{cheevd} this without you. I would also like to thank the other Part III theory students, who have made this year a~much more enjoyable experience -- I wish you all good luck in the future!

And lastly, I would like to thank my friends and family for supporting me not just this year, but throughout all four years I have spent in Cambridge. There are a few special people who have always been here for me and for that they have my eternal gratitude.

\leftskip=0pt\rightskip=0pt
\vfill
\pagebreak

% -------------------------------
% ABSTRACT
% -------------------------------
\cleardoubleoddpage 
\thispagestyle{plain}

\vspace*{1cm}

\begin{center}
    \Large
    \textbf{Abstract}
\end{center}
\addcontentsline{toc}{chapter}{Abstract}

\leftskip=2.25cm\rightskip=2.25cm
Recently, the mapping approach to surface hopping (MASH) was proposed as a method to simulate the non-adiabatic dynamics of two-level systems. It was shown that the method possesses many desirable qualities, both theoretically and through numerical simulations. We explain this success by proving that, out of similar mapping methods, MASH dynamics is unique in guaranteeing correct thermalisation, but that many different \enquote{estimators} can be used on top of it. We also show that MASH can successfully calculate multi-time correlation functions, which can be used for the simulation of 2D spectra. We also generalise an, in principle more accurate, technique known as the quantum jump procedure to these calculations and show that, contrary to expectations, it does not improve the results.

\leftskip=0pt\rightskip=0pt
\vfill
\pagebreak

% -------------------------------
% TABLE OF CONTENTS
% -------------------------------
\cleardoubleoddpage 
\begingroup % Local scope for the following commands

% Define the style for the TOC, LOF, and LOT
%\setstretch{1} % Uncomment to modify line spacing in the ToC
%\hypersetup{linkcolor=blue} % Uncomment to set the colour of links in the ToC
\setlength{\textheight}{230\hscale} % Manually adjust the height of the ToC pages

% Turn on compatibility mode for the etoc package
\etocstandarddisplaystyle % "toc display" as if etoc was not loaded
\etocstandardlines % "toc lines" as if etoc was not loaded

\tableofcontents % Output the table of contents

\endgroup

\vfill
\pagebreak

% -------------------------------
% NOTATION
% -------------------------------
\cleardoubleoddpage 
\thispagestyle{plain}

\begin{center}
    \Large
    \textbf{List of abbreviations}
\end{center}
\addcontentsline{toc}{chapter}{List of abbreviations}

\begin{center}
    \begin{tabular}{ L{3cm} J{9cm} }
     MASH & The original mapping approach to surface hopping from \cite{mannouchMappingApproachSurface2023}. Without qualification, this refers to both the dynamics and the estimators. We shall always use \enquote{MASH dynamics} to refer purely to the former. \\
     & \\
     ms-MASH & multi-state MASH \cite{runesonMultistateMappingApproach2023} \\
     & \\
     unSMASH & uncoupled spheres MASH \cite{lawrenceSizeconsistentMultistateMapping2024a} \\
     & \\
     FSSH & fewest switches surface hopping \\
     & \\
     QCLE & quantum-classical Liouville equation \\
     & \\
     CPA & classical path approximation \\
     & \\
     (T)CF & (time) correlation function \\
     & \\
     NACV & non-adiabatic coupling vector\\
    \end{tabular}
\end{center}

\vspace*{0.2cm}

\begin{center}
    \Large
    \textbf{List of mathematical symbols}
\end{center}
\addcontentsline{toc}{chapter}{List of mathematical symbols}

\begin{center}
    \begin{tabular}{ L{3cm} J{9cm} }
     $\Tr{\cdot}$ & trace over both nuclear and electronic variables \\
     & \\
     $\tr{\cdot}$ & trace over electronic variables only \\
     & \\
     $\hat{\mathcal{A}}$ & operator acting on both nuclear and electronic variables \\
     & \\
     $\hat{A}$ & Operator acting only on the electronic variables. The partial Wigner transform of $\hat{\mathcal{A}}$ will therefore be denoted $\hat{A}_W$. \\
     & \\
    $A(\vec{S})$ & the spin--mapped MASH version of $\hat{A}$ \\
    & \\
    $P_+$, $P_-$ & populations of the upper and lower adiabatic states \\
    & \\
    $P_1$, $P_2$ & populations of the two \emph{diabatic} states \\
    & \\
    $\sigma_x$, $\sigma_y$, $\sigma_z$ & Pauli matrices and the corresponding operators. Without qualification, these mean the adiabatic operators -- otherwise, we shall distinguish $\sigma_i^\dia$ and $\sigma_i^\adia$. \\
    & \\
    $C_{xy}(t)$ & The TCF between $\sigma_x$ and $\sigma_y(t)$. Similarly for other combinations of $x$, $y$ and $z$. \\
    & \\
    $C_{+-}(t)$ & The TCF between $P_+$ and $P_-(t)$. We shall also use the mixed version $C_{+x}(t)$ for the TCF between $P_+$ and $\sigma_x(t)$. \\
    & \\
    $\vec{p}_t, \vec{q}_t, \vec{S}_t$ & time-evolved (extended) phase space variables $\vec{p}, \vec{q}$ and $\vec{S}$ \\
    \end{tabular}
\end{center}

\vfill
\pagebreak

% -------------------------------

% Start of the main document content (resets page numbering)
\mainmatter 
\setchapterstyle{kao} % Choose the default chapter heading style

\setcounter{chapter}{-1}

% !TEX root = main.tex

\chapter{Introduction}
\label{chapter:intro}

Some chemical processes involve a change of electronic state and therefore require a \emph{non-adiabatic} description. These include biological processes like vision or photosynthesis -- and generally any process involving photoexcitation. In fact, non-adiabatic methods can be used whenever there is a separation between \enquote{quantum} and \enquote{classical} degrees of freedom\sidenote{Importantly, in models of the condensed phase, the environment can usually be treated classically.} -- electronic states are one example, but high frequency vibrations or even spin states could be treated in the same way.

Non-adiabatic dynamics is much more difficult than adiabatic dynamics. In the latter, the nuclei and electrons are effectively decoupled: we can first solve the electronic problem and then move the nuclei appropriately; in fact, the nuclei can easily be treated classically. One might hope that non-adiabatic dynamics would be the same, the only difficult part being the calculation of excited states.\sidenote{In fact, accuracy of excited state calculations is the main obstacle for current non-adiabatic simulations. One might argue that rigorously correct dynamics is useless if we cannot get the correct potential energy surfaces in the first place. However, there is progress on the electronic structure side. \cite{gonzalezQuantumChemistryDynamics2021} Additionally, a significant body of work uses model systems, where this problem disappears.} This is sadly not the case -- the electrons will \enquote{act back} on the nuclei to change the dynamics, so a simple separation is no longer possible. Intuitively, in order to hop to an excited state, the electrons take kinetic energy from the nuclei.

The conceptually easiest and most satisfying solution would be to solve the time-dependent Schrödinger equation for \emph{both} the electrons and the nuclei. This is only possible exactly\sidenote{\autoref{appendix:chebyshev} discusses some exact methods.} for small gas-phase molecules, though formally exact methods exist that can be extended to tens of atoms.\sidenote{These include MCTDH \cite{beckMulticonfigurationTimedependentHartree2000} and full multiple spawning \cite{curchodFullInitioMultiple2020}. Specialised techniques like HEOM \cite{tanimuraNumericallyExactApproach2020} can also be used for some classes of problems.} However, any quantum propagation method will suffer from an exponential scaling problem and it would never be possible to apply it to condensed-phase systems.

We would therefore much rather propagate the nuclei classically and only treat the electrons in a quantum way. The general framework for this problem, \emph{mixed quantum-classical dynamics},\sidenote{This is not to be confused with \emph{semiclassical dynamics}, which tries to incorporate quantum effects like tunnelling into nuclear dynamics. In mixed quantum-classical dynamics, the nuclei show no quantum behaviour.} was laid down in Kapral and Ciccotti's seminal paper \cite{kapralMixedQuantumclassicalDynamics1999}, where the \emph{quantum-classical Liouville equation} (QCLE) was derived. Even though this is a major step forward in physical rigour, it does not yield a tractable method\sidenote{The reason is that the QCLE cannot be solved using independent trajectories. \cite{kapralMixedQuantumclassicalDynamics1999,kellyMappingQuantumclassicalLiouville2012} The alternative, \enquote{coupled trajectories}, means that we would have to do simulations of $N$ replicas of the system, all of which would be interacting with each other. \cite{martensSurfaceHoppingConsensus2016,martensSurfaceHoppingMomentum2019a,pieroniNonadiabaticDynamicsCoupled2021}} for solving real-world non-adiabatic problems. A~more approximate method is thus needed.

The quantum system is inherently \enquote{fuzzy} and we could therefore imagine that the back-action of the electrons on the nuclei could be represented by an appropriate stochastic force. This is the reasoning behind \emph{fewest switches surface hopping} (FSSH). This method was introduced by Tully \cite{tullyMolecularDynamicsElectronic1990} in a relatively \textit{ad hoc} manner,\sidenote{It is possible to establish a link between FSSH and the QCLE, but several approximations are required. \cite{kapralSurfaceHoppingPerspective2016,subotnikCanWeDerive2013}} but it has become popular due to its simplicity and ability to correctly describe many non-adiabatic phenomena.

A novel non-adiabatic dynamics method, MASH, was recently developed by Mannouch and Richardson. \cite{mannouchMappingApproachSurface2023} MASH possesses many desirable qualities -- e.g., it is rigorously derivable from the QCLE and thermalises correctly in the long-time limit. In essence, one could use this method in an MD simulation and get a good result \enquote{out of the box}. Its main flaw is that it only works for two-level systems.

MASH was subsequently adapted by Runeson and Manolopoulos \cite{runesonMultistateMappingApproach2023} into a multi-state method we shall refer to as ms-MASH.\sidenote[][*0]{The method was originally called multi-state MASH, but it is also referred to as MISH in \cite{lawrenceSizeconsistentMultistateMapping2024a} and \cite{richardsonNonadiabaticDynamicsMapping2025a}.} While ms-MASH still thermalises correctly, it is not consistent with the QCLE. A QCLE-compliant multi-state version of MASH, unSMASH, was proposed in~\cite{lawrenceSizeconsistentMultistateMapping2024a}, but it loses some of the qualities that made two-level MASH so desirable.

MASH and ms-MASH have already been used to model real-world problems. For example, MASH is able to recover Marcus theory from the spin--boson model \cite{lawrenceRecoveringMarcusTheory2024}, ms-MASH can simulate the Fenna--Matthews--Olson complex \cite{runesonMultistateMappingApproach2023,runesonExcitonDynamicsMapping2024} and charge transport in semiconductors \cite{runesonChargeTransportOrganic2024}, and unSMASH can simulate the ultra-fast dynamics of cyclobutanone following photoexcitation \cite{lawrenceSizeconsistentMultistateMapping2024a}.

The fundamental concepts behind MASH are summarised in Chapter~\ref{chapter:background-theory}. The chapter contains extensive references to Appendix~\ref{appendix:theory-detail}, where a review of mixed quantum-classical dynamics and more details about MASH can be found.

Despite the success of MASH and ms-MASH, it is unclear what makes them work so well in the first place. In Chapter~\ref{chapter:uniqueness}, we show that from a class of similar methods, MASH dynamics is the only way to ensure correct thermalisation. However, there is great freedom in choosing the estimators, which explains why both MASH and ms-MASH can give similar results.

In Chapter~\ref{chapter:quantum-jump}, we discuss whether it is possible to calculate multi-time correlation functions with MASH. For simplicity, we only focus on 2D~experiments with a single \enquote{pulse}. We propose a modification to the quantum jump procedure for multi-time correlation functions and show that -- surprisingly -- it does not improve the results of the calculation.

Some additional results are collected in the Appendices. Appendix~\ref{appendix:integrators} contains a derivation of a 4th order \enquote{symplectic} integrator for MASH and Appendix~\ref{appendix:variance} contains a discussion of the observed difference in the statistical errors of MASH and ms-MASH. Finally, Appendix~\ref{appendix:impl-details} introduces a method for calculating MASH correlation functions in Liouville space.

% !TEX root = main.tex

\chapter{Background theory}
\label{chapter:background-theory}

In this chapter, we shall review the fundamental ideas behind MASH. We start by separating a general system into \enquote{quantum} and \enquote{classical} parts. To go beyond the Born--Oppenheimer approximation, \emph{derivative couplings} need to be introduced. Then, we define the diabatic basis and give an overview of the model systems used in this work. We shall also
see how experimental observables can be represented by correlation functions.

We shall then restrict our discussion purely to 2--level systems.\sidenote{Real-world problems usually have infinitely many quantum states -- however, in cases where MASH is expected to be useful, only a few low-energy ones will contribute to the dynamics. If there are many states close together in energy (this is common in the condensed phase), mean-field methods like Ehrenfest dynamics may be more appropriate. However, ms-MASH can still be used in such cases.~\cite{runesonChargeTransportOrganic2024}} With all these ideas in place, we discuss the description of such systems by the Bloch sphere and introduce \emph{spin mapping} as a way to associate states with regions on the sphere. 

There are two components of MASH that work together -- the \emph{dynamics} and the \emph{estimators}. The dynamics gives us a set of trajectories, but the estimators are needed to calculate observables from them. We shall see that MASH and ms-MASH differ only in the estimators -- this is an important point that forms the basis of the next chapter.

\section{Beyond the Born--Oppenheimer approximation}
\label{sec:beyond-BO}

The full system of electrons and nuclei can be described by a wavefunction $\Psi(\vec{r},\vec{R})$. However, we would much rather like to think in terms of separate electronic states, so we need to decompose the wavefunction into electronic and nuclear parts. When the Schrödinger equation in this representation is derived, \emph{non-adiabatic coupling} terms appear, which are given by derivatives of the electronic states with respect to the nuclear coordinates.

The full Hamiltonian for electrons and nuclei can be written as
\[
    \hat{H} = \hat{T}_\nuc + \hat{T}_\el + \hat{V}(\vec{r},\vec{R}).
\]
We then define the electronic Hamiltonian
\[
\hat{H}_\el(\vec{R}) = \hat{T}_\el + \hat{V}(\vec{r},\vec{R}),
\]
which depends \emph{parametrically} on the nuclear coordinates $\vec{R}$. The electronic wavefunctions $\phi_i(\vec{r};\vec{R})$ -- again with a parametric dependence on $\vec{R}$ -- are the eigenfunctions of the electronic Hamiltonian with energies $E_{\el,i}(\vec{R})$.

We can expand the full wavefunction $\Psi(\vec{r},\vec{R})$ as a linear combination of the electronic eigenfunctions with $\vec{R}$-dependent coefficients $\chi_i (\vec{R})$,
\begin{equation}
    \label{eq:born-huang-expansion}
    \Psi(\vec{r},\vec{R}) = \sum_i \chi_i (\vec{R}) \ket{\phi_i}.
\end{equation}
We can also think about this as representing the wavefunction not as a function of electronic and nuclear coordinates, but as an infinite-dimensional vector
\[
    \vec{\chi} = \begin{pmatrix}
        \chi_1 (\vec{R}) \\
        \chi_2 (\vec{R}) \\
        \vdots
    \end{pmatrix},
\]
where each electronic state $i$ gets its own nuclear wavefunction $\chi_i (\vec{R})$.

The Schrödinger equation in this representation (see \autoref{sec:detail-beyond-BO} for the derivation) is given by\sidenote{Note that this equation is still exact (at least if we do not discard the continuum states). The Born-Oppenheimer approximation corresponds to neglecting the non-adiabatic couplings, so that the equations for different $\chi_i (\vec{R})$ decouple from each other.}
\begin{equation}
    \label{eq:se-with-gauge-potential}
    \sum_I -\frac{\hbar^2}{2M_I} \left( \pder{}{q_I} + \vec{d}^I \right)^2 \vec{\chi} + \vec{V}\vec{\chi} = E\vec{\chi},
\end{equation}
where $\vec{d}^I$ is a matrix\sidenote{Equivalently, we could consider it an operator $\hat{d}^I$.} of non-adiabatic couplings
\[
    d_{ij}^I = \bra{\phi_i} \pder{}{q_I} \ket{\phi_j}
\]
\marginnote{Note that we have defined the non-adiabatic couplings as derivatives with respect to a single coordinate. In electronic structure literature, it is common to instead consider the gradient $\vec{\nabla}_I$ with respect to all three coordinates of a single nucleus -- see \autoref{sec:detail-beyond-BO}.}
and $\vec{V}$ is the diagonal potential energy matrix
\[
    V_{ij} = E_{\el,i}(\vec{R}) \, \delta_{ij} .
\]

To conclude, non-adiabatic effects can be fully accounted for by the \emph{derivative couplings} $\vec{d}^I$, whose effect is to change the (mechanical) momentum from $-\imag\hbar \partial_I$ to $-\imag\hbar (\partial_I + \vec{d}^I)$. The latter term is a non-diagonal matrix, which leads to coupling between different electronic states.

\marginnote{
\begin{kaocounter}{Pauli matrices}
    \begin{align*}
    \sigma_x & = \begin{pmatrix} 0 & 1 \\ 1 & 0\end{pmatrix} \\
    \sigma_y & = \begin{pmatrix} 0 & -\imag \\ \imag & 0\end{pmatrix} \\
    \sigma_z & = \begin{pmatrix} 1 & 0 \\ 0 & -1\end{pmatrix}
    \end{align*}
    We shall occasionally use $\sigma_0$ to denote the identity matrix, to make it manifest that the four matrices together span the vector space of Hermitian 2$\times$2 matrices.    
\end{kaocounter}
}

\section{Two-level systems -- diabatic and adiabatic bases}
\label{sec:diabatic}

We now have a way to separate nuclear and electronic variables in the \emph{adiabatic} basis. The adiabatic basis functions depend on nuclear positions, which leads to a coupling term in the momentum. However, the potential energy is diagonal.

We can also define the \emph{diabatic basis}, which has position-independent basis functions.\sidenote{A way to think about them physically is that the electrons follow adiabatic states when the nuclei move slowly and diabatic states when they move fast.} This comes at the cost of a non-diagonal potential energy matrix, but the mechanical momentum is simply given by $p$ without any coupling terms. This makes it the preferred choice for numerical simulations. For real-world electronic problems truncated to a few states, the diabatic basis does generally not exist -- it is not possible to fully remove the position dependence of the adiabatic basis functions.\sidenote{The truncation of the infinite amount of electronic states is key here. If we included all the states, we could use any basis we wanted, even a trivially diabatic one. \cite{domckeConicalIntersectionsElectronic2004a} Additionally, model problems with finitely many states can easily be defined in the diabatic basis.} However, even in such cases, a \enquote{locally diabatic} basis is used in simulations rather than the adiabatic one. \cite{plasserSurfaceHoppingDynamics2012,geutherTimeReversibleImplementationMASH2025}

For two--level systems, we shall denote the diabatic states $\ket{1}$ and $\ket{2}$. The diabatic potential energy matrix,
\[
\vec{V}^{\dia} =
\begin{pmatrix}
    V_{11} & V_{12} \\
    V_{21} & V_{22}
\end{pmatrix},
\]
is Hermitian and in most cases, we shall take it to be real.\sidenote{This is in principle not necessary and the formulae given here generalise easily to the complex case.} We can thus express it as a linear combination of the identity and the diabatic Pauli matrices,\sidenote{The matrices $\sigma_i^\dia$ have the form of Pauli matrices when expressed in the diabatic basis. For example, \[\sigma_z^\dia = \ket{1}\bra{1} - \ket{2}\bra{2}.\]
We shall also be using the adiabatic Pauli matrices, e.g. \[\sigma_z^\adia = \ket{+}\bra{+} - \ket{-}\bra{-}.\]
Note that $\sigma_i^\adia$ is \emph{not} the same as $\sigma_i^\dia$ transformed to the adiabatic basis.}
\[
    \vec{V}^{\dia} = \bar{V}(\vec{q}) + \kappa(\vec{q})\sigma_z^\dia + \Delta(\vec{q})\sigma_x^\dia .
\]
The Hamiltonian in the diabatic basis is then
\begin{equation}
    \label{eq:H-dia}
    \hat{H}^\dia = \sum_{i=1}^f \frac{p_i^2}{2m} + \bar{V}(\vec{q}) + \kappa(\vec{q})\,\hat{\sigma}_z^\dia + \Delta(\vec{q})\,\hat{\sigma}_x^\dia ,
\end{equation}
where we have used hats to emphasise operators acting on the electronic states. Of course, in a fully quantum treatment, the nuclear positions and momenta would also be operators.

The adiabatic basis is given by diagonalising the potential energy matrix at each value of $\vec{q}$.
\sidenote{A numerically stable way of doing this is outlined in \autoref{sec:analytical-2x2-matrix}.}
The eigenvalues of $\vec{V}^{\dia}$ are given by $E_{\pm} = \bar{V}(\vec{q}) \pm V_z(\vec{q})$ with $V_z(\vec{q}) \equiv  \sqrt{\kappa(\vec{q})^2+\Delta(\vec{q})^2}$.
The eigenfunctions corresponding to $E_{\pm}$ are denoted by $\ket{+}$ and $\ket{-}$. Since these are position-dependent, they give rise to the non-adiabatic coupling operator (see \autoref{sec:beyond-BO})
\[
    \bra{a} \hat{d}^i(\vec{q}) \ket{b} = \bra{a} \pder{}{q_i} \ket{b}.
\]
It can be shown that $d_{++}^i = d_{--}^i = 0$ and $d_{+-}^i = -d_{-+}^i$, so we can use the \emph{non-adiabatic coupling vector} $\vec{d}$ with components
\[
    d_i = \bra{+} \pder{}{q_i} \ket{-}
\]
to write
\[
    \hat{d}^i = \imag d_i \, \hat{\sigma}_y^\adia .
\]
The adiabatic Hamiltonian is therefore
\begin{equation}
    \label{eq:H-adia}
    \hat{H}^\adia = \sum_{i=1}^f \frac{(p_i + \hbar d_i \, \hat{\sigma}_y^\adia )^2}{2m} + \bar{V}(\vec{q}) + V_z(\vec{q}) \, \hat{\sigma}_z^\adia .
\end{equation}

\marginnote{
    \begin{kaocounter}{Tully's model I}
        \label{box:tully-i}
        \begin{align*}
            \bar{V}(x) &= 0 \\
            \kappa(x) & = A\tanh(Bx) \\
            \Delta(x) & = C\eul^{-Dx^2}
        \end{align*}
        This is a modified version from~\cite{ananthSemiclassicalDescriptionElectronically2007}. The parameters are $A=0.01$, $B=1.6$, $C=0.005$, $D=1.0$. We shall generally take the particle mass to be $m=2000$, unless specified otherwise.
    \end{kaocounter}
}

\begin{marginfigure}
    \centering
    \begin{tikzpicture}
    \pgfplotsset{compat=1.18}
    \begin{axis}[enlarge x limits=false,
                 tick align=outside,
                 xlabel=$x$,
                 ylabel=$E$,
                 width=0.75\textwidth,
                 scale only axis,
                 ytick={-0.01, 0, 0.01},
                 yticklabels={$-0.01$, $0$, $0.01$},
                 scaled y ticks = false,
                 xlabel style={font=\footnotesize},
                 ylabel style={font=\footnotesize,yshift=-2.3ex},
                 yticklabel style = {font=\footnotesize,xshift=0.5ex},
                 xticklabel style = {font=\footnotesize,yshift=0.5ex}
                ]

        \addplot[mark=none, line width=0.9pt, densely dotted] table[x index = 0, y index = 1] {data/tully-i.dat};
        \addplot[mark=none, line width=0.9pt, densely dotted] table[x index = 0, y index = 2] {data/tully-i.dat};
        \addplot[mark=none, line width=1.0pt, Maroon] table[x index = 0, y index = 3] {data/tully-i.dat};
        \addplot[mark=none, line width=1.0pt, Maroon] table[x index = 0, y index = 4] {data/tully-i.dat};
        
    \end{axis}

\end{tikzpicture}
    \caption{Energy levels of Tully's model~I. Full lines represent the adiabatic states, while dotted lines are the diabatic energies $V_{11}(x)$ and $V_{22}(x)$.}
    \label{fig:tully-i}
\end{marginfigure}

\marginnote{
    \begin{kaocounter}{Tully's model II}
        \label{box:tully-ii}
        \begin{align*}
            \bar{V}(x) &= -\frac 1 2 \left( A\eul^{-Bx^2} - \varepsilon \right) \\
            \kappa(x) & = -\bar{V}(x) \\
            \Delta(x) & = C\eul^{-Dx^2}
        \end{align*}
        The parameters are $A=0.1$, $B=0.28$, $C=0.015$, $D=0.05$, $\varepsilon = 0.05$.
    \end{kaocounter}
}

\begin{marginfigure}
    \centering
    \begin{tikzpicture}
    \pgfplotsset{compat=1.18}
    \begin{axis}[enlarge x limits=false,
                 tick align=outside,
                 xlabel=$x$,
                 ylabel=$E$,
                 width=0.75\textwidth,
                 scale only axis,
                 ytick={-0.05, 0, 0.05},
                 yticklabels={$-0.05$, $0$, $0.05$},
                 scaled y ticks = false,
                 xlabel style={font=\footnotesize},
                 ylabel style={font=\footnotesize,yshift=-3.0ex},
                 yticklabel style = {font=\footnotesize,xshift=0.5ex},
                 xticklabel style = {font=\footnotesize,yshift=0.5ex}
                ]

        \addplot[mark=none, line width=0.9pt, densely dotted] table[x index = 0, y index = 1] {data/tully-ii.dat};
        \addplot[mark=none, line width=0.9pt, densely dotted] table[x index = 0, y index = 2] {data/tully-ii.dat};
        \addplot[mark=none, line width=1.0pt, Maroon] table[x index = 0, y index = 3] {data/tully-ii.dat};
        \addplot[mark=none, line width=1.0pt, Maroon] table[x index = 0, y index = 4] {data/tully-ii.dat};
        
    \end{axis}

\end{tikzpicture}
    \caption{Energy levels of Tully's model~II. Full lines represent the adiabatic states, while dotted lines are the diabatic energies $V_{11}(x)$ and $V_{22}(x)$.}
\end{marginfigure}

We can illustrate the difference with the Tully models, which are one dimensional scattering models for common non-adiabatic situations. \cite{tullyMolecularDynamicsElectronic1990} Tully's model I (see Box~\ref{box:tully-i}) represents a single avoided crossing. If we initialise the particle on the left in diabat $\ket{1}$ -- where it coincides with the adiabat $\ket{-}$ -- and it moves through the crossing fast enough, it will stay on the same diabat and end up in the upper state. If the particle is initially too slow, there is a possibility of reflection. In a fully quantum treatment, the particle would be able to tunnel through, but MASH is not able to capture that.

Tully's model II (Box~\ref{box:tully-ii}) has a more complicated structure with two crossings. At low initial momenta, this is a very difficult model for quantum-classical methods due to interference effects between the two crossing points and due to the possibility of a resonance in the upper state. \cite{tullyMolecularDynamicsElectronic1990,mannouchMappingApproachSurface2023}

We shall also be using the spin--boson model (Box~\ref{box:spin-boson}), which represents a two-level system in a condensed phase environment; physically, it might describe an electron transfer process. The environment (or \emph{bath}) is modelled by a set of harmonic oscillators with a characteristic \emph{spectral density}, which become the classical degrees of freedom in our treatment.

For the spin--boson model, the adiabatic states do not make much sense as observables. Rather, we shall be interested in the diabatic populations $\ket{1}\bra{1}$ and $\ket{2}\bra{2}$, which would correspond to the amount of reactants and products in the electron transfer reaction.

\pagebreak

\begin{marginfigure}
    \centering
    \begin{tikzpicture}
    \pgfplotsset{compat=1.18}
    \begin{axis}[ymin=-2.5, ymax=2.0,
                 xmin=-3.2, xmax=3,
                 xlabel=reaction coordinate,
                 ylabel=$E$,
                 width=0.85\textwidth,
                 scale only axis,
                 ticks=none
                ]

        \addplot[mark=none, line width=1pt, Maroon, domain=-3:1] {x^2 + 2.0*x - 1.0}
            node[pos=0.2,above, right=0.5pt,Maroon] {$\ket{2}$};

        \addplot[mark=none, line width=1pt, RoyalBlue, domain=-0.414:2.414] {x^2 - 2.0*x + 1.0}
            node[pos=0.8,below, right=0.5pt,RoyalBlue] {$\ket{1}$};

        \draw[dashed, line width=0.9pt] ({axis cs:-2,-2}) -- ({axis cs:2,-2});
        \draw[<->, line width=0.75pt] ({axis cs:1,-2}) -- ({axis cs:1,0})
            node[pos=0.5,right] {$2\varepsilon$};
        
    \end{axis}

\end{tikzpicture}
    \caption{The \emph{diabatic} states of the spin boson model plotted with respect to the reaction coordinate $\sum_i c_i q_i$ of the electron transfer reaction. All the other modes act as a thermal bath and could be replaced by a friction term. This is not apparent from Box~\ref{box:spin-boson} -- the transformation to a reaction coordinate description is described in \cite{gargEffectFrictionElectron1985} and \cite{thossSelfconsistentHybridApproach2001}.}
    \label{fig:spin-boson}
\end{marginfigure}

\small

\begin{kaocounter}{Spin--boson model}
    \footnotesize
    This is a multi-dimensional model with nuclear coordinates $q_1,q_2, \dots, q_f$. We shall take $f=100$ and the mass $m=1$ for all nuclear modes.
    \label{box:spin-boson}
    \begin{align*}
        \bar{V}(\vec{q}) &= \frac 1 2 \sum_{i=1}^f \omega_i^2 q_i^2  \\
        \kappa(\vec{q}) & = \varepsilon + \sum_{i=1}^f c_i q_i \\
        \Delta(\vec{q}) & = \Delta
    \end{align*}
    The parameters $\varepsilon$~(energy bias) and $\Delta$~(coupling strength) can be changed at will, but $\omega_i$ and $c_i$ are sampled from the Debye spectral density,
    \[
        J(\omega) = \frac{\Lambda}{2}\, \frac{\omega \, \omega_c}{\omega^2 + \omega_c^2},
    \]
    discretised into $f$ modes using the scheme presented in \cite{craigProtonTransferReactions2007}. We shall initialise the nuclear degrees of freedom from the thermal Wigner distribution (Box~\ref{box:thermal-wigner}), which introduces the inverse temperature $\beta$ as another parameter. This is all consistent with the treatment in \cite{mannouchMappingApproachSurface2023} and \cite{runesonMultistateMappingApproach2023}.
\end{kaocounter}

\normalsize

\section{Observables can be represented by correlation functions}
\label{sec:observables-cfs}

We need to think about what we actually \enquote{measure} in simulations. Generally, we follow the time dependence of the expectation value of some observable $\hat{\mathcal{B}}$. In quantum mechanics, this can be written as
\begin{align*}
    \avg{\mathcal{B}}(t) & = \bra{\Psi(t)} \hat{\mathcal{B}} \ket{\Psi(t)} \\
                         & = \bra{\Psi} \eul^{\imag \hat{H}t/\hbar} \hat{\mathcal{B}} \eul^{-\imag \hat{H}t/\hbar} \ket{\Psi} \\
                         & = \bra{\Psi} \hat{\mathcal{B}}(t) \ket{\Psi} \\
                         & = \Tr{\rho \hat{\mathcal{B}}(t)},
\end{align*}
where we have used the time-evolved Heisenberg operator $\hat{\mathcal{B}}(t)$ and the initial density matrix $\rho$. Note that this has the form of a correlation function $C_{\rho\mathcal{B}}(t)$.

\begin{marginfigure}
    \centering
    \begin{tikzpicture}[
    %Global config
    >=latex,
    line width=1pt,
]

\draw (0.2, 0) -- (4, 0);

\draw[Maroon, fill=Maroon] (0.55,0) -- (0.55,1) -- node [midway,above,align=center,black] {$\rho$} (0.75,1) -- (0.75, 0) -- cycle;

\draw[domain=0.75:3.5, smooth, variable=\x, line width=0.7pt] plot ({\x}, {0.3-0.2*sin(280*\x) + 0.15*sin(521*\x)});

\draw[Maroon, fill=Maroon] ({3.5}, {0.3-0.2*sin(280*3.5) + 0.15*sin(521*3.5)}) circle(0.05) node[right,black] {$\mathcal{B}(t)$};

\draw[->, line width=0.7pt] (0.75, 0.85) -- node [midway,above,align=center] {$t$} (3.5, 0.85);
\draw[densely dotted, line width=0.7pt] (3.5, 0) -- (3.5, 1);

\end{tikzpicture}
    \caption{A simple experiment. We initialise the system in state $\rho$ and measure $\avg{\mathcal{B}}$ as a function of $t$.}
\end{marginfigure}

This result also has a straightforward experimental interpretation -- we initialise the system in state $\rho$, we wait for a period of time $t$ and then we measure the observable $\hat{\mathcal{B}}$. In fact, if our measurements do not disturb the system, we can generate the whole time series $C_{\rho\mathcal{B}}(t)$ in a single measurement session.

We shall generally use factorised initial conditions,\sidenote{This is not much of a limitation -- any $\rho$ can be written as a sum of factorised terms, e.g. by using the basis of electronic Pauli matrices.} where the initial density matrix can be written as a product of nuclear and electronic parts, $\rho = \rho_\nuc \hat{A}$. For example, we could initialise Tully's model I as a wavepacket fully in state $\ket{-}$. Very often, the operator $\hat{\mathcal{B}}$ will only act on the electronic states, so we shall write it as $\hat{B}$ and the corresponding correlation function will be
\[
    C_{AB}(t) = \Tr{\rho_\nuc \hat{A}\hat{B}(t)},
\]
where the $\rho_\nuc$ is implicit in the notation $C_{AB}$.

\section{The classical path approximation and the Bloch sphere}

\begin{marginfigure}
    \centering
    \input{images/bloch-sphere.tex}
    \caption{A trajectory for Tully's model~I in the CPA. The particle starts at $x=-10$ and moves with momentum $p=45$ to the right. The initial wavefunction is $\ket{\psi}=0.954\ket{-} + 0.3\ket{+}$. It starts in the lower hemisphere, undergoing \enquote{Larmor precession} when the states are not coupled. When it moves through the avoided crossing, it quickly changes hemispheres and continues precessing once it has leaved the coupling region. Compare with Figure~\ref{fig:tully-i}.}
    \label{fig:bloch-sphere}
\end{marginfigure}

The easiest version of quantum-classical dynamics is when we need not consider the back-action of the electrons on the nuclei at all. For example, the nuclear kinetic energy could be so large that electronic transitions do not have any effect on it. This means that the nuclei move along a path $\vec{q}=\vec{q}(t)$ and the electrons simply follow their motion -- i.e. they evolve according to the (time-dependent) electronic Hamiltonian
\[
\hat{H}_\el^\dia (t) = \kappa[\vec{q}(t)]\,\hat{\sigma}_z^\dia + \Delta[\vec{q}(t)]\,\hat{\sigma}_x^\dia
\]
or equivalently, its adiabatic version
\[
\hat{H}_\el^\adia (t) = \frac {\hbar \vec{p}(t) \cdot \vec{d}[\vec{q}(t)]} {m} \, \hat{\sigma}_y^\adia + V_z[\vec{q}(t)] \, \hat{\sigma}_z^\adia
\]
obtained by only keeping the electronic operators in \eqref{eq:H-adia}. This is known as the \emph{classical path approximation} (CPA).

The electronic dynamics of this system is equivalent to a spin--1/2 particle in a time-dependent magnetic field -- the energy separation $V_z$ is proportional to the magnetic field strength and the coupling term $\hbar\vec{p}\cdot\vec{d}/m$ is related to the transverse field that causes transitions between states. It is well known that a spin--1/2 particle can be described by a unit vector moving on the surface of the \emph{Bloch sphere} (see Figure~\ref{fig:bloch-sphere}).

The mapping between the \emph{adiabatic} wavefunction $\psi = c_+ \ket{+} + c_- \ket{-}$ and the Bloch sphere coordinates $S_x$, $S_y$ and $S_z$ is given by\sidenote{It is useful to note that the spin coordinates are equal to the expectation values of the corresponding quantum operators $\hat{\sigma}_x^\adia$, $\hat{\sigma}_y^\adia$ and $\hat{\sigma}_z^\adia$. Also, it is easily checked that for a normalised wavefunction, $S_x^2 + S_y^2 + S_z^2 = 1$.}
\begin{subequations}
    \label{eq:wf-to-spin}
    \begin{align}
        S_x & = 2\,\Re[c_+^*c_-] \\
        S_y & = 2\,\Im[c_+^*c_-] \\
        S_z & = |c_+|^2 - |c_-|^2
    \end{align}
\end{subequations}

\marginnote{The Bloch sphere can be described by the usual spherical coordinates $\theta$ and $\phi$, where we take
\begin{align*}
    S_x & = \sin\theta \cos \phi \\
    S_y & = \sin\theta \sin \phi \\
    S_z & = \cos\theta .
\end{align*}
}

The dynamics of the spin variables can be derived from the time-dependent Schrödinger equation \cite{mannouchMappingApproachSurface2023} and is given by\sidenote{
    Note that these are entirely equivalent to the Schrödinger equation -- in the simulation, we can either propagate the spin (as is done in \cite{mannouchMappingApproachSurface2023}) or the wavefunction coefficients (like \cite{runesonMultistateMappingApproach2023}). This is not the case in the unSMASH method, where the (generalised) spin variables have to be used for propagation. \cite{lawrenceSizeconsistentMultistateMapping2024a}
}
\begin{subequations}
    \label{eq:cpa}
    \begin{align}
        \dot S_x & = \frac{2\vec{p} \cdot \vec{d}}{m} \, S_z - \frac{2V_z(\vec{q})}{\hbar} \, S_y \\
        \dot S_y & = \frac{2V_z(\vec{q})}{\hbar} \, S_x \\
        \dot S_z & = -\frac{2\vec{p} \cdot \vec{d}}{m} \, S_x
    \end{align}
\end{subequations}

\section{Spin mapping -- representing states by regions of the Bloch sphere}
\label{sec:spin-mapping}

So far, every quantum state\sidenote{Note that the mapping deals with \emph{states}, not wavefunctions. Two wavefunctions that only differ by phase are the same state and are thus mapped to the same point on the Bloch sphere. In other words, there is a redundancy in the wavefunction description.} corresponds to a point on the Bloch sphere. This is perfect for purely electronic dynamics, but in order to describe quantum-classical dynamics, we shall instead represent a state by a \emph{region} of the Bloch sphere. We shall also move from wavefunctions to density matrices.\sidenote{See \autoref{sec:detail-density-matrix} for a review of density matrices.}

This is very similar to representing a wavepacket by a probability distribution in phase space\sidenote{See \autoref{sec:detail-wigner-transform} for the phase space formulation of quantum mechanics.} -- in fact, it is exactly the same thing, once the Winger transform is defined for finite systems. However, this is quite involved and we shall content ourselves with an informal description.\sidenote{The formal way has been leveraged to construct other mapping approaches -- see \cite{runesonGeneralizedSpinMapping2020}.}

In MASH, we represent the state $\hat{P}_- \equiv \ket{-}\bra{-}$ by the whole lower hemisphere of the Bloch sphere. Similarly, $\hat{P}_+\equiv \ket{+}\bra{+}$ is represented by the upper hemisphere. Since the lower hemisphere contains points with $S_z<0$, we can formally write the MASH mapping in terms of the Heaviside function,
\[
    \hat{P}_- \mapsto h(-S_z)
\]
and similarly
\[
    \hat{P}_+ \mapsto h(S_z).
\]

When we run a MASH simulation, we sample points from this initial distribution and evolve each of them according to \eqref{eq:cpa}. Finally, we use the evolved set of points to evaluate correlation functions.\sidenote{See Figure~\ref{fig:MASH-master} for a~visualisation of MASH trajectories.}

Let us try to think about estimating the $P_-$--$P_+$ correlation function, $C_{-+}(t)$. In this case, we initialise the electronic state in the state $\hat{P}_-$ and see how much is transferred to $\hat{P}_+$ after time $t$. The initial state thus corresponds to the lower hemisphere $h(-S_z)$ -- we would sample the initial spin variables uniformly from this region. After time $t$, we obtain an evolved set of points on the Bloch sphere. Naively, we could estimate the final population by counting how many points are in the upper hemisphere, which would correspond to the integral
\[
    \int \dd\vec{S}\, h(S_z)h(S_z(t)),
\]
where
\begin{equation}
    \int \dd \vec{S} \, \cdots = \frac{1}{2\pi} \int_0^\pi \dd\theta \, \sin\theta \int_0^{2\pi} \dd \phi \, \cdots
\end{equation}
is the integral over the Bloch sphere.\sidenote{Note that this is \emph{not} the usual integral over the sphere -- it has a factor of $2\pi$ rather than $4\pi$. This is so that it correctly reproduces the trace of the 2$\times$2 identity matrix, $\int \dd\vec{S} =\Tr{\hat{I}}=2$.}

It turns out this formula is inherently flawed and to rescue it, we need to introduce a \emph{weighting factor} of $2|S_z|$,\sidenote{This factor was introduced in a somewhat \textit{ad hoc} way in \cite{mannouchMappingApproachSurface2023}. We explain in \autoref{sec:detail-MASH-pauli-matrices} that it does arise quite naturally from the $\hat{\sigma}_z \mapsto \sgn(S_z)$ prescription.}
\[
    C^\MASH_{++}(t) =\int \dd\vec{S}\, 2|S_z|h(S_z)h(S_z(t)).
\]

Correlation functions for other observables can also be defined with their respective weighting factors (Box~\ref{box:MASH-estimators}). The estimators for ms-MASH are summarised in Box~\ref{box:ms-MASH-estimators} -- this form does not appear in \cite{runesonMultistateMappingApproach2023}, but is derived in \autoref{sec:detail-msMASH-spin-variables}. The estimators for general observables including nuclear coordinates are discussed in \autoref{sec:detail-MASH-estimators}.

\marginnote{
    \begin{kaocounter}{MASH estimators}
        \label{box:MASH-estimators}
        \begin{align*}
            C_{++}(t)  & = \int\dd\vec{S} \, 2|S_z| \, h(S_z) \, h(S_z(t)) \\
            C_{+x}(t)  & = \int\dd\vec{S} \, 2  h(S_z) \, S_x(t) \\
            C_{x+}(t)  & = \int\dd\vec{S} \, 2  S_x \, h(S_z(t)) \\
            C_{xx}(t)  & = \int\dd\vec{S} \, 3  S_x S_x(t)
        \end{align*}
        For the population of the lower adiabat, change $S_z\to -S_z$ and for $\sigma_y$, change $x\to y$.
    \end{kaocounter}
}

\marginnote{
    \begin{kaocounter}{ms-MASH estimators}
        \label{box:ms-MASH-estimators}
        \begin{align*}
            C_{++}(t)  & = \int\dd\vec{S} \, h(S_z)       \left( \frac 1 2 + S_z(t) \right) \\
            C_{+x}(t)  & = \int\dd\vec{S} \, h(S_z)    \, 2 S_x(t) \\
            C_{x+}(t)  & = \int\dd\vec{S} \, \sgn(S_x)    \left( \frac 1 2 + S_z(t) \right) \\
            C_{xx}(t)  & = \int\dd\vec{S} \, \sgn(S_x) \, 2 S_x(t)
        \end{align*}
    \end{kaocounter}
}

\section{MASH: the mapping approach to surface hopping}
\label{sec:MASH-dynamics}

MASH is a combination of ideas from spin mapping and surface hopping. Like spin mapping, the dynamics of every single trajectory is deterministic, and the \enquote{fuzzy} nature of the quantum system is represented by sampling over the Bloch sphere. However, the dynamics is inspired by the surface hopping idea of propagating the nuclei on the potential energy surface corresponding to the \emph{active state} -- the adiabatic state with largest $|c_n|^2$.

The advantage of this approach is that it is consistent with the CPA and the QCLE (like other mapping approaches), but thanks to the surface-hopping terms, it thermalises correctly in the long-time limit. Practically, it is at least as good as FSSH and very often better -- both in terms of accuracy and in terms of computational cost.~\cite{mannouchMappingApproachSurface2023}

\marginnote{
    \begin{kaocounter}{The Gaussian wavepacket}
        \label{box:gaussian-wavepacket}
        \vspace*{1em}
        We shall generally initialise the Tully models with a wavepacket centred at $q_0$ and with average momentum $p_0$. The wavefunction is given by
        \[
            \psi(x) = \left(\frac{\gamma}{\pi}\right)^{1/4} \eul^{-\frac \gamma 2 (x-q_0)^2} \eul^{\imag p_0 x / \hbar}.
        \]
        The corresponding Wigner distribution can be obtained using \eqref{eq:wigner-transform-pure-state},
        \[
            \rho_W(p,q) = 2 \, \eul^{-\gamma(q-q_0)^2} \, \eul^{-(p-p_0)^2/\gamma\hbar^2} .
        \]
    \end{kaocounter}
}

\marginnote{
    \begin{kaocounter}{The thermal Wigner distribution}
        \label{box:thermal-wigner}
        \vspace*{1em}
        The spin--boson model has to be initialised in thermal equilibrium with inverse temperature $\beta$.
        This corresponds to the state $\rho = \eul^{-\beta \hat{H}} / Z$. For a harmonic oscillator, the Wigner transform of this density operator can be obtained as \cite{tannorIntroductionQuantumMechanics2007}
        \[
            \rho_W(p,q) = 2\zeta \eul^{-\zeta (p^2 + m^2\omega^2 q^2) / m\omega \hbar}
        \]
        with $\zeta = \tanh(\hbar \omega \beta / 2)$. For the spin-boson model, we sample each oscillator from this distribution independently.
    \end{kaocounter}
}

Just like in any other mixed quantum-classical approach, the nuclear degrees of freedom are initialised from the Wigner distribution corresponding to $\rho_\nuc$. In our models, this corresponds to sampling the initial $p$ and $q$ from appropriate normal distributions (see Box~\ref{box:gaussian-wavepacket} and~\ref{box:thermal-wigner}).

For the electronic variables, we need to evaluate the integral $\int\dd\vec{S}\cdots$ to get the correlation functions. This is achieved by a Monte Carlo method with uniform\sidenote{The statistical error of this method is explored in \autoref{appendix:variance}. Importance sampling is also possible -- see \cite{lawrenceRecoveringMarcusTheory2024} for an example.} \emph{full-sphere sampling} of the spin coordinates. Thus, for each trajectory, the electronic state is initialised as a random point on the Bloch sphere -- the estimators then take care of weighting each contribution correctly.

In MASH, the electronic and the nuclear subsystems are propagated differently. The electronic state for a given trajectory simply evolves according to \eqref{eq:cpa} with the instantaneous values of $\vec{p}$ and $\vec{q}$. The nuclear degrees of freedom evolve so as to conserve the \emph{MASH energy function}
\begin{equation}
    \label{eq:mash-energy}
    E(\vec{p}, \vec{q}, \vec{S}) = \sum_i \frac{p_i^2}{2m} + \bar{V}(\vec{q}) + V_z(\vec{q})\,\sgn (S_z) .
\end{equation}
This says that if the electrons are mainly in the upper state, $\sgn (S_z)=1$ and the potential energy is the one corresponding to the upper surface, $\bar{V}(\vec{q}) + V_z(\vec{q})$. If the lower state is more populated, the nuclei propagate on the lower surface, $\bar{V}(\vec{q}) - V_z(\vec{q})$. This \enquote{quantisation}, common to MASH and FSSH, is ensured by the sign function.

The energy expression allows us to derive the MASH force -- which we actually use in the simulation\sidenote{We explain how the MASH equations of motion are propagated in \autoref{appendix:integrators}.} -- by requiring that it is conserved during time propagation. This gives \cite{mannouchMappingApproachSurface2023}
\begin{equation}
    \label{eq:mash-force}
    F_i (\vec{q}, \vec{S}) = -\pder{\bar{V}}{q_i} - \pder{V_z}{q_i}\,\sgn (S_z) + 4V_z(\vec{q}) d_i(\vec{q}) \, S_x \delta(S_z).
\end{equation}
The first two terms are simply due to propagation on the active surface. The delta function term comes from differentiating the sign function and is responsible for \emph{momentum rescaling} when the active surface changes.

% !TEX root = main.tex

\chapter{Uniqueness of MASH}
\label{chapter:uniqueness}

MASH has many properties one would want from a non-adiabatic dynamics method -- correctness in the CPA, consistency with the QCLE and correctness in the long-time limit. We would like know whether MASH is in any way a \enquote{unique} method possessing these properties -- that would explain why it works so well in cases where other methods fail.

We shall consider methods that use the general mapping\sidenote{We use the Pauli matrix basis rather than $P_+$ and $P_-$, as is common in the MASH literature. The conversion between these can be found in \autoref{sec:detail-MASH-pauli-matrices}.}
\begin{align*}
    \sigma_0 & \mapsto 1 \\
    \sigma_x & \mapsto S_x \\
    \sigma_y & \mapsto S_y \\
    \sigma_z & \mapsto f(S_z)
\end{align*}
and show that while we can still make the resulting method correct in the CPA and in the short-time limit, only MASH dynamics is able to correctly recover the long-time equilibrium distribution. In this prescription, $f(x)=\sgn(x)$ corresponds to MASH.\sidenote{Note that we could think about $f(x) = \tanh(\mu x)$ as a kind of \enquote{smooth MASH}, since $f(x)$ then approaches the sign function as $\mu\to\infty$.}
On physical grounds, we would expect $f(x)$ to be an odd non-decreasing function -- the former follows from the reflection symmetry of the Bloch sphere and the latter seems natural if both $\sigma_z$ and $f(S_z)$ represent the population difference between the two states.

We shall first see how estimators can be found for a general $f(x)$ and show that $f(x)=\sgn(x)$ is the only physically reasonable choice that recovers the correct long-time limit. We shall then see that by using this prescription for the dynamics, but allowing freedom in the estimators, other \enquote{versions of MASH} can be found that regain some or all of the properties of original MASH. Finally, we shall see that MASH itself can use this arbitrariness to correctly estimate non-adiabatic rates.

\section{Estimators correct in the CPA can always be found}
\label{sec:all-estimators-are-cpa-consistent}

Let us firstly show that estimators consistent with the CPA can be found for any $f(x)$. The coherence--coherence estimator can clearly be kept from MASH (see Box~\ref{box:MASH-estimators}) with the weighting factor $W_{CC}=3$.

For the $\sigma_z$--$\sigma_i$ correlation function, we can write
\[
    \int \dd \vec{S}\, W f(S_z) S_i(t),
\]
with a yet unknown weighting factor $W$ independent of $\vec{S}$. Following the same method as Appendix B in \cite{mannouchMappingApproachSurface2023}, we can show that\sidenote[][*-2]{Two things can be noted. Firstly, this excludes functions where the integral equals zero -- e.g. any even function will never work. Secondly, for $f(u)=\sgn(u)$, this gives $W=2$, which agrees with MASH.}
\[
    W = \frac{2}{\int_{-1}^1 \dd u \, uf(u)} .
\]

Since the weighting factor is a constant, we can also use the time-reversed estimator to obtain
\[
    \int \dd \vec{S}\, W  S_i \, f(S_z(t))
\]
for the $\sigma_i$--$\sigma_z$ correlation function.

This gives us two conflicting estimators for $C_{zz}(t)$, one of which corresponds to MASH and the other to ms-MASH.\sidenote{Or MASH with $2|S_z(t)|$ as a weighting factor, which was proposed in \cite{mannouchMappingApproachSurface2023}.} We can choose either one.

One possible choice of population--population estimators is therefore
\begin{align*}
    C_{II}(t) & = \int \dd\vec{S} = 2\\
    C_{zI}(t) & = \int \dd\vec{S} \, f(S_z) = 0\\
    C_{Iz}(t) & = \int \dd\vec{S} \, W S_z(t) \\
    C_{zz}(t) & = \int \dd\vec{S} \, W f(S_z)\, S_z(t)
\end{align*}
which constistently maps
\begin{align*}
    \hat{I} &\mapsto 1 \\
    \hat{\sigma}_z &\mapsto f(S_z) \\
    \hat{\sigma}_z(t) &\mapsto W S_z(t).
\end{align*}

For $f(u)=\sgn(u)$, this gives us a mixture of ms-MASH and MASH-like estimators. Specifically, we obtain the population--population CFs from ms-MASH, but the coherence--population and coherence--coherence CFs from MASH. In fact, the next sections will show that this mixed method satisfies both the QCLE and is correct in the long-time limit.\sidenote{Importantly, this means that the only thing standing between ms-MASH and the QCLE is the coherence--population estimator \[C_{x+}(t) = \int\dd\vec{S} \, \sgn(S_x) \left(\frac 1 2 + S_z(t) \right).\] However, giving up this estimator also means giving up unitarity.}

Note also that the original MASH estimators are not covered by this method, since they do not represent the identity consistently in this way. We could obtain them by requiring that the populations are estimated by $[1+ f(\pm S_z)]/2$, but the resulting integrals seem to be intractable except in the special case $f(u) = \sgn(u)$. We shall therefore not attempt to do so.

\section{Almost all mappings are correct in the short-time limit}
\label{sec:all-estimators-are-qcle-consistent}

The MASH energy function is now given by
\[
    E(\vec{p}, \vec{q}, \vec{S}) = \sum_i \frac{p_i^2}{2m} + \bar{V}(\vec{q}) + V_z(\vec{q})\,f (S_z),
\]
which leads to the MASH force
\[
    F_i (\vec{q}, \vec{S}) = -\pder{\bar{V}}{q_i} - \pder{V_z}{q_i}\,f (S_z) + 2V_z(\vec{q}) d_i(\vec{q}) \, S_x f'(S_z).
\]

Consistency with the QCLE -- which is equivalent to correctness in the $t\to 0$ limit -- is intuitively a statement that the estimators are consistent with the dynamics generated by this force. 

In \autoref{appendix:qcle}, we show that -- for the estimators in the previous section -- any $f(x)$ works, as long as it satisfies four conditions:
\[
    \int_{-1}^1 \dd u\, f(u) = 0,
\]
\[
    \int_{-1}^1 \dd u\, uf^2(u) = 0,
\]
\[
    \int_{-1}^1 \dd u\, u^2 f(u) = 0
\]
and
\[
    \int_{-1}^1 \dd u\, f^2(u) = 2.
\]
The first three are satisfied by any odd function and the last of them can be achieved by scaling $f(x)$ in an appropriate way.\sidenote{This is related to \enquote{changing the radius of the spin sphere} in other mapping approaches. \cite{mannouchMappingApproachSurface2023,amatiDetailedBalanceNonadiabatic2023}}

To summarise our results so far, if $f(x)$ is an appropriately scaled odd function, we have estimators that satisfy both the CPA and the QCLE.

\section{Only MASH can thermalise correctly}

Following \cite{amatiDetailedBalanceMixed2023}, we can prove that smooth MASH with the mapping $\sigma_z \mapsto f(S_z)$ conserves the classical Boltzmann distribution\sidenote{For a discussion of the long-time limit, see \autoref{sec:detail-long-time}. The proof in \cite{amatiDetailedBalanceMixed2023} assumes ergodicity, it does not show that the correct long-time limit is attained in low-dimensional microcanonical models (which would be quite a miracle).} if we can show that
\[
    \frac{\int_{-1}^1 \dd u \, f(u)\,\eul^{-\alpha f(u)}}{\int_{-1}^1 \dd u \, \eul^{-\alpha f(u)}} = -\frac{\sinh\alpha}{\cosh\alpha},
\]
where $\alpha\equiv \beta V_z$. This is an expression for the long-time limit of the $C_{Iz}(t)$ correlation function -- it was shown that it is sufficient if this one is reproduced correctly.

We can integrate with respect to $\alpha$ to obtain
\[
    \int_{-1}^1 \dd u \, \eul^{-\alpha f(u)} = \lambda \cosh \alpha
\]
with an integration constant $\lambda$. Setting $\alpha = 0$ then gives $\lambda=2$.

We shall now prove that this identity can only be satisfied by a function whose absolute value is one everywhere on the interval $(-1,1)$.\sidenote[][*0]{Except, of course, on a set of measure zero.} Expanding into powers of $\alpha$ gives the requirement
\[
    \int_{-1}^1 \dd u \, f(u)^n = \begin{cases}
        0 & n\;\text{odd}\\
        2 & n\;\text{even}
    \end{cases}
\]

Suppose that $|f(u)|<1$ everywhere on the interval. Then,
\[
    \int_{-1}^{1} \dd u\, f(u)^2 = \int_{-1}^{1} \dd u\, |f(u)|^2 <\int_{-1}^{1} \dd u = 2,
\]
which is a contradiction. Thus $|f(u)|\geq 1$ on at least some subinterval.

Suppose that $|f(u)|>1+\varepsilon$ on an interval of length $L$. Then, as $n\to\infty$,
\[
    \int_{-1}^{1} \dd u\, f(u)^{2n} = \int_{-1}^{1} \dd u\, |f(u)|^{2n} \geq L(1+\varepsilon)^{2n} \to \infty,
\]
since the integrand is everywhere positive, so the contribution cannot be cancelled by any other part of the function. But a sequence diverging to infinity cannot be equal to two everywhere. This holds for any $\varepsilon > 0$ and we must therefore have $|f(u)|=1$, as required.

Since $(-1)^n = -1$ for odd $n$, all the odd order integrals vanish if
\[
    \int_{-1}^{1} \dd u\, f(u) = 0.
\]

These conditions are satisfied by the sign function, as we already knew from \cite{amatiDetailedBalanceMixed2023}. Infinitely many other functions exist which satisfy these requirements; however, their relevance is unclear.\sidenote{For example, the comb-like function in Figure~\ref{fig:comb-function}.} In any case, the sign function is the only non-decreasing function out of the possibilities.\sidenote{A non-decreasing function can make at most one \enquote{jump} from $-1$ to $1$. This jump has to be at $x=0$ for the integral to vanish.}

\begin{marginfigure}
    \centering
    \begin{tikzpicture}
    \draw[->] (0,-1.4) -- (0,1.4) node[above] {$y$};
    \draw[->] (-1.4,0) -- (1.4,0) node[right] {$x$};

    \draw[line width=0.9pt, Maroon] (-1,-1) -- (-0.5,-1) -- (-0.5,1) -- (0,1) -- (0,-1) -- (0.5,-1) -- (0.5,1) -- (1,1) node[right] {$f(x)$};
\end{tikzpicture}
    \caption{An unphysical $f(x)$ that conserves the Boltzmann distribution. Mathematically, we could write it as $f(x) = \sgn(4x^3 - x).$}
    \label{fig:comb-function}
\end{marginfigure}

There is at least one caveat to the above proof -- it assumed that the function $f(u)$ is used both in the dynamics and as an estimator for $\sigma_z$. However, we have seen with ms-MASH that other functions can be used as estimators -- we shall tackle that problem below. Before that, we note that the proof still shows that MASH \emph{dynamics} is the only (physical) way to correctly reproduce the partition function
\[
    \int_{-1}^1 \dd u \, \eul^{-\alpha f(u)} = 2 \cosh \alpha.
\]

Let us therefore think about the related problem of using $f(S_z)=\sgn(S_z)$ for the dynamics, but a different function $g(S_z)$ as an estimator for $\sigma_z$. This leads to the condition
\[
    \int_{-1}^1 \dd u \, g(u)\,\eul^{-\alpha \sgn(u)} = -2 \sinh \alpha.
\]
Expanding the $\sinh$ in terms of exponentials gives us two necessary and sufficient conditions for this identity to hold,
\[
    \int_0^1 g(t) \, \dd t = 1
\]
and
\[
    \int_{-1}^0 g(t) \, \dd t = -1 .
\]
This is satisfied by $\sgn(u)$, but also by many other functions -- in particular by $g(u)=2u$. Remember that the prescription $\sigma_z(t) \mapsto 2S_z(t)$ is used for the final observable in the ms-MASH estimators. This confirms the results of~\cite{runesonMultistateMappingApproach2023} that ms-MASH is also correct in the long-time limit.\sidenote{It also confirms that MASH with $2|S_z(t)|$ as the weighting factor gives the correct long-time limit, which means that MASH satisfies detailed balance as $t\to\infty$. \cite{amatiDetailedBalanceMixed2023}}

\section{Many other estimators are consistent with the CPA}
\label{sec:other-cpa-estimators}

We shall now show that even in the case $f(x)=\sgn(x)$, we have significant freedom in the choice of estimators. In the original paper \cite{mannouchMappingApproachSurface2023}, the estimators are derived in what almost seems to be an \textit{ad hoc} way -- we simply \enquote{guess} the weighing factors and show that they are consistent with the CPA. This procedure is highly non-unique -- both MASH and ms-MASH estimators are exact in the CPA and we can presumably invent many other ones.

For example, we could also apply the weighting factor $\lambda|S_z|$ to the coherence--coherence or population--coherence CFs. We can convince ourselves by simple integration that
\[
    C^{\mathrm{CPA}}_{xx}(t) = \int \dd \vec{S}\, 8|S_z|S_x S_x(t)
\]
and that\sidenote{Interestingly, this one gives the \enquote{natural} estimator $C_{zx}(t) \mapsto \int \dd \vec{S}\, 3S_z S_x(t)$.}
\[
    C^{\mathrm{CPA}}_{+x}(t) = \int \dd \vec{S}\, 3|S_z|h(S_z) S_x(t).
\]
By time reversibility, the former suggests that $8|S_z(t)|$ would be an equally good weighting factor for $C_{xx}(t)$ and the latter gives us an estimator for the coherence--population correlation function,
\[
    C^{\mathrm{CPA}}_{x+}(t) = \int \dd \vec{S}\, 3|S_z(t)|S_x h(S_z(t)).
\]

While the form of these estimators is very different compared to MASH or ms-MASH, they all give very similar results for both the Tully models and for the spin--boson model in various parameter regimes -- the results are collected in Appendix~\ref{appendix:all-estimators}.

\section{ms-MASH estimates derivatives consistently}
\label{sec:consistent-derivatives}

We shall now see whether MASH and ms-MASH are able to consistently estimate derivatives of correlation functions. This is not just a theoretical question -- it was inspired by the relationships between the side--side and flux--side correlation functions from rate theory. \cite{henriksenTheoriesMolecularReaction2018}

For a process where the electronic state changes from $\ket{-}$ to $\ket{+}$, the rate can be measured by following the correlation function $C_{-+}(t) = \Tr{\rho_\nuc \hat{P}_- \hat{P}_+(t)}$. This is similar to the side--side correlation function -- the initial operator puts us in the reactant region and the final operator measures the quantity of products. Its derivative, the flux--side correlation function, would tell us the rate directly.

\begin{marginfigure}[*0]
    \begin{tikzpicture}

    \matrix (m) [matrix of math nodes,row sep=3em,column sep=4em,minimum width=2em] {
        C_{zz}(t) & C^\MASH_{zz}(t) \\
        \dot{C}_{zz}(t) & ??? \\};
    \path[-stealth]
        (m-1-1) edge node [left] {$\der{}{t}$} (m-2-1)
                edge node [above] {\small \textsc{mash}} (m-1-2)
        (m-2-1) edge node [below] {\small \textsc{mash}} (m-2-2)
        (m-1-2) edge node [right] {$\der{}{t}$} (m-2-2);
    
\end{tikzpicture}
\end{marginfigure}
In MASH, it is somewhat easier to consider the $\sigma_z$--$\sigma_z$ correlation function rather than the population--population one. We can take two approaches to estimating the derivative. Either we estimate $C_{zz}(t)$ first and then differentiate or we first find the exact quantum mechanical expression for $\dot{C}_{zz}(t)$ and use that in the MASH estimators.

By some algebraic manipulation, we can find the exact derivative
\begin{align*}
    \dot{C}_{zz}(t) & = \Tr{\rho_\nuc \sigma_z \dot\sigma_z(t)} \\
                    & = \Tr{\rho_\nuc \sigma_z \frac{-\vec{p}(t)\cdot\vec{d}(t) - \vec{d}(t)\cdot\vec{p}(t)}{m} \sigma_x(t)},
\end{align*}
where $\vec{p}$ and $\vec{d}$ are still quantum-mechanical operators, so they do not commute with each other. Taking the partial Wigner transform and the classical limit of $\vec{p}(t)\cdot\vec{d}(t) + \vec{d}(t)\cdot\vec{p}(t)$ simply gives $2\vec{p}_t\cdot\vec{d}(\vec{q}_t)$.

\marginnote{For spin boson and similar models, it might be more natural to use the diabatic correlation function $\Tr{\rho_\nuc \sigma^\dia_z \sigma^\dia_z(t)}$. Its derivative is easily shown to be $\Tr{\rho_\nuc \sigma^\dia_z \, 2\Delta(\vec{q}(t)) \sigma^\dia_y(t)}$. The results are similar to the adiabatic case in the case of the spin boson model -- ms-MASH reproduces the derivative perfectly, whereas MASH gives small discrepancies. (Data not shown.)}

The estimator for the exact $\dot{C}_{zz}(t)$ is the same in MASH and ms-MASH and is given by
\[
    \frac{1}{(2\pi\hbar)^f}\int \dd\vec{p}\,\dd\vec{q} \, \rho(\vec{p},\vec{q}) \int \dd \vec{S} \, 2\,\sgn(S_z) S_x(t) \, \frac{-2\vec{p}_t\cdot\vec{d}(\vec{q}_t)}{m}.
\]

It is easily seen that differentiating\sidenote{The derivative of $S_z(t)$ is given by~\eqref{eq:cpa}.} the ms-MASH estimator for $C_{zz}(t)$ (given in \autoref{sec:detail-MASH-pauli-matrices}) results in exactly this expression -- ms-MASH is consistent with regard to estimating derivatives. Intuitively, this is a~consequence of unitarity -- we can think about derivatives as converting populations to coherences, and since ms-MASH treats these on an equal footing, it is automatically consistent.

On the other hand, differentiating the MASH estimator for $C_{zz}(t)$ yields
\[
    \frac{1}{(2\pi\hbar)^f}\int \dd\vec{p}\,\dd\vec{q} \, \rho(\vec{p},\vec{q}) \int \dd \vec{S} \, 4S_z \delta(S_z(t))S_x(t) \, \frac{-2\vec{p}_t\cdot\vec{d}(\vec{q}_t)}{m}.
\]
This looks completely different from the correct version, but despite the seeming inconsistency, they give virtually identical results in most cases -- in fact, it has to be so if the MASH correlation functions represent the exact ones accurately.

\marginnote{From the MASH derivative estimator, we can also obtain the prescription
\[
    C_{zx}(t) \mapsto \int \dd \vec{S} \, 4S_z \delta(S_z(t))S_x(t)
\]
by removing the nuclear observables. Simple integration confirms that it is exact in the CPA -- yet another example of such an estimator! However, using this one in practice would be quite difficult, owing to the delta function at time $t$. The time-reversed version
\[
    C_{xz}(t) \mapsto \int \dd \vec{S} \, 4\delta(S_z)S_xS_z(t)
\]
would be more practical, as it corresponds to sampling from the equator of the Bloch sphere. In this case, the trajectory would make a \enquote{half--hop} at the beginning, so we would need to make sure to rescale its momentum in the proper way.
}

\begin{figure}[H]
    \centering
    \begin{tikzpicture}
    \begin{axis}[enlarge x limits=false,
                 tick align=outside,
                 xlabel=$t$,
                 ylabel=$\mathrm{d}C_{zz}(t)/\mathrm{d}t$,
                 width=0.75\textwidth,
                 scale only axis
                ]

        \addplot[mark=none, line width=1pt, dashed] table[x index = 0, y index = 1] {data/derivatives-p25-exact.dat};
        \addplot[mark=none, line width=1.3pt, RoyalBlue]  table[x index = 0, y index = 1] {data/derivatives-p25-MASH.dat};
        \addplot[mark=none, line width=1.3pt, Maroon]   table[x index = 0, y index = 2] {data/derivatives-p25-MASH.dat};
        
        \addplot[mark=none, line width=0.5pt] coordinates {(1200,-0.006)(1310,-0.006)(1310,-0.0052)(1200,-0.0052)(1200,-0.006)};
        \addplot[mark=none, line width=0.5pt] coordinates {(1200,-0.0052)(850,-0.004)};
        \addplot[mark=none, line width=0.5pt] coordinates {(1200,-0.006)(850,-0.00638)};

        \addplot[mark=none, line width=0.5pt] coordinates {(1350,-0.0002)(1650,-0.0002)(1650,0.0008)(1350,0.0008)(1350,-0.0002)};
        \addplot[mark=none, line width=0.5pt] coordinates {(1350,-0.0002)(1700,-0.001)};
        \addplot[mark=none, line width=0.5pt] coordinates {(1650,-0.0002)(2443,-0.001)};

        \coordinate (inset1top) at (axis cs:850,-0.004);
        \coordinate (inset1bot) at (axis cs:850,-0.00638);

        \coordinate (inset2right) at (axis cs:1700,-0.001);
        \coordinate (inset2left)  at (axis cs:2443,-0.001);

    \end{axis}

    \pgfmathsetmacro{\insetheight}{distance("inset1top","inset1bot")}

    \begin{axis}[at={(inset1top)},anchor={outer north east},scale only axis,ticks=none,height=\insetheight]
        \addplot[mark=none, line width=1pt] table[x index = 0, y index = 1] {data/derivatives-p25-exact-closeup1.dat};
        \addplot[mark=none, line width=1pt, RoyalBlue]  table[x index = 0, y index = 1] {data/derivatives-p25-MASH-closeup1.dat};
        \addplot[mark=none, line width=1pt, Maroon]   table[x index = 0, y index = 2] {data/derivatives-p25-MASH-closeup1.dat};
    \end{axis}

    \pgfmathsetmacro{\insetwidth}{distance("inset2right","inset2left")}

    \begin{axis}[at={(inset2right)},anchor={outer north west},scale only axis,ticks=none,width=\insetwidth]
        \addplot[mark=none, line width=1pt] table[x index = 0, y index = 1] {data/derivatives-p25-exact-closeup2.dat};
        \addplot[mark=none, line width=1pt, RoyalBlue]  table[x index = 0, y index = 1] {data/derivatives-p25-MASH-closeup2.dat};
        \addplot[mark=none, line width=1pt, Maroon]   table[x index = 0, y index = 2] {data/derivatives-p25-MASH-closeup2.dat};
    \end{axis}

\end{tikzpicture}
    \caption{The derivative of $C_{zz}(t)$ for Tully's model II with $p_0=25$, $q_0=-15$ and $\gamma=0.5$. Insets show that the MASH estimator for the exact derivative (red) is not universally more or less accurate than the derivative of the MASH $C_{zz}(t)$ (blue). In this case, the agreement with exact results (black) is almost perfect.}
    \label{fig:plot-derivatives-p25}
\end{figure}
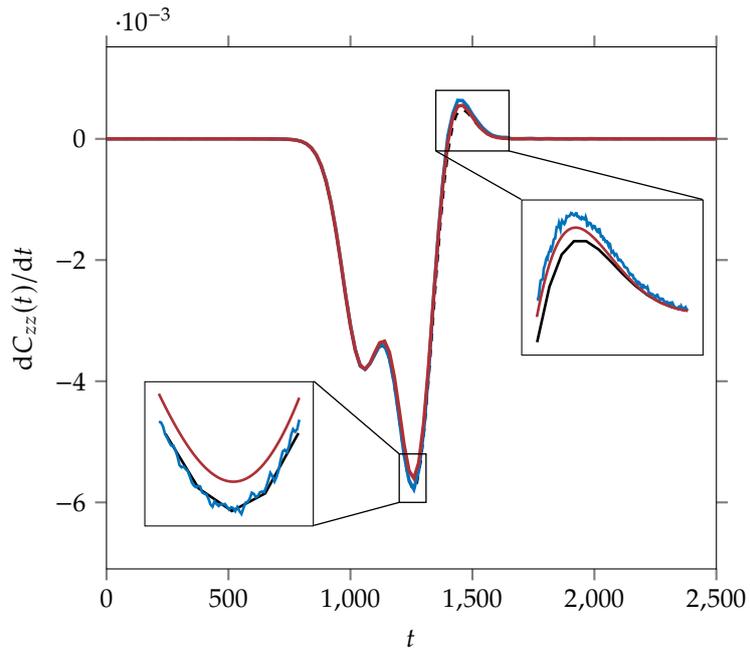

All of the above reasoning is confirmed numerically for Tully's model~II in Figures~\ref{fig:plot-derivatives-p10} and~\ref{fig:plot-derivatives-p25}. The former shows a very difficult case for MASH ($p_0=10$) and it is thus not surprising that the two estimators disagree. For $p_0=25$, where MASH works much better, the discrepancy is negligible. It cannot be said that one estimator is better than the other -- we can find regions that are better reproduced by either one.

The derivative of the MASH correlation function has been recently used to estimate electron transfer rates from the spin-boson model and has been shown to agree with Marcus theory in the appropriate limit. \cite{lawrenceRecoveringMarcusTheory2024} This is another confirmation that if MASH provides a good description of the dynamics in the first place, the derivative will be correct as well.

\begin{figure}[H]
        \centering
        \begin{tikzpicture}
    \begin{axis}[enlarge x limits=false,
                 tick align=outside,
                 xlabel=$t$,
                 ylabel=$\mathrm{d}C_{zz}(t)/\mathrm{d}t$,
                 width=0.75\textwidth,
                 scale only axis
                ]

        \addplot[mark=none, line width=1pt, dashed] table[x index = 0, y index = 1] {data/derivatives-p10-exact.dat};
        \addplot[mark=none, line width=1.3pt, RoyalBlue]  table[x index = 0, y index = 1] {data/derivatives-p10-MASH.dat};
        \addplot[mark=none, line width=1.3pt, Maroon]   table[x index = 0, y index = 2] {data/derivatives-p10-MASH.dat};
        
    \end{axis}

\end{tikzpicture}
        \caption{The derivative of $C_{zz}(t)$ for Tully's model II with $p_0=10$, $q_0=-15$ and $\gamma=0.5$. The estimator for the exact derivative (red) and the derivative of MASH $C_{zz}(t)$ (blue) are compared with the exact quantum-mechanical results (black) obtained using Chebyshev propagation. The derivative of ms-MASH $C_{zz}(t)$ coincides perfectly with the red curve, as it ought to. The results are an average over $10^8$ trajectories.}
        \label{fig:plot-derivatives-p10}
\end{figure}
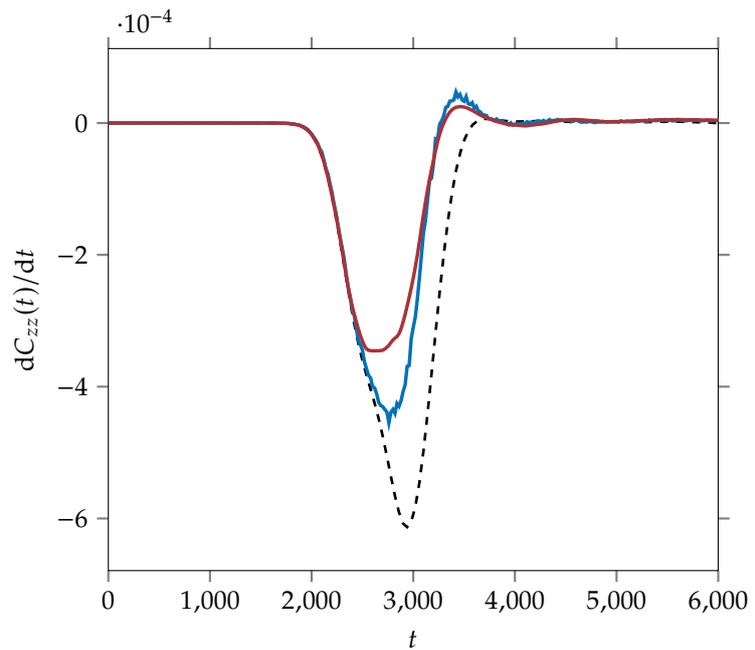

% !TEX root = main.tex

\chapter{Multi-time TCFs -- does a quantum jump help?}
\label{chapter:quantum-jump}

\begin{marginfigure}
    \centering
    \begin{tikzpicture}[
    %Global config
    >=latex,
    line width=1pt,
]

\draw (0.2, 0) -- (4.2, 0);

\draw[Maroon, fill=Maroon] (0.55,0) -- (0.55,1) -- node [midway,above,align=center,black] {$\rho$} (0.75,1) -- (0.75, 0) -- cycle;

\draw[domain=0.75:2, smooth, variable=\x, line width=0.7pt, densely dotted] plot ({\x}, {0.3-0.2*sin(280*\x) + 0.15*sin(521*\x)});

\draw[Maroon, fill=Maroon] (2,0) -- (2,1) -- node [midway,above,align=center,black] {$\hat{\mathcal{U}}$} (2.2,1) -- (2.2, 0) -- cycle;

\draw[domain=2.2:3, smooth, variable=\x, line width=0.7pt] plot ({\x}, {0.35-0.16*sin(380*\x) - 0.11*sin(821*\x)});

\draw[Maroon, fill=Maroon] ({3}, {0.35-0.16*sin(380*3) - 0.11*sin(821*3)}) circle(0.05)  node[right,black] {$\mathcal{B}(t_0+t_1)$};

\draw[->, line width=0.7pt] (0.75, 0.85) -- node [midway,above,align=center] {$t_0$} (2, 0.85);
\draw[->, line width=0.7pt] (2.2, 0.85) -- node [midway,above,align=center] {$t_1$} (3, 0.85);
\draw[densely dotted, line width=0.7pt] (3, 0) -- (3, 1);

\tikzset{shift={(0,-1.85)}}

\draw (0.2, 0) -- (4.2, 0);

\draw[Maroon, fill=Maroon] (0.55,0) -- (0.55,1) -- node [midway,above,align=center,black] {$\rho$} (0.75,1) -- (0.75, 0) -- cycle;

\draw[domain=0.75:2.5, smooth, variable=\x, line width=0.7pt, densely dotted] plot ({\x}, {0.3-0.2*sin(280*\x) + 0.15*sin(521*\x)});

\draw[Maroon, fill=Maroon] (2.5,0) -- (2.5,1) -- node [midway,above,align=center,black] {$\hat{\mathcal{U}}$} (2.7,1) -- (2.7, 0) -- cycle;

\draw[domain=2.7:3.5, smooth, variable=\x, line width=0.7pt] plot ({\x}, {0.3-0.16*sin(380*\x) + 0.15*sin(721*\x)});

\draw[Maroon, fill=Maroon] ({3.5}, {0.3-0.16*sin(380*3.5) + 0.15*sin(721*3.5)}) circle(0.05);

\draw[->, line width=0.7pt] (0.75, 0.85) -- node [midway,above,align=center] {$t_0$} (2.5, 0.85);
\draw[->, line width=0.7pt] (2.7, 0.85) -- node [midway,above,align=center] {$t_1$} (3.5, 0.85);
\draw[densely dotted, line width=0.7pt] (3.5, 0) -- (3.5, 1);

\tikzset{shift={(0,-1.85)}}

\draw (0.2, 0) -- (4.2, 0);

\draw[Maroon, fill=Maroon] (0.55,0) -- (0.55,1) -- node [midway,above,align=center,black] {$\rho$} (0.75,1) -- (0.75, 0) -- cycle;

\draw[domain=0.75:3, smooth, variable=\x, line width=0.7pt, densely dotted] plot ({\x}, {0.3-0.2*sin(280*\x) + 0.15*sin(521*\x)});

\draw[Maroon, fill=Maroon] (3,0) -- (3,1) -- node [midway,above,align=center,black] {$\hat{\mathcal{U}}$} (3.2,1) -- (3.2, 0) -- cycle;

\draw[domain=3.2:4, smooth, variable=\x, line width=0.7pt] plot ({\x}, {0.3-0.16*sin(380*\x) + 0.15*sin(1021*\x)});

\draw[Maroon, fill=Maroon] ({4}, {0.3-0.16*sin(380*4) + 0.15*sin(1021*4)}) circle(0.05);

\draw[->, line width=0.7pt] (0.75, 0.85) -- node [midway,above,align=center] {$t_0$} (3, 0.85);
\draw[->, line width=0.7pt] (3.2, 0.85) -- node [midway,above,align=center] {$t_1$} (4, 0.85);
\draw[densely dotted, line width=0.7pt] (4, 0) -- (4, 1);

\end{tikzpicture}
    \caption{A two-dimensional experiment. We initialise the system in state $\rho$ and apply $\hat{\mathcal{U}}$ pulses at various times $t_0$. We then measure $\avg{\mathcal{B}}$ during $t_1$ evolution. The dotted line represents that $\avg{\mathcal{B}}$ also evolves during $t_0$, but we do not measure that.}
\end{marginfigure}

Some experiments are too complicated to be described by the simple TCFs in \autoref{sec:observables-cfs} -- instead of just measuring $\mathcal{B}$ over time, we might apply a pulse in the middle of the measurement. In~most cases, this is equivalent to acting on the wavefunction by a (generally non-Hermitian) operator $\hat{\mathcal{U}}$. Thus, we start with $\ket{\Psi}$, evolve for a time~$t_0$ and apply $\hat{\mathcal{U}}$ to get
\[
    \hat{\mathcal{U}}\eul^{-\imag\hat{H}t_0/\hbar}\ket{\Psi}.
\]
We then evolve through $t_1$ and measure the expectation value of $\hat{\mathcal{B}}$,
\[
    \bra{\Psi} \eul^{\imag\hat{H}t_0/\hbar} \hat{\mathcal{U}}^\dagger \eul^{\imag\hat{H}t_1/\hbar} \hat{\mathcal{B}} \eul^{-\imag\hat{H}t_1/\hbar} \hat{\mathcal{U}} \eul^{-\imag\hat{H}t_0/\hbar} \ket{\Psi} .
\]
This unwieldy expression can be written as
\[
    C_{\rho\mathcal{U}\mathcal{B}}(t_0,t_1) = \Tr{\rho \hat{\mathcal{U}}^\dagger(t_0) \hat{\mathcal{B}}(t_0+t_1) \hat{\mathcal{U}}(t_0)} ,
\]
a two-dimensional correlation function. This corresponds directly\sidenote{A different form for multi-time TCFs that involves commutators is used in non-linear optical spectroscopy \cite{videlaMatsubaraDynamicsApproximation2023} -- however, the methods in this chapter could be generalised to these TCFs as well by expanding out the commutators and applying the operators at appropriate times.} to what would be measured in a simple 2D NMR experiment like COSY.\sidenote{For COSY specifically , we would have $\rho=\sigma_x$ and $\hat{\mathcal{B}} = \sigma_x + \imag \sigma_y$, which is the magnetisation in $x$ and $y$ directions. The pulse would be \[\hat{\mathcal{U}} = \frac{1}{\sqrt{2}}\begin{pmatrix}
1 & -\imag \\ -\imag & 1 \end{pmatrix}, \] which represents a rotation by $\pi/2$ on the Bloch sphere.} 

We shall investigate whether such experiments can also be simulated with MASH. There are two avenues one could take -- either use the operator $\hat{\mathcal{U}}$ to act directly on the spin variables, or evolve the density matrix by the quantum jump procedure introduced in~\cite{mannouchMappingApproachSurface2023}.\sidenote{The quantum jump procedure is explained in \autoref{sec:detail-MASH-schrodinger-picture}.} We shall see that both approaches give equivalent results, but the former requires many fewer trajectories for convergence, so would be preferred in practice.

Note that experimentally, we can measure continuously in $t_1$, but we need to repeat the whole experiment for each increment of $t_0$. Since we shall be simulating the physical acquisition of the spectrum, this will increase the computational cost dramatically.

\section{The quantum jump procedure can improve MASH to QCLE accuracy}

We know that MASH is correct in the short-time limit. Thus, we could increase accuracy by splitting a simulation of a long time $t$ into shorter time periods $t_1 + t_2 + \cdots + t_n=t$ and restarting MASH every time. In order to know what the \enquote{state} is at time $t_1$, we need to directly evolve the density matrix. Luckily, we can rewrite the MASH dynamics and estimators to do so, as explained in \autoref{sec:detail-MASH-schrodinger-picture}.

Once we can directly evolve the density matrix, the path to QCLE accuracy is clear. We first evolve $\rho$ through $t_1$ to obtain $\rho_1$. Then, we start a new simulation in state $\rho_1$ and evolve through $t_2$ to obtain $\rho_2$. We repeat this again and again until we reach time $t$.

In practice, starting a new simulation corresponds to keeping the nuclear coordinates the same\sidenote[][*0]{Thanks to the delta functions in \eqref{eq:density-matrix-MASH-evolution}.} and resampling the spin from the Bloch sphere. This is known as a \emph{quantum jump}. Note that even though we resample the electronic variables, their history is encoded in the density matrix $\rho$ through terms depending on $\vec{S}$ and $\vec{S}_t$ from the previous step.

\section{A naive way to evaluate multi-time correlation functions -- no free lunch}

We would like to estimate the multi-time correlation function
\[
    C_{\rho\mathcal{U}\mathcal{B}}(t_0,t_1) = \Tr{\rho \hat{\mathcal{U}}^\dagger(t_0) \hat{\mathcal{B}}(t_0+t_1) \hat{\mathcal{U}}(t_0)} .
\]
Assuming factorised initial conditions and electronic operators only, we can simplify this to
\[
    C_{AUB}(t_0,t_1) = \Tr{\rho_\nuc \hat{A} \hat{U}^\dagger(t_0) \hat{B}(t_0+t_1) \hat{U}(t_0)} .
\]

For two-time correlation functions, we have used the prescription
\begin{align*}
    \hat{A} & \mapsto A(\vec{S}) \\
    \hat{B}(t) & \mapsto B(\vec{S}_t).
\end{align*}
We might thus assume that the multi-time correlation function could simply be estimated as
\[
    \int \dd\vec{S}\, W_{AUB}(\vec{S}) \, A(\vec{S}) \, U^\dagger(\vec{S}_{t_0}) \, B(\vec{S}_{t_1}) \, U(\vec{S}_{t_0}).
\]
However, this approach is doomed to fail -- in the classical path approximation, this integral will reduce to products of two-time correlation functions rather than the proper multi-time one.\sidenote{Therefore, we cannot even obtain an expression for $W_{AUB}$.} Additionally, it would be very miraculous if we could obtain any multi-time correlation function from just a single set of trajectories.

\section{Applying the pulse directly to the spin variables}

We can try to directly simulate the 2D experiment by evolving through $t_0$, applying the pulse and then evolving through $t_1$. In an exact scheme, the pulse would correspond to changing the wavefunction $\ket{\psi}$ instantaneously into $\hat{\mathcal{U}} \ket{\psi}$.

With MASH, we have an \enquote{electronic wavefunction} that we propagate, so we could similarly do this replacement for every trajectory. It is not \textit{a priori} obvious that it would work\sidenote{After all, the MASH \enquote{electronic wavefunction} is best thought of as a probability distribution in an extended phase space, rather than a true quantum object.} -- we explain some reasoning behind it in \autoref{sec:detail-time-dep-pulse}.

\section{Multi-time TCFs with the quantum jump procedure}

We can also rewrite the multi-time correlation function as\sidenote{$\rho_S(t_0)$ is the density matrix evolved through $t_0$. See \autoref{sec:detail-MASH-schrodinger-picture}.}
\[
    C_{\rho\mathcal{U}\mathcal{B}}(t_0,t_1) = \Tr{\hat{\mathcal{U}}\rho_S(t_0) \hat{\mathcal{U}}^\dagger \hat{\mathcal{B}}(t_1) },
\]
which suggests to evolve the density matrix through $t_0$ first, then apply the operator $\hat{\mathcal{U}}$ to the density matrix by the prescription
\[
    \rho \to \hat{\mathcal{U}}\rho \hat{\mathcal{U}}^\dagger
\]
and finally evolve the new density matrix for a time $t_1$. This means that we could easily resample the variables before the $t_1$ evolution, starting in the state $\hat{\mathcal{U}}\rho_S(t_0) \hat{\mathcal{U}}^\dagger$. This is a slightly modified version of the quantum jump procedure.\sidenote{Its practical implementation is discussed in \autoref{sec:impl-multi-time-tcfs}.}

\section{To jump or not to jump?}

We shall try to estimate multi-time TCFs using the two methods described above. Two questions need to be answered -- whether either of these methods works and if so, whether one offers an improvement over the other. To this end, we shall use the Tully models, for which exact results can be obtained using Chebyshev propagation.\sidenote{See \autoref{appendix:chebyshev} for an overview of the method.} There are infinitely many possible choices for the operators $\hat{\mathcal{U}}$ and $\hat{\mathcal{B}}$. We shall use the very simple\sidenote{In spectroscopy, the pulses would usually be operators that generate coherences from populations. Luckily, MASH can simulate the evolution of coherences surprisingly well (see \autoref{appendix:all-estimators}), so there is hope that it could describe more realistic scenarios.}
\[
    \hat{\mathcal{U}} = \hat{\sigma}_x^\adia ,
\]
which corresponds to swapping the wavefunctions of the $\ket{+}$ and $\ket{-}$ states. For the operator $\hat{\mathcal{B}}$, we shall entertain the possibilities of both electronic and nuclear operators. We shall use low initial momenta, which is quite a stringent test for any quantum-classical method, including MASH. This will hopefully allow us to better expose the limits of the approach.

\begin{figure*}[h!]
    \centering
    \begin{tikzpicture}[
    subcaption/.style={anchor=north, font=\small}
    ]

    \begin{groupplot}[
            group style={
                group name=plotGroup,
                group size=3 by 1,
                horizontal sep=1.0cm,
                x descriptions at=edge bottom,
                y descriptions at=edge left, 
            },
            width=0.3\textwidth,
            height=0.3\textwidth,
            enlargelimits=false,
            tick align=outside,
            xlabel=$t_0$,
            ylabel=$t_1$,
            scale only axis,
            axis on top,
            xtick={0, 1, 2, 3},
            xticklabels={$0$, $10000$, $20000$, $30000$},
            ytick={0, 1, 2, 3},
            yticklabels={$0$, $10000$, $20000$, $30000$}
            ]
            
        \nextgroupplot

       \addplot graphics[xmin=0, xmax=3, ymin=0, ymax=3] {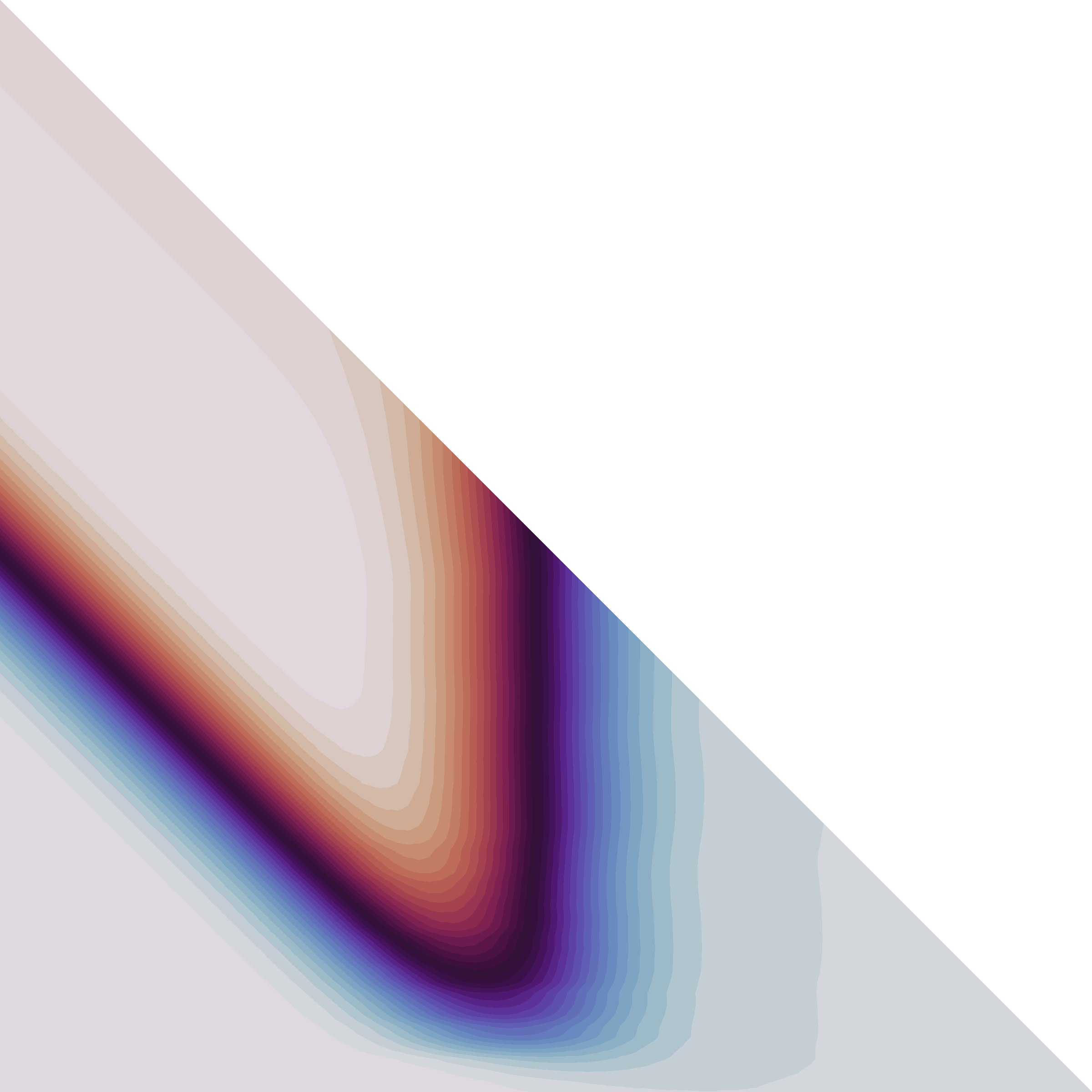};

       \nextgroupplot
        
       \addplot graphics[xmin=0, xmax=3, ymin=0, ymax=3] {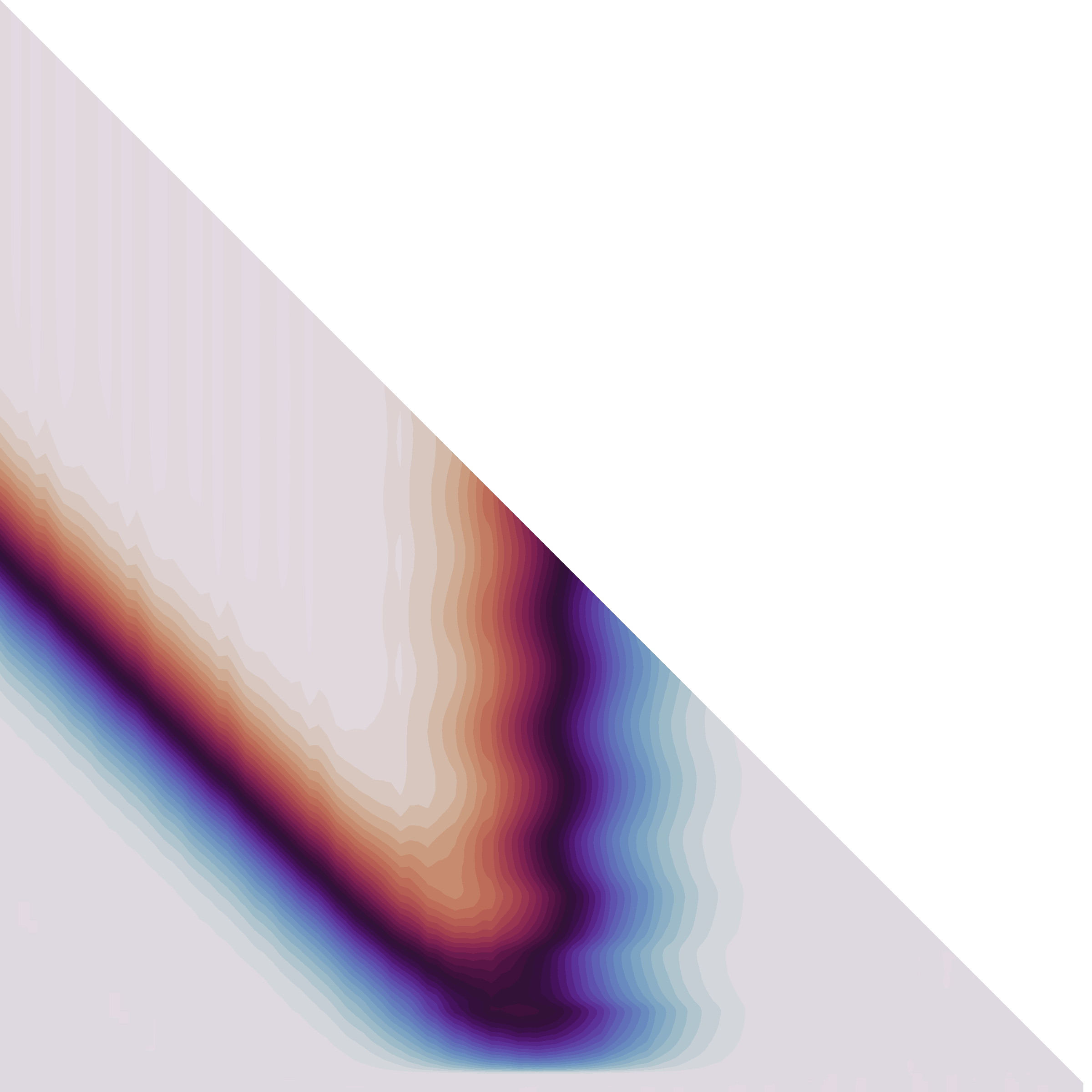};

       \nextgroupplot
        
       \addplot graphics[xmin=0, xmax=3, ymin=0, ymax=3] {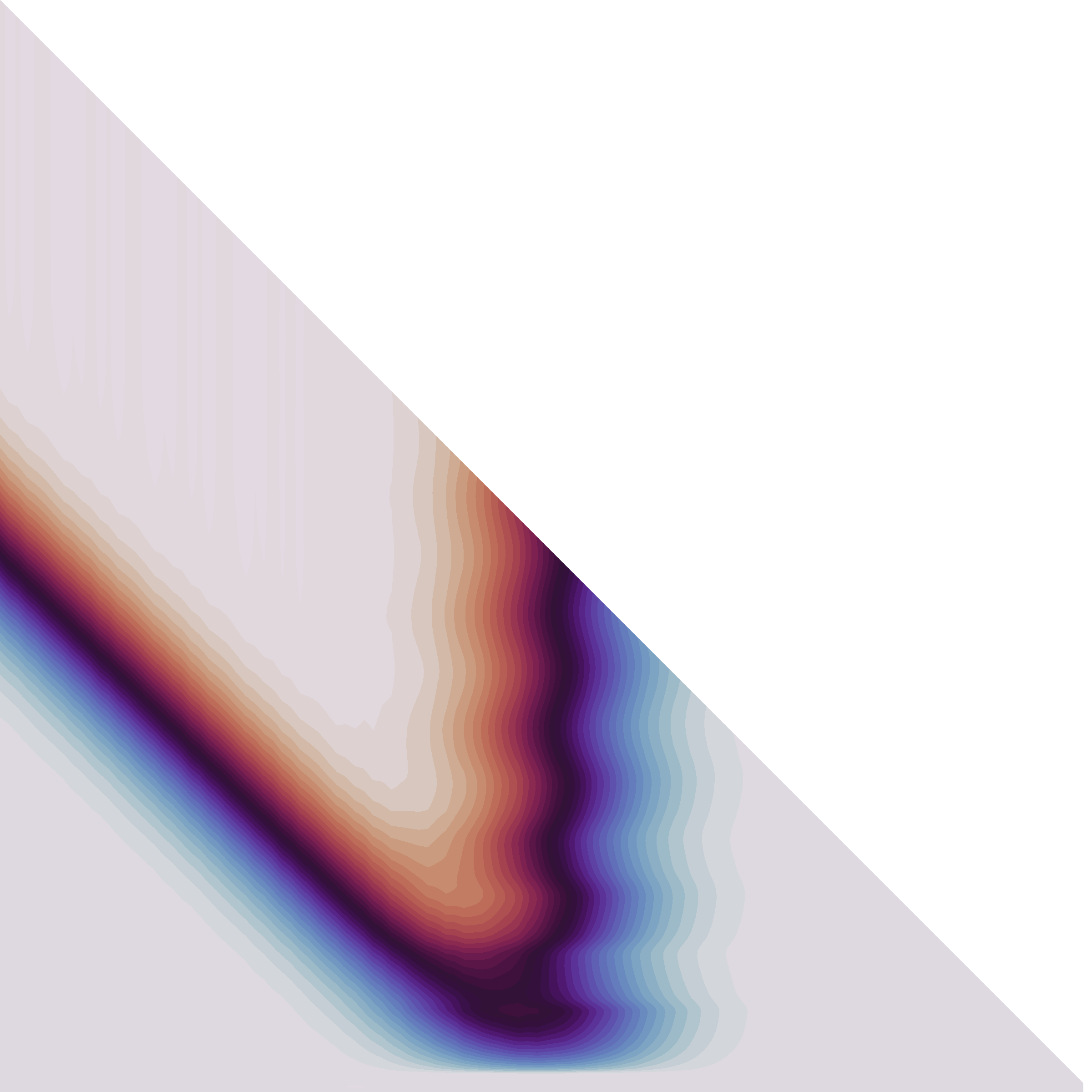};

       \coordinate (topright) at (rel axis cs:1.2,1);

    \end{groupplot}

    \begin{axis}[%
        hide axis,
        scale only axis,
        height=0.3\textwidth,
        % width=0.01\textwidth,
        at={(topright)},
        anchor=north east,
        point meta min=0.0,
        point meta max=1.0,
        colormap/twilight,
        colorbar right
      ]

        \addplot [draw=none] coordinates {(0,0) (1,1)};
    \end{axis}

    \node[subcaption] at (plotGroup c1r1.below south) {(a) exact};
    \node[subcaption] at (plotGroup c2r1.below south) {(b) quantum jump};
    \node[subcaption] at (plotGroup c3r1.below south) {(c) direct};

\end{tikzpicture}
    \caption{Multi-time TCF with $\hat{A}=\hat{P}_-$ and $\hat{\mathcal{B}}=h(x)$ for Tully's model~I with $q_0=-15$, $p_0=8$, $m=8000$ and $\gamma=0.1$. If $t_0$ is large, the $\sigma_x$ pulse is applied once the particle has already reflected and $\avg{\mathcal{B}}=0$. If the pulse is applied before the particle gets to the coupling region ($t_0\lessapprox 12000$), transmission is guaranteed. When the particle is close to $x=0$, it has almost no momentum left and the pulse sets up a resonance in the upper state. This corresponds to the black vertical region around $t_0 = 15000$. Results are an average over $2\times10^5$ trajectories.}
    \label{fig:plot-MT-1}
\end{figure*}
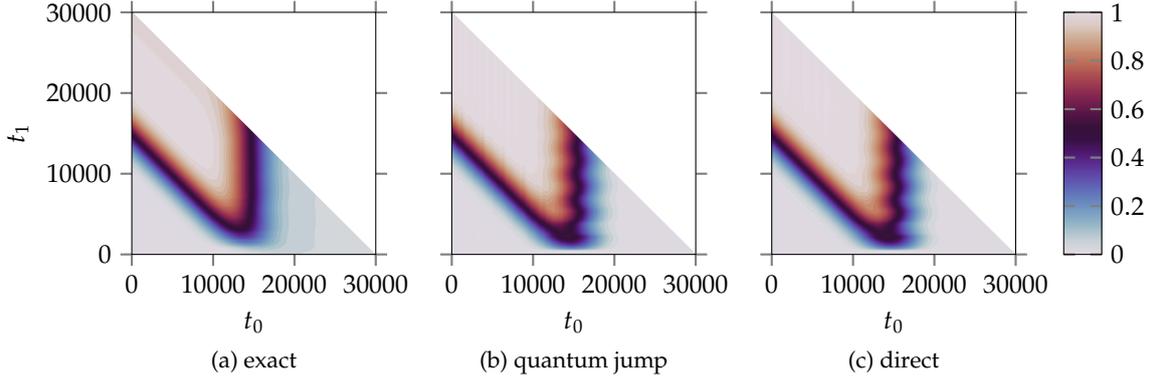

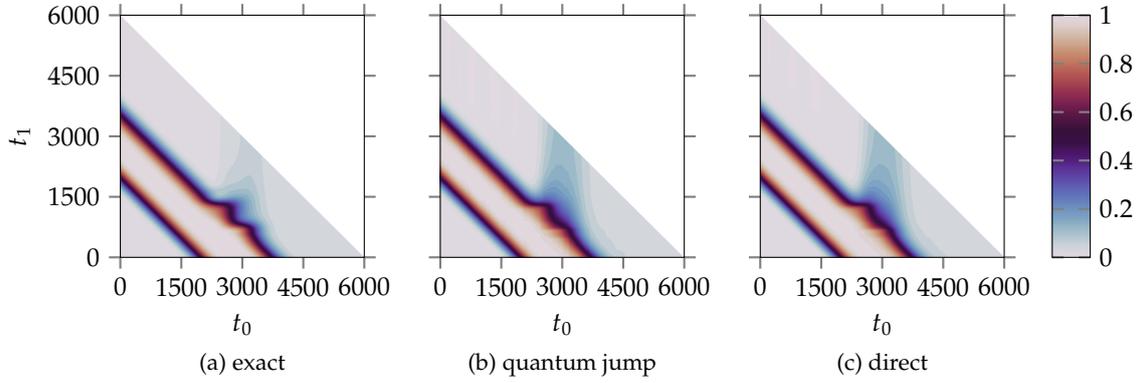
\begin{figure*}[h!]
    \centering
    \begin{tikzpicture}[
    subcaption/.style={anchor=north, font=\small}
    ]

    \begin{groupplot}[
            group style={
                group name=plotGroup,
                group size=3 by 1,
                horizontal sep=1.0cm,
                x descriptions at=edge bottom,
                y descriptions at=edge left, 
            },
            width=0.3\textwidth,
            height=0.3\textwidth,
            enlargelimits=false,
            tick align=outside,
            xlabel=$t_0$,
            ylabel=$t_1$,
            scale only axis,
            axis on top,
            xtick={0, 1.5, 3, 4.5, 6},
            xticklabels={$0$, $1500$, $3000$, $4500$, $6000$},
            ytick={0, 1.5, 3, 4.5, 6},
            yticklabels={$0$, $1500$, $3000$, $4500$, $6000$}
            ]
            
        \nextgroupplot

        \addplot graphics[xmin=0, xmax=6, ymin=0, ymax=6] {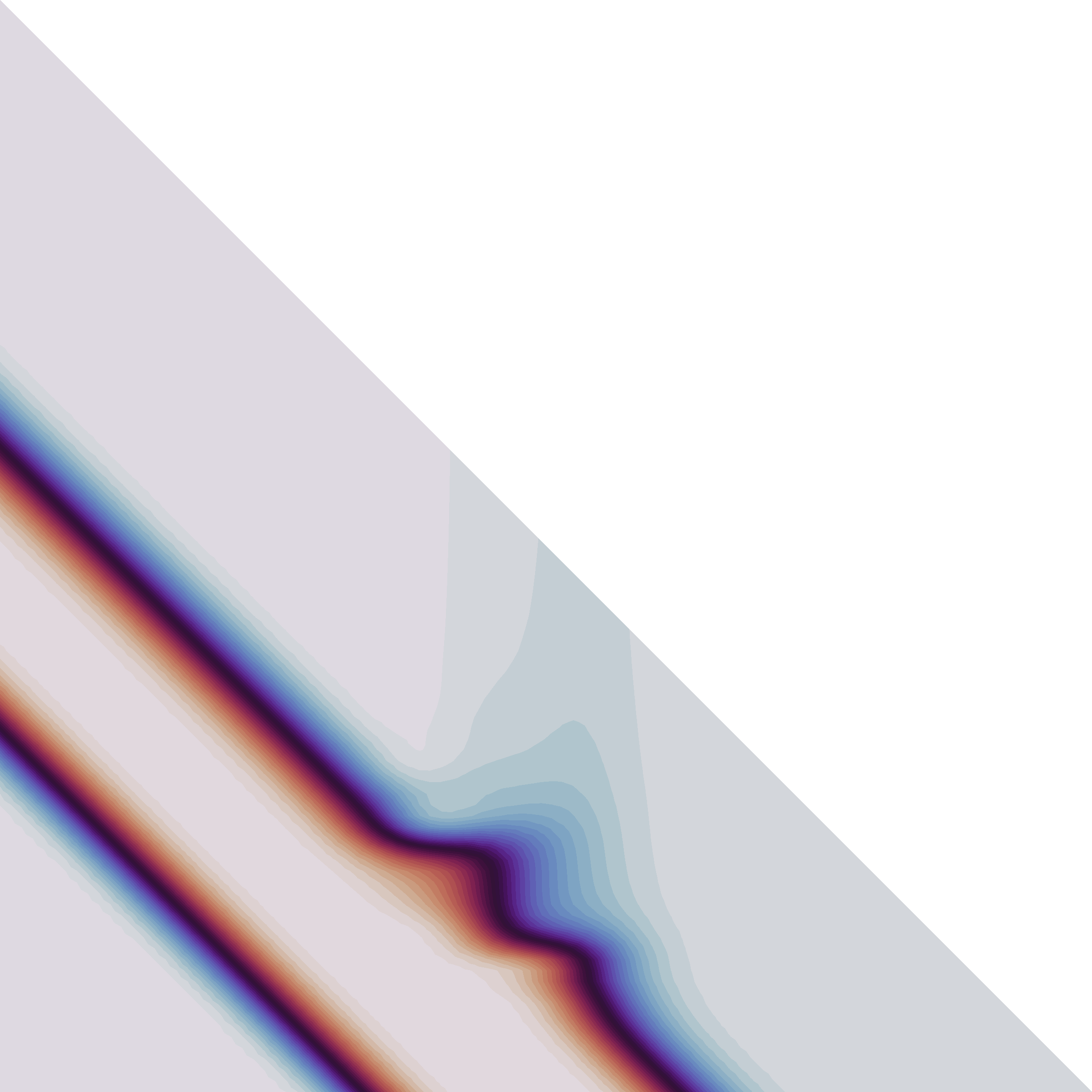};

       \nextgroupplot
        
       \addplot graphics[xmin=0, xmax=6, ymin=0, ymax=6] {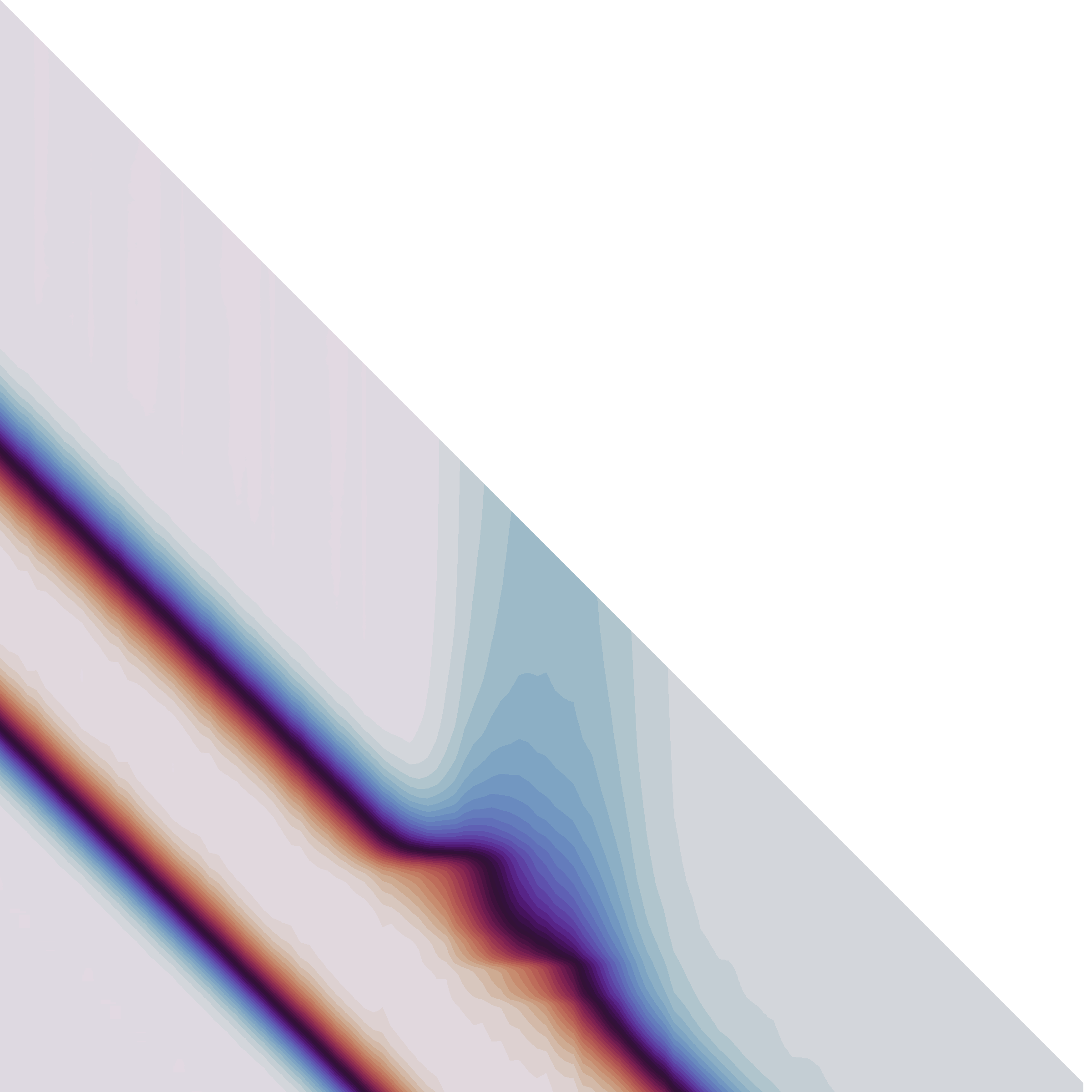};

       \nextgroupplot
        
       \addplot graphics[xmin=0, xmax=6, ymin=0, ymax=6] {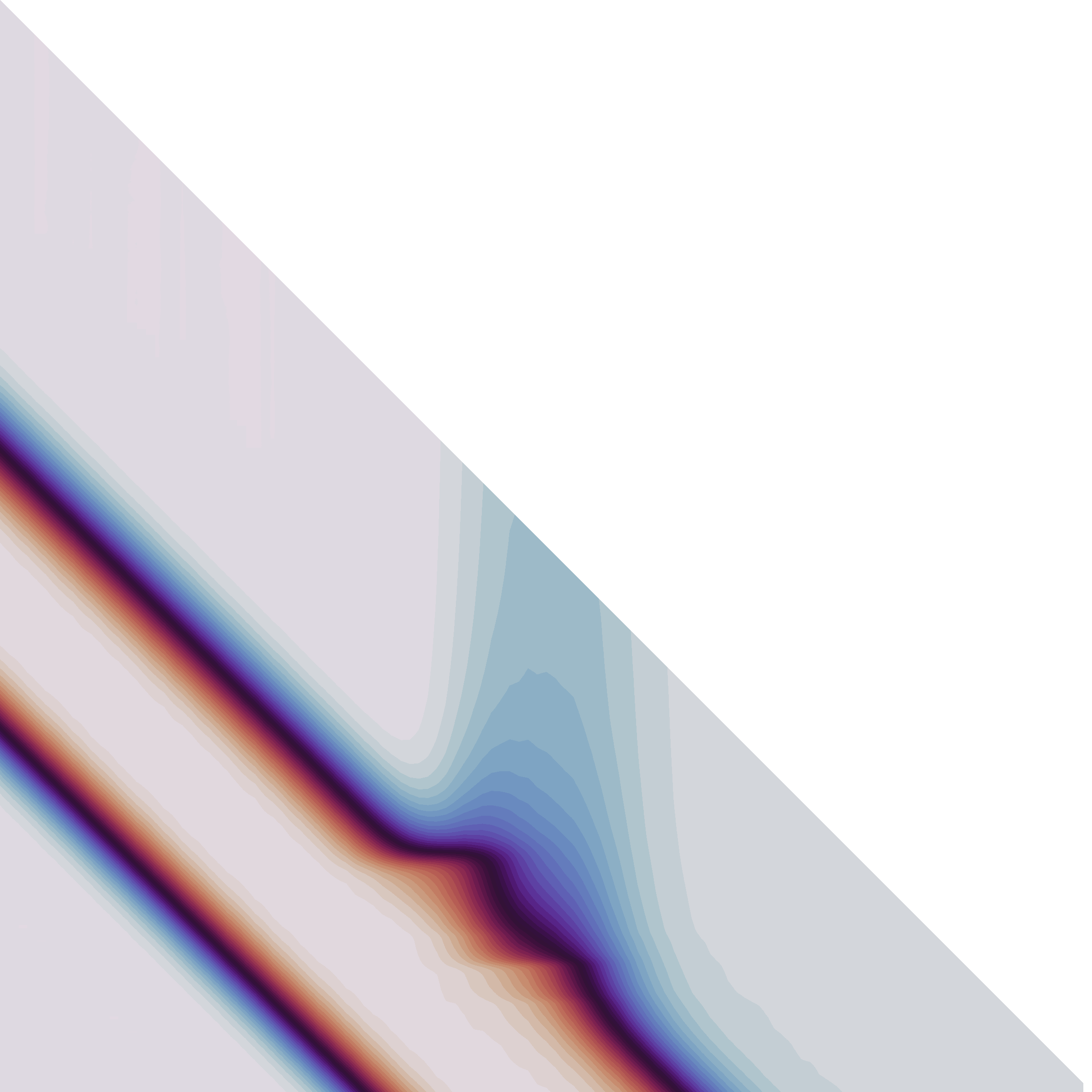};

       \coordinate (topright) at (rel axis cs:1.2,1);

    \end{groupplot}

    \begin{axis}[%
        hide axis,
        scale only axis,
        height=0.3\textwidth,
        % width=0.01\textwidth,
        at={(topright)},
        anchor=north east,
        point meta min=0.0,
        point meta max=1.0,
        colormap/twilight,
        colorbar right
      ]

        \addplot [draw=none] coordinates {(0,0) (1,1)};
    \end{axis}

    \node[subcaption] at (plotGroup c1r1.below south) {(a) exact};
    \node[subcaption] at (plotGroup c2r1.below south) {(b) quantum jump};
    \node[subcaption] at (plotGroup c3r1.below south) {(c) direct};

\end{tikzpicture}
    \caption{Multi-time TCF with $\hat{A}=\hat{P}_-$ and $\hat{\mathcal{B}}=h(5-x)h(5+x)$ for Tully's model~II with $q_0=-15$, $p_0=10$ and $\gamma=0.5$. The two black diagonal lines correspond to the particle entering and leaving the central region. The two triangular features on the second line reflect the particle staying in the upper-state well for a prolonged period of time. Results are an average over $4\times10^5$ trajectories.}
    \label{fig:plot-MT-2}
\end{figure*}

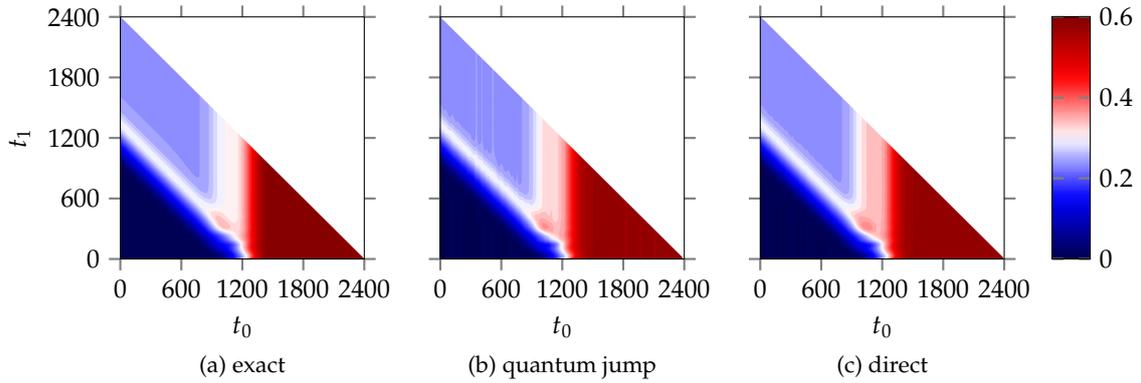
\begin{figure*}[h!]
    \centering
    \begin{tikzpicture}[
    subcaption/.style={anchor=north, font=\small}
    ]

    \begin{groupplot}[
            group style={
                group name=plotGroup,
                group size=3 by 1,
                horizontal sep=1.0cm,
                x descriptions at=edge bottom,
                y descriptions at=edge left, 
            },
            width=0.3\textwidth,
            height=0.3\textwidth,
            enlargelimits=false,
            tick align=outside,
            xlabel=$t_0$,
            ylabel=$t_1$,
            scale only axis,
            axis on top,
            xtick={0, 0.6, 1.2, 1.8, 2.4},
            xticklabels={$0$, $600$, $1200$, $1800$, $2400$},
            ytick={0, 0.6, 1.2, 1.8, 2.4},
            yticklabels={$0$, $600$, $1200$, $1800$, $2400$}
            ]
            
        \nextgroupplot

        \addplot graphics[xmin=0, xmax=2.4, ymin=0, ymax=2.4] {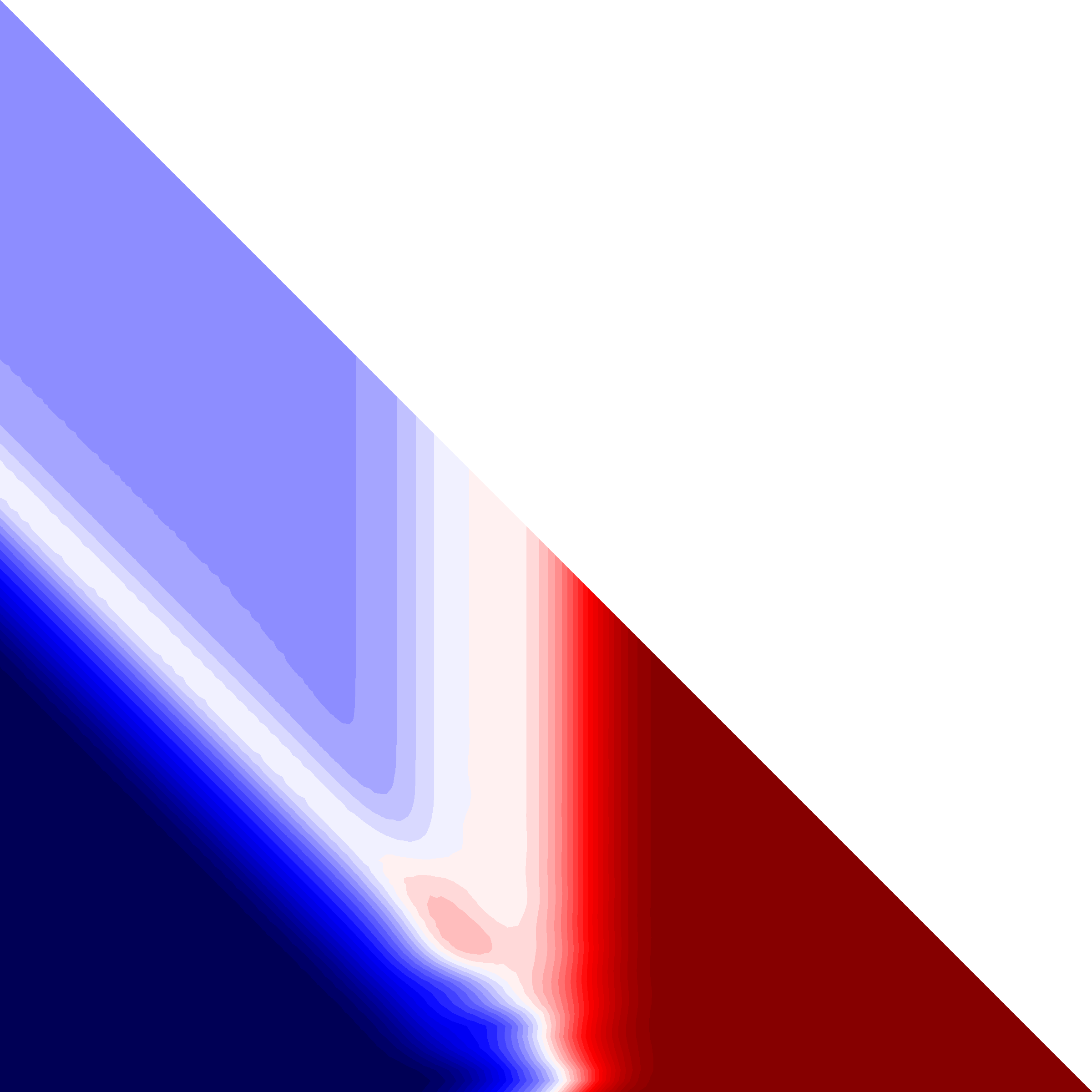};

       \nextgroupplot
        
       \addplot graphics[xmin=0, xmax=2.4, ymin=0, ymax=2.4] {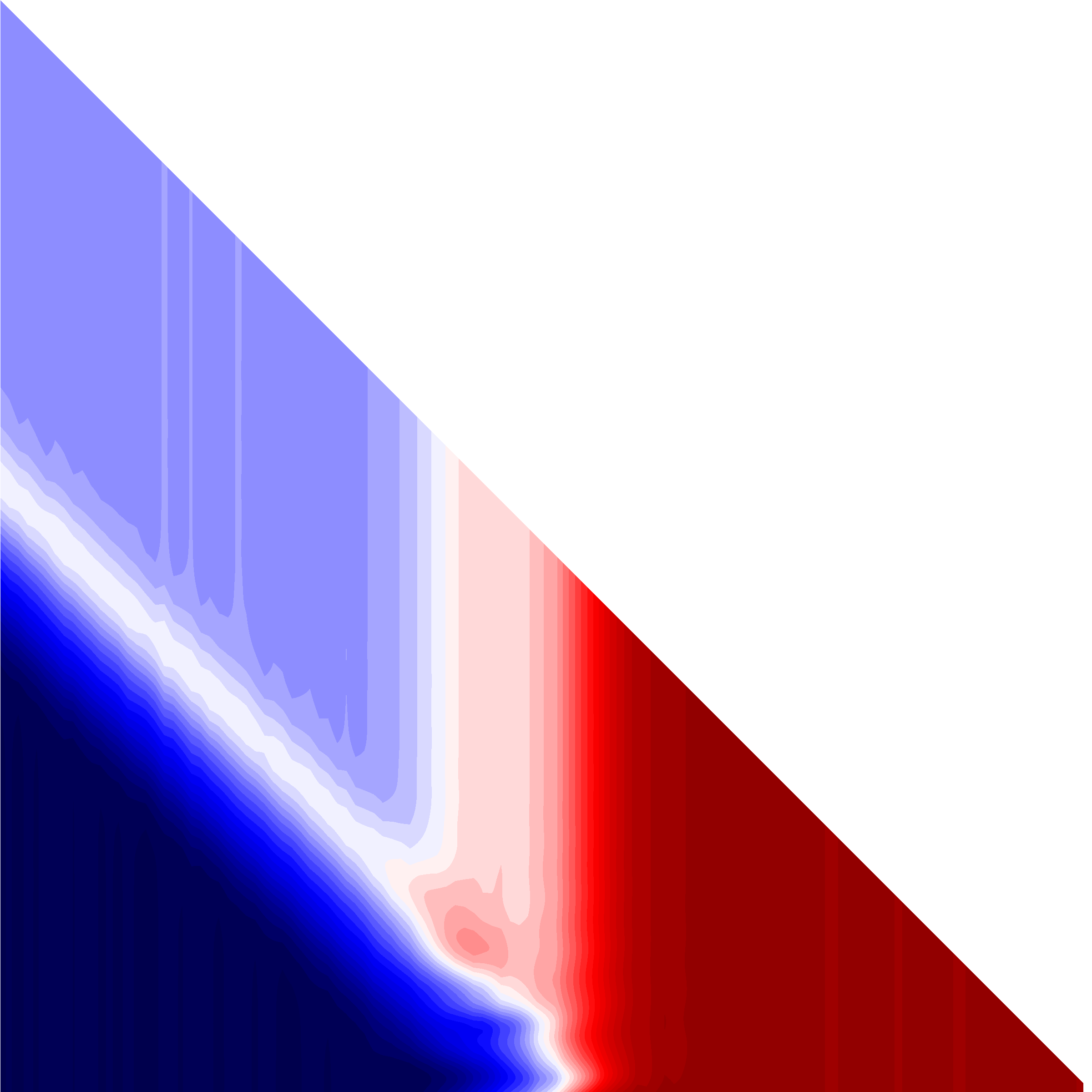};

       \nextgroupplot
        
       \addplot graphics[xmin=0, xmax=2.4, ymin=0, ymax=2.4] {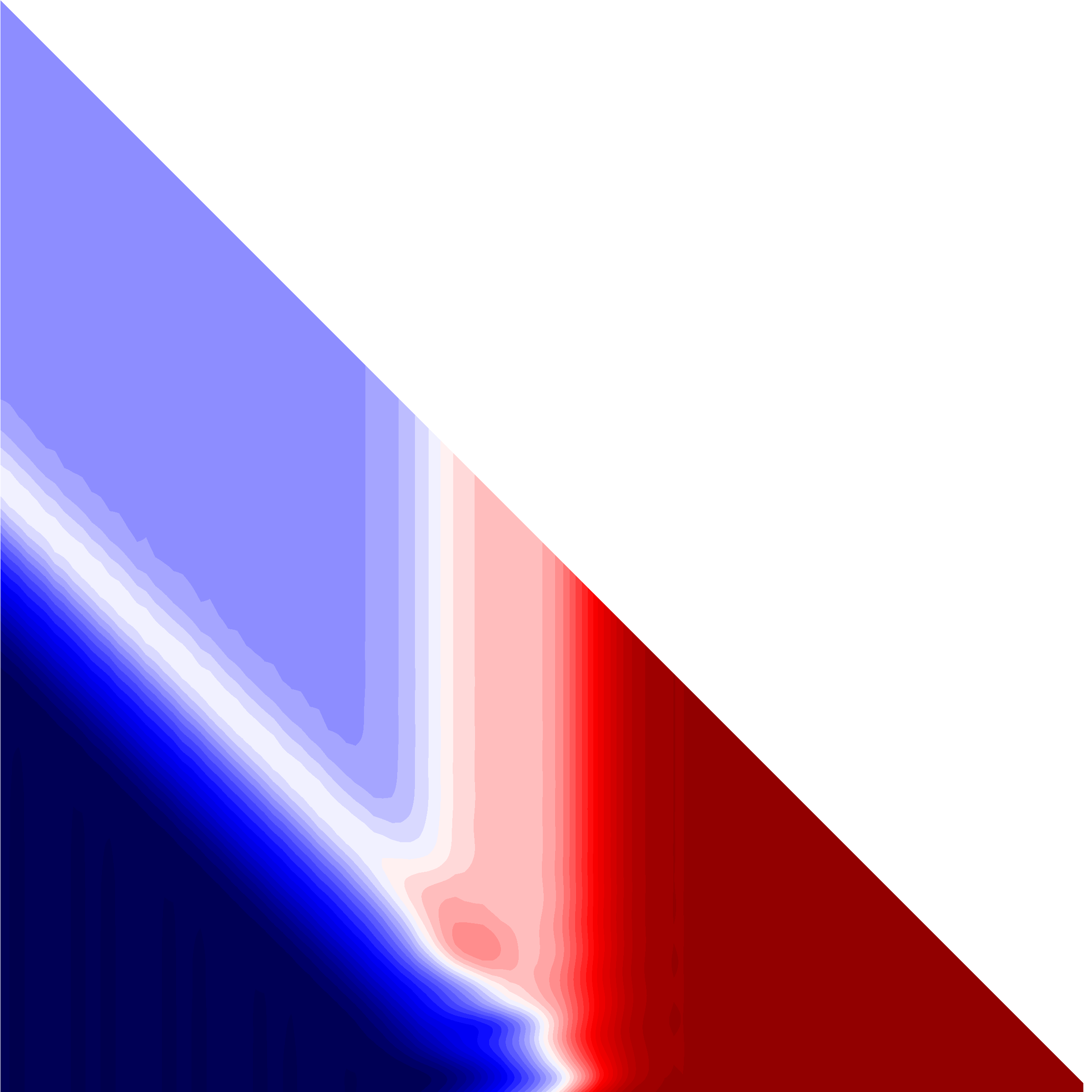};
       
       \coordinate (topright) at (rel axis cs:1.2,1);

    \end{groupplot}

    \begin{axis}[%
        hide axis,
        scale only axis,
        height=0.3\textwidth,
        % width=0.01\textwidth,
        at={(topright)},
        anchor=north east,
        point meta min=0.0,
        point meta max=0.6,
        colormap/seismic,
        colorbar right
      ]

        \addplot [draw=none] coordinates {(0,0) (1,1)};
    \end{axis}

    \node[subcaption] at (plotGroup c1r1.below south) {(a) exact};
    \node[subcaption] at (plotGroup c2r1.below south) {(b) quantum jump};
    \node[subcaption] at (plotGroup c3r1.below south) {(c) direct};

\end{tikzpicture}
    \caption{Multi-time TCF with $\hat{A}=\hat{P}_+$ and $\hat{\mathcal{B}}=\hat{P}_+$ for Tully's model~II with $q_0=-15$, $p_0=25$ and $\gamma=0.5$. The vertical section $t_0=0$ corresponds to the $C_{-+}(t)$ correlation function, the horizontal section $t_1=0$ to $C_{+-}(t)$. Results are an average over $7.5\times10^5$ trajectories -- the larger statistical error of the quantum jump procedure can be seen as banding in (b).}
    \label{fig:plot-MT-3}
\end{figure*}

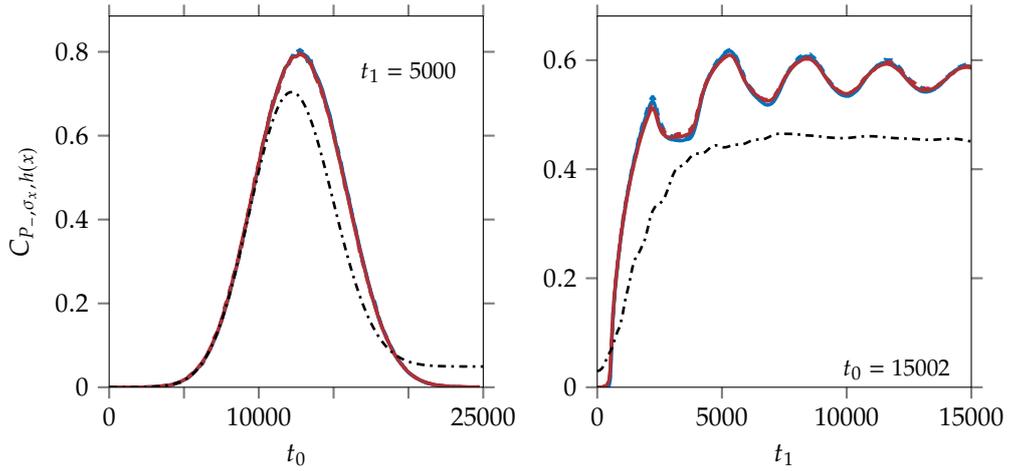
\begin{figure*}[h!]
    \centering
    \begin{tikzpicture}[
    subcaption/.style={anchor=north, font=\small}
    ]

    \begin{groupplot}[
        group style={
            group name=plotGroup,
            group size=2 by 1,
            horizontal sep=1.5cm
        },
        width=0.46\textwidth,
        height=0.46\textwidth,
        enlarge x limits=false,
        ymin=0,
        tick align=outside,
        scale only axis,
        axis on top,
        groupplot ylabel=$C_{P_-, \sigma_x, h(x)}$
        ]
        
    \nextgroupplot[scaled x ticks = false,
                   xlabel=$t_0$,
                   xtick={0, 5000, 10000, 15000, 20000, 25000},
                   xticklabels={$0$, , $10000$, , , $25000$}]

    \addplot[mark=none, line width=1.3pt, RoyalBlue, dashed]  table[x index = 0, y index = 1] {data/slice-1.dat};
    \addplot[mark=none, line width=1.3pt, RoyalBlue]          table[x index = 0, y index = 2] {data/slice-1.dat};
    \addplot[mark=none, line width=1.3pt, Maroon, dashed]     table[x index = 0, y index = 3] {data/slice-1.dat};
    \addplot[mark=none, line width=1.3pt, Maroon]             table[x index = 0, y index = 4] {data/slice-1.dat};
    \addplot[mark=none, line width=1pt, dashdotted]               table[x index = 0, y index = 1] {data/slice-1-exact.dat};

    \node[subcaption] at ({rel axis cs:0.8,0.9}) {$t_1=5000$};

    \nextgroupplot[scaled x ticks = false,
                   xlabel=$t_1$,
                   xtick={0, 5000, 10000, 15000},
                   xticklabels={$0$, $5000$, $10000$, $15000$}]
    
    \addplot[mark=none, line width=1.3pt, RoyalBlue, dashed]  table[x index = 0, y index = 1] {data/slice-4.dat};
    \addplot[mark=none, line width=1.3pt, RoyalBlue]          table[x index = 0, y index = 2] {data/slice-4.dat};
    \addplot[mark=none, line width=1.3pt, Maroon, dashed]     table[x index = 0, y index = 3] {data/slice-4.dat};
    \addplot[mark=none, line width=1.3pt, Maroon]             table[x index = 0, y index = 4] {data/slice-4.dat};
    \addplot[mark=none, line width=1pt, dashdotted]               table[x index = 0, y index = 1] {data/slice-4-exact.dat};

    \node[subcaption] at ({rel axis cs:0.8,0.1}) {$t_0=15002$};

    \end{groupplot}

\end{tikzpicture}
    \caption{Horizontal (left) and vertical (right) slice through Figure~\ref{fig:plot-MT-1}. The MASH (blue) and ms-MASH (red) results are shown both with the direct method (solid) and the quantum jump procedure (dashed). Exact results (black dash-dotted) were obtained using Chebyshev propagation.}
    \label{fig:slice-1}
\end{figure*}

\begin{figure*}[h!]
    \centering
    \begin{tikzpicture}[
    subcaption/.style={anchor=north, font=\small}
    ]

    \begin{groupplot}[
        group style={
            group name=plotGroup,
            group size=2 by 1,
            horizontal sep=1.5cm
        },
        width=0.46\textwidth,
        height=0.46\textwidth,
        enlarge x limits=false,
        ymin=0,
        tick align=outside,
        scale only axis,
        axis on top,
        groupplot ylabel=$C_{P_-, \sigma_x, P_+}$
        ]
        
    \nextgroupplot[xlabel=$t_0$]

    \addplot[mark=none, line width=1.3pt, RoyalBlue, dashed]  table[x index = 0, y index = 1] {data/slice-6.dat};
    \addplot[mark=none, line width=1.3pt, RoyalBlue]          table[x index = 0, y index = 2] {data/slice-6.dat};
    \addplot[mark=none, line width=1.3pt, Maroon, dashed]     table[x index = 0, y index = 3] {data/slice-6.dat};
    \addplot[mark=none, line width=1.3pt, Maroon]             table[x index = 0, y index = 4] {data/slice-6.dat};
    \addplot[mark=none, line width=1pt, dashdotted]               table[x index = 0, y index = 1] {data/slice-6-exact.dat};

    \node[subcaption] at ({rel axis cs: 0.8,0.1}) {$t_1=325$};

    \nextgroupplot[xlabel=$t_1$, xtick={0, 400, 800, 1200}]
    
    \addplot[mark=none, line width=1.3pt, RoyalBlue, dashed]  table[x index = 0, y index = 1] {data/slice-3.dat};
    \addplot[mark=none, line width=1.3pt, RoyalBlue]          table[x index = 0, y index = 2] {data/slice-3.dat};
    \addplot[mark=none, line width=1.3pt, Maroon, dashed]     table[x index = 0, y index = 3] {data/slice-3.dat};
    \addplot[mark=none, line width=1.3pt, Maroon]             table[x index = 0, y index = 4] {data/slice-3.dat};
    \addplot[mark=none, line width=1pt, dashdotted]               table[x index = 0, y index = 1] {data/slice-3-exact.dat};

    \node[subcaption] at ({rel axis cs: 0.8,0.1}) {$t_0=1041$};

    \end{groupplot}

\end{tikzpicture}
    \caption{Horizontal (left) and vertical (right) slice through Figure~\ref{fig:plot-MT-3}. Labelling is the same as Figure~\ref{fig:slice-1}.}
    \label{fig:slice-3}
\end{figure*}
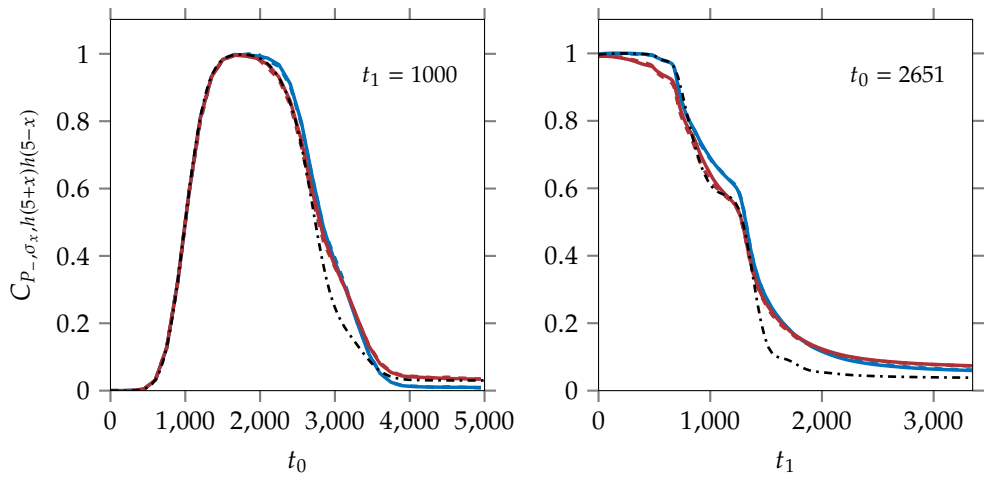

The resulting two-dimensional correlation functions are shown in Figures~\ref{fig:plot-MT-1}, \ref{fig:plot-MT-2} and~\ref{fig:plot-MT-3}. They were only measured until a given total time $t=t_0+t_1$, which gives the plots triangular shape.\sidenote{This is to limit the grid size for exact propagation. Also, nothing much interesting is happening by then; the particle has either left the coupling region or is stuck in a long-lived resonance.} Both MASH and ms-MASH were used, but the contour plots are visually indistinguishable, so only the latter is shown. Sections through interesting values of $t_0$ and $t_1$ are shown in Figures~\ref{fig:slice-1}, \ref{fig:slice-2} and~\ref{fig:slice-3} for both MASH and ms-MASH.

\marginnote[*1]{Data shown in Figures~\ref{fig:plot-MT-1} and \ref{fig:slice-1}.}
For Tully's model~I, we have used the final operator $\hat{\mathcal{B}}=h(x)$, which measures how much of the wavefunction made it to the other side. The discrepancies observed between the exact and MASH CFs are likely the result of the upper-state resonance generated by the pulse. However, it is apparent from Figure~\ref{fig:slice-1} that both MASH and ms-MASH lead to identical results and the quantum jump procedure does not improve them.

\marginnote[*1]{Data shown in Figures~\ref{fig:plot-MT-2} and \ref{fig:slice-2}.}
For Tully's model~II, the operator $\hat{\mathcal{B}}=h(5-x)h(5+x)$ measures how much of the wavefunction is \enquote{stuck} in the central region. Again, there is no difference between the direct method and the quantum jump procedure, but a difference between MASH and ms-MASH can be seen in Figure~\ref{fig:slice-2}. The better performance of ms-MASH is observed generally for Tully's model~II (see \autoref{appendix:all-estimators}). Note that both methods struggle with describing the resonance created at $t_0 \approx 3000$ -- here, both predict a decay rate that is too slow (see Figure~\ref{fig:slice-2}).

\marginnote[*1]{Data shown in Figures~\ref{fig:plot-MT-3} and \ref{fig:slice-3}.}
For Tully's model~II, we also look at the electronic operator $\hat{\mathcal{B}}=\hat{P}_+$. Here, a slight difference between the direct method and the quantum jump procedure is observed -- as expected, the latter does lead to somewhat better results. However, even when the pulse is applied when the particle is near $x\approx 0$ -- where greatest improvement is expected -- the difference is quite small.\sidenote[][*-4]{We can compare this to the improvement for two-time correlation functions in \autoref{appendix:single-jump}, where a single quantum jump applied when $x\approx 0$ is able to significantly increase the accuracy of MASH.}

\pagebreak

\pagebreak

\begin{figure*}[h!]
    \centering
    \begin{tikzpicture}[
    subcaption/.style={anchor=north, font=\small}
    ]

    \begin{groupplot}[
        group style={
            group name=plotGroup,
            group size=2 by 1,
            horizontal sep=1.5cm
        },
        width=0.46\textwidth,
        height=0.46\textwidth,
        enlarge x limits=false,
        ymin=0,
        tick align=outside,
        scale only axis,
        axis on top,
        groupplot ylabel=$C_{P_-, \sigma_x, h(5+x)h(5-x)}$
        ]
        
    \nextgroupplot[xlabel=$t_0$]

    \addplot[mark=none, line width=1.3pt, RoyalBlue, dashed]  table[x index = 0, y index = 1] {data/slice-5.dat};
    \addplot[mark=none, line width=1.3pt, RoyalBlue]          table[x index = 0, y index = 2] {data/slice-5.dat};
    \addplot[mark=none, line width=1.3pt, Maroon, dashed]     table[x index = 0, y index = 3] {data/slice-5.dat};
    \addplot[mark=none, line width=1.3pt, Maroon]             table[x index = 0, y index = 4] {data/slice-5.dat};
    \addplot[mark=none, line width=1pt, dashdotted]               table[x index = 0, y index = 1] {data/slice-5-exact.dat};

    \node[subcaption] at ({rel axis cs:0.8,0.9}) {$t_1=1000$};

    \nextgroupplot[xlabel=$t_1$]
    
    \addplot[mark=none, line width=1.3pt, RoyalBlue, dashed]  table[x index = 0, y index = 1] {data/slice-2.dat};
    \addplot[mark=none, line width=1.3pt, RoyalBlue]          table[x index = 0, y index = 2] {data/slice-2.dat};
    \addplot[mark=none, line width=1.3pt, Maroon, dashed]     table[x index = 0, y index = 3] {data/slice-2.dat};
    \addplot[mark=none, line width=1.3pt, Maroon]             table[x index = 0, y index = 4] {data/slice-2.dat};
    \addplot[mark=none, line width=1pt, dashdotted]               table[x index = 0, y index = 1] {data/slice-2-exact.dat};

    \node[subcaption] at ({rel axis cs:0.8,0.9}) {$t_0=2651$};

    \end{groupplot}

\end{tikzpicture}
    \caption{Horizontal (left) and vertical (right) slice through Figure~\ref{fig:plot-MT-2}. Labelling is the same as Figure~\ref{fig:slice-1}.}
    \label{fig:slice-2}
\end{figure*}

\pagebreak

To summarise, there is virtually no difference between the direct method and the quantum jump, aside from the fact that the latter inherently increases the statistical error, which manifests as \enquote{banding} in the contour plots. Thus, not jumping is clearly the preferred choice -- it is computationally simpler and requires fewer trajectories without sacrificing accuracy. There is a slight improvement for electronic operators suggested by Figure~\ref{fig:slice-3}, but probably not enough to justify the increased computational cost. Compared to exact results, both MASH and ms-MASH generally work quite well. Some discrepancies can be observed, but these would likely happen with any approximate quantum-classical method.

% !TEX root = main.tex

\chapter{Conclusions and further work}
\label{chapter:conclusion}

In this work, we have shown that the surface hopping prescription $\sigma_z \mapsto \sgn(S_z)$ used by MASH is unique in ensuring the correct long-time limit for a class of similar methods. We have also shown that one can build many different methods by using different \enquote{estimators} on top of this prescription. This explains why both MASH and ms-MASH can be similarly successful. In fact, MASH itself uses this arbitrariness to correctly estimate rates of non-adiabatic processes.

We therefore do not expect that some kind of \enquote{slightly improved MASH} would be possible -- the result suggests that a better method would have to come from a different approach to non-adiabatic dynamics. It also does not seem that a \enquote{best set of estimators} can be found, though it might be interesting to explore why some work better than others for certain models. Instead, we expect that improvements to MASH will come from reducing the computational cost even further -- e.g. by approximations to the quantum jump scheme or by improved sampling of the Bloch sphere to speed up convergence.

We have also investigated whether MASH can be used to calculate multi-time correlation functions. We have shown that this is possible by simulating the physical acquisition of the corresponding spectrum. While we have only considered simple experiments with a single pulse, the method could be easily generalised to more complicated versions. This would also include multi-time TCFs that appear in 2D optical spectroscopy.

We have modified the quantum jump scheme for the calculation of multi-time TCFs and shown that it yields negligible improvement compared to applying the pulses directly to the spin variables. This suggests that the direct method -- which is significantly less expensive than the quantum jump -- can be used for these simulations.

The main goal for MASH in this space is a simulation of a \enquote{realistic} 2D spectrum. While the same method could in principle be used, less expensive methods will likely have to be developed to make such simulations practical. This has already been achieved for other non-adiabatic methods, e.g. using partial linearisation \cite{mannouchPartiallyLinearizedSpinmapping2022} or the generalised quantum master equation \cite{sayerEfficientFormulationMultitime2024}.

% Start of appendices (labels by letters)
\appendix

\pagelayout{wide} % No margins
\addpart{Appendices}
\pagelayout{margin} % Restore margins

% !TEX root = main.tex

\chapter[Background theory -- a companion]{Background theory:\\a companion}
\label{appendix:theory-detail}

\section{Beyond the Born--Oppenheimer approximation}
\label{sec:detail-beyond-BO}

In order to apply the formalism of mixed quantum-classical dynamics, we must first separate the nuclear and electronic degrees of freedom in the \enquote{full} molecular Hamiltonian,\sidenote{Of course, this Hamiltonian does not include spin-dependent effects or the coupling to the electromagnetic field. In fact, spin-orbit coupling may be important at conical intersections -- which is one of the main situations where non-adiabatic dynamics is useful -- and the naive Born--Oppenheimer approximation breaks down in the presence of a magnetic field. \cite{domckeConicalIntersectionsElectronic2004a}}
\[
\hat{H} = \hat{T}_\nuc + \hat{T}_\el + \hat{V}(\vec{r},\vec{R}),
\]
where we have separated the kinetic energy into nuclear and electronic parts. We then define the electronic Hamiltonian
\[
\hat{H}_\el(\vec{R}) = \hat{T}_\el + \hat{V}(\vec{r},\vec{R}),
\]
which depends \emph{parametrically} on the nuclear coordinates $\vec{R}$. The electronic wavefunctions $\phi_i(\vec{r};\vec{R})$ -- again with a parametric dependence on $\vec{R}$ -- are then the eigenfunctions of the electronic Hamiltonian with energies $V_{\el,i}(\vec{R})$.

We can expand the full wavefunction $\Psi(\vec{r},\vec{R})$ as a linear combination of the electronic eigenfunctions with $\vec{R}$--dependent coefficients $\chi_i (\vec{R})$,
\[
    \Psi(\vec{r},\vec{R}) = \sum_i \chi_i (\vec{R}) \ket{\phi_i}.
\]
Substituting the expansion \eqref{eq:born-huang-expansion} into the Schrödinger equation yields, after multiplication by $\bra{\phi_j}$ and integrating over the electronic coordinates,
\begin{equation}
    \label{eq:se-with-nac}
    \left[ \hat{T}_\nuc + V_{\el,j}(\vec{R}) \right] \chi_j(\vec{R}) - \sum_i \hat{\Lambda}_{ji} \, \chi_i(\vec{R})
    = E \chi_j(\vec{R}),
\end{equation}
where
\[
    \hat{\Lambda}_{ji} = \delta_{ji} \hat{T}_\nuc - \bra{\phi_j} \hat{T}_\nuc \ket{\phi_i}_\el
\]
is the non-adiabatic coupling operator, which acts only on the nuclear coordinates. \cite{domckeConicalIntersectionsElectronic2004a}

We can see that if we make the approximation $\hat{\Lambda}_{ji} = 0$,\sidenote{The coupling term is inversely proportional to the energy difference $V_{\el,j}-V_{\el,i}$, which justifies the Born--Oppenheimer approximation for widely separated electronic states. This can also be shown in a different way using perturbation theory with $(m/M)^{1/4}$ as the small parameter. \cite{domckeConicalIntersectionsElectronic2004a,bornZurQuantentheorieMolekeln1927}} we obtain
\[
    \left[ \hat{T}_\nuc + V_{\el,j}(\vec{R}) \right] \chi_j(\vec{R}) = E \chi_j(\vec{R}),
\]
which is the well known Born--Oppenheimer approximation -- the nuclear wavefunction $\chi_j$ only evolves on the potential energy surface given by $V_{\el,j}(\vec{R})$.

In order to make progress with \eqref{eq:se-with-nac}, we need to invoke the full form of the nuclear kinetic energy,
\[
    \hat{T}_\nuc = \sum_I -\frac{\hbar^2}{2M_I} \vec{\nabla}_I^2 .
\]
We can evaluate the action of $\hat{\Lambda}_{ji}$ on $\chi_i(\vec{R})$ by two applications of the product rule,
\[
    \hat{\Lambda}_{ji}\,\chi_i(\vec{R}) = \sum_I \frac{\hbar^2}{2M_I} \Big\{ \bra{\phi_j} \vec{\nabla}_I^2 \ket{\phi_i} \chi_i(\vec{R}) +  2  \bra{\phi_j} \vec{\nabla}_I \ket{\phi_i} \cdot \vec{\nabla}_I\chi_i(\vec{R}) \Big\}.
\]
Thus, we can see that the non-adiabatic effects are expressed through \emph{derivative} couplings -- elements of the form
\[
    \vec{F}_{ji}^I = \bra{\phi_j} \vec{\nabla}_I \ket{\phi_i},
\]
which are the \emph{vector} couplings (since the gradient $\vec{\nabla}_I$ is a vector) and
\[
    G_{ji}^I = \bra{\phi_j} \vec{\nabla}_I^2 \ket{\phi_i},
\]
the \emph{scalar} couplings (since the Laplacian $\vec{\nabla}_I^2$ is a scalar).

We can then write the nuclear Schrödinger equation as
\[
    \sum_I -\frac{\hbar^2}{2M_I} \left( \vec{\nabla}_I + \vec{F}^I \right)^2 \vec{\chi} + \vec{V}\vec{\chi} = E\vec{\chi},
\]
where $\vec{\chi} = \left(\chi_1 \; \chi_2 \; \cdots \; \right)^T$ is the nuclear wavefunction expressed as a vector in the basis of adiabatic electronic states and $\vec{V}$ is a diagonal matrix of the adiabatic electronic energies $V_{ij}=V_{\el,i}(\vec{R})\,\delta_{ij}$.

The non-adiabatic coupling appears as a term modifying the kinetic energy by an extra potential -- this is very similar to what happens in the presence of a magnetic field.

In electronic structure literature, we consider the gradients $\vec{\nabla}_I$. However, in mixed quantum-classical dynamics, we will generally work with individual coordinates and consider the derivatives $\pder{}{q_I}$ with respect to the nuclear coordinate $q_I$ instead. Additionally, instead of $\vec{F}^I$, we shall call the non-adiabatic coupling matrix $\vec{d}^I$ -- the letter $F$ is reserved for the force. By the same reasoning as above, this will lead to the equation
\[
    \sum_I -\frac{\hbar^2}{2M_I}\left( \pder{}{q_I} + \vec{d}^I \right)^2 \vec{\chi} + \vec{V}\vec{\chi} = E\vec{\chi},
\]
which can also be rewritten as
\begin{equation}
    \label{eq:SE-exact-separated}
    \sum_I \frac{1}{2M_I}\left( p_I - \imag\hbar\vec{d}^I \right)^2 \vec{\chi} + \vec{V}\vec{\chi} = E\vec{\chi}.
\end{equation}

Note that we will generally work in coordinates scaled so that every degree of freedom has the same mass $M$, which is taken to be some kind of \enquote{average nuclear mass}.

\section{Density matrices and the partial trace}
\label{sec:detail-density-matrix}

In order to describe quantum-classical systems, we shall represent our system by a \emph{density matrix} rather than a wavefunction. This will allow us to move to a phase space description of the nuclear system and there, we will finally be able to take the classical limit.\sidenote{Note that even in the adiabatic case, the Born--Oppenheimer approximation by itself does not say anything about the classical limit -- the nuclei are still quantum objects. However, if the electronic state does not change, taking the classical limit is quite trivial.} We shall also introduce the \emph{partial trace} and show that it generally leads to \emph{mixed states} that cannot be described purely by wavefunctions.

If a system is described by a wavefunction $\ket{\psi}$, its density matrix (or density operator) is given by\sidenote{Two things can be noted about this expression. Firstly, if our wavefunctions are vectors, this is just the vector outer product $\vec{\rho} = \vec{\psi} \vec{\psi}^\dagger$. Secondly, $\rho^2 = \ket{\psi}\braket{\psi}{\psi}\bra{\psi} = \ket{\psi}\bra{\psi} = \rho$. In fancy language, the density matrix of a pure state is idempotent. (We have to normalise our wavefunctions for this to hold -- e.g. the identity matrix is not a pure state.)}
\[
    \rho = \ket{\psi}\bra{\psi}.
\]
Differentiating this expression with respect to $t$ and remembering the Schrödinger equation yields the \emph{Liouville--von Neumann equation},
\begin{equation}
    \label{eq:liouville-von-neumann}
    \dot\rho = -\frac{\imag}{\hbar} [\hat{H}, \rho].
\end{equation}

So far, we have not gained anything -- this equation is equivalent to the Schrödinger equation, but is more difficult to solve, since we are dealing with matrices rather than vectors. However, density matrices shine when we want to decompose our system into parts -- this is usually the case when we have a small system of interest and a huge environment it interacts with (imagine a molecule in solution). Describing the system by a wavefunction would be next to impossible; however, with a density matrix, we will be able to \enquote{trace out} the environment degrees of freedom and only concentrate on the system.\sidenote{Of course, an exact description of the full system+environment problem would not be any easier with density matrices. But the power of this approach is that it allows us to make physically justified approximations that would not be possible in the wavefunction picture.}

For a very simple example, suppose that our system of interest has two possible states, $\ket{1}$ and $\ket{2}$, and our environment also has two states, $\ket{E}$ and $\ket{E^*}$. The complete wavefunction would thus be a superposition of \emph{four} states,\sidenote{We can choose any system state to go with any environment state, which gives us $2\times2=4$ states in total. In general, this construction is known as the \emph{tensor product}.} $\ket{1E}$, $\ket{1E^*}$, $\ket{2E}$ and $\ket{2E^*}$. Suppose that the complete wavefunction is
\[
    \ket{\psi} = \frac{1}{\sqrt{2}} \ket{1E} + \frac{\eul^{\imag\alpha}}{\sqrt{2}} \ket{2E^*} ,
\]
which could be written as the vector $(1/\sqrt{2} \quad 0 \quad 0 \quad \eul^{\imag\alpha}/\sqrt{2})^T$. The corresponding density matrix is
\[
    \rho = \begin{pmatrix}
        1/2 & 0 & 0 & \eul^{-\imag\alpha}/2 \\
        0   & 0 & 0 & 0 \\
        0   & 0 & 0 & 0 \\
        \eul^{\imag\alpha}/2 & 0 & 0 & 1/2
    \end{pmatrix}.
\]
Now, we do not care what the environment is doing, so we would like to treat the elements like $\ket{1E}\bra{1E}$ and $\ket{1E^*}\bra{1E^*}$ as the \enquote{same} state. The formal way to do this is a \emph{partial trace} over the environment, which gives us the reduced density matrix
\begin{marginfigure}
    \begin{tikzpicture}[
    %Global config
    >=latex,
    line width=1pt,
    %Styles
    Brace/.style={
        decorate,
        decoration={
            brace,
            raise=-7pt
        }
    },
    Matrix/.style={
        matrix of nodes,
        % text height=2.5ex,
        % text depth=0.75ex,
        % text width=3.25ex,
        align=center,
        %left delimiter=\{,
        %right delimiter=\},
        % column sep=5pt,
        % row sep=5pt,
        % nodes={draw=black!10}, % Uncoment to see the square nodes.
        nodes in empty cells,
    },
    DL/.style={
        left delimiter=[,
        right delimiter=],
        inner sep=-2pt,
    },
    DG/.style={
        line cap= round,
        line width =15pt,
        opacity=0.2,
    }
]

\matrix[Matrix] at (0,0) (M){ % Matrix contents
    $\frac 1 2$ & 0 & 0 & $\frac {e^{-i\alpha}} 2$ \\
    0 & 0 & 0 & 0 \\
    0 & 0 & 0 & 0 \\
    $\frac {e^{i\alpha}} 2$ & 0 & 0 & $\frac 1 2$ \\
};

\matrix[Matrix] at (2.7,0) (M1){ % Matrix contents
    $\frac 1 2$ & 0 \\
    0 & $\frac 1 2$ \\
};

\begin{scope}[on background layer] 
    %FOR MATRIX M
    %To delimit internal braces
    \node[DL,fit=(M-1-1)(M-4-4)](subM-1){};
    \node[DL,fit=(M1-1-1)(M1-2-2)](subM1-1){};
    % For line sectors
    \draw[DG,red](M-1-1.center) --(M-2-2.center);
    \draw[DG,red](M-1-3.center) --(M-2-4.center);
    \draw[DG,red](M-3-3.center) --(M-4-4.center);
    \draw[DG,red](M-3-1.center) --(M-4-2.center);

    % For line sectors
    \draw[DG,red](M1-1-1.center) -- (M1-1-1.center);
    \draw[DG,red](M1-1-2.center) -- (M1-1-2.center);
    \draw[DG,red](M1-2-1.center) -- (M1-2-1.center);
    \draw[DG,red](M1-2-2.center) -- (M1-2-2.center);
    
    % arrow
    \draw[->,line width=1pt,black] (M.east) -- (M1.west) ;  
\end{scope}

\end{tikzpicture}
\end{marginfigure}
\[
    \rho^{\mathrm{sys}}_{s,s'} = \tr{\rho} = \sum_{e=E,E^*} \rho_{se, s'e} .
\]
Performing the partial trace leads to
\[
    \rho^{\mathrm{sys}} = \begin{pmatrix}
        1/2 & 0 \\
        0   & 1/2
    \end{pmatrix}.
\]

We can note several things about this. Firstly, it tells us that a half of the system is in state $\ket{1}$ and the other half in state $\ket{2}$ -- this was hopefully expected. Secondly, we have lost the phase factor $\eul^{\imag\alpha}$ -- by tracing out the environment, we have lost some of the coherence.\sidenote{This is not as important for large classical environments, where decoherence is very fast.} And thirdly -- the reduced density matrix does not correspond to a wavefunction;\sidenote{We can either note that it is not idempotent or that it has a non-zero determinant, whereas the outer product of vectors always has determinant zero.} it is a \emph{mixed state}!

\section{The Wigner transform -- quantum mechanics in phase space}
\label{sec:detail-wigner-transform}

Just like classical mechanics, quantum mechanics can be formulated in several different \enquote{arenas}. \cite{schwichtenbergNononsenseClassicalMechanics2020} The most widely known is the Hilbert space formulation, where the states $\ket{\psi}$ live in a Hilbert space and are acted on by operators like the Hamiltonian. However, we can also formulate quantum mechanics in real space,\sidenote{Quantum mechanics in real space is known as \emph{pilot wave theory} or \emph{Bohmian mechanics}. Its classical equivalent is Newtonian mechanics and it might therefore seem to be the ideal candidate for developing trajectory-based methods. Even though many people have tried, no such method is currently in common use.} configuration space\sidenote{The configuration space or \emph{path integral} formulation -- the quantum equivalent of Lagrangian mechanics -- has proven to be very useful in the disparate realms of QFT and semiclassical dynamics.} or even in phase space. The last of these will be the most useful for discussing mixed quantum--classical dynamics.

In classical mechanics, phase space is useful for discussing probability densities -- this is crucial in statistical mechanics. If the system is described by a probability distribution $\rho(p, q),$\sidenote{Note the subtle foreshadowing by also using $\rho$ as a symbol for the density matrix.} the average value of an observable $A(p,q)$ is given by the phase space average
\begin{equation}
    \label{eq:classical-avg}
    \avg{A}_{\mathrm{cl}} = \frac{1}{2\pi\hbar}\iint \dd p \dd q \, \rho(p, q)A(p,q).
\end{equation}
In quantum mechanics, the average value of an observable $\hat{A}$ is
\[
    \avg{A}_{\mathrm{QM}} = \bra{\psi} \hat{A} \ket{\psi},
\]
which can also be written as\sidenote{While the identity with $\psi$ only makes sense for pure states, writing it in terms of $\rho$ extends it to mixed states as well.}
\[
    \avg{A}_{\mathrm{QM}} = \Tr{\, \ket{\psi}\bra{\psi} \hat{A}} = \Tr{\rho \hat{A}}.
\]

In the phase space formulation, we replace quantum operators by their \emph{Wigner transforms}, which are functions of $p$ and $q$. These are defined so that the quantum average has the form of a classical average \eqref{eq:classical-avg},
\[
    \avg{A}_{\mathrm{QM}} = \frac{1}{2\pi\hbar}\iint \dd p \dd q \, \rho_W(p, q)A_W(p,q).
\]
\marginnote{Note that the phase space average is often written in terms of the Wigner quasi-probability distribution,
\[f_W(p,q) = \frac{1}{2\pi\hbar} \rho_W(p, q).\] The prefix quasi- is used because it can in general have negative regions.}

The Wigner transform of the operator $\hat{A}$ is given by\sidenote{$\ket{q}$ is the position eigenstate \[\hat{q}\ket{q} = q\ket{q}.\]}
\[
    A_W(p, q) = \int_{-\infty}^{\infty} \dd s \, \eul^{\imag ps/\hbar} \bra{q-\frac{s}{2}} \hat{A} \ket{q+\frac{s}{2}}.
\]
The Wigner transform of the density operator is given in the same way and for a pure state $\rho = \ket{\psi}\bra{\psi}$,
\begin{equation}
    \label{eq:wigner-transform-pure-state}
    \rho_W(p, q) = \int_{-\infty}^{\infty} \dd s \, \eul^{\imag ps/\hbar} \, \psi \left(q-\frac{s}{2}\right) \, \psi^* \left(q+\frac{s}{2}\right) .
\end{equation}

We shall need to know that the Wigner transform of a product is
\[
    (\hat{A}\hat{B})_W = A_W \eul^{-\imag\hbar \Lambda/2} B_W,
\]
where $\Lambda$ is an operator defined by \marginnote{
    The operator $\Lambda$ is usually written as $\overleftarrow{\pder{}{p}}\,\overrightarrow{\pder{}{q}} - \overleftarrow{\pder{}{q}}\,\overrightarrow{\pder{}{p}}$, where the arrow shows the direction in which the derivative acts.
}
\[
    f \, \Lambda \, g = \pder{f}{p}\pder{g}{q} - \pder{f}{q}\pder{g}{p},
\]
which is the negative of the \emph{Poisson bracket} $\{f,g\}$ in classical mechanics.

\section{Time evolution and the classical limit}
\label{sec:detail-classical-limit}

Taking the Wigner transform of the Liouville--von Neumann equation \eqref{eq:liouville-von-neumann} gives
\begin{align*}
    \dot{\rho}_W & = -\frac{\imag}{\hbar} \left( H_W \eul^{-\imag\hbar \Lambda/2} \rho_W  - H_W \eul^{\imag\hbar \Lambda/2} \rho_W \right) \\
                 & = -\frac{2}{\hbar} H_W \sin \left( \frac {\hbar \Lambda} {2} \right) \rho_W.
\end{align*}
In the classical limit $\hbar \to 0$, this becomes\sidenote{$\sin\frac{\hbar\Lambda}{2} = \frac{\hbar\Lambda}{2} + \mathcal{O}(\hbar^3)$}
\[
    \dot{\rho}_W = - H_W \, \Lambda \, \rho_W + \mathcal{O}(\hbar^2),
\]
which is the classical \emph{Liouville equation}\sidenote{This is usually written in terms of the Poisson bracket, $\dot\rho = \{H,\rho\}$.} -- we have recovered classical equations of motion in the Hamiltonian formalism!

We also need to look at what happens to time-evolved operators $\hat{A}(t) = \eul^{\imag \hat{H} t/\hbar} \hat{A} \eul^{-\imag \hat{H} t/\hbar} $ in the classical limit. Since these obey the Heisenberg equation of motion,
\[
    \der{}{t} \hat{A}(t) = \frac{\imag}{\hbar} [\hat{H}, \hat{A}(t)],
\]
we can use the same method as above to write
\[
    \dot{A}(p,q;t)_W = H_W \, \Lambda \, A(p,q;t)_W + \mathcal{O}(\hbar^2).
\]
In fact, since all evolution is now classical, we obtain\sidenote{A possible proof is to expand $A_W$ as a Taylor series in $p$ and $q$ and apply the classical time evolution to each individual term.}
\[
    A(p,q;t)_W \to A_W(p_t, q_t),
\]
so that we can simply run trajectories for a period of time $t$ and evaluate the observable at the final phase space point $(p_t,q_t)$.

For \emph{mixed quantum-classical mechanics}, we follow the same procedure, but we instead of the full Wigner transform, we use a \emph{partial} Wigner transform. Our system has two parts -- nuclear and electronic. We want to keep the latter quantum, but transform the former into phase space.\sidenote{This is conceptually similar to the partial trace, but the Wigner transform does not lose any information.} Thus, if $\hat{\mathcal{A}}$ is the full nuclear--electronic operator, after the partial Wigner transform, it will become $\hat{A}(\vec{p},\vec{q})$ -- a function of the nuclear phase space coordinates $\vec{p}$ and $\vec{q}$, but an operator in the electronic space.

Performing the necessary algebra\sidenote{Note that properly, the limit of large nuclear mass is taken, rather than $\hbar\to 0$. See \cite{kapralMixedQuantumclassicalDynamics1999} for details.} leads to the \emph{quantum-classical Liouville equation} (QCLE),
\begin{align}
    \label{eq:qcle}
    \pder{\hat{\rho}_W(t)}{t} = -\frac{\imag}{\hbar} \left[ \hat{H}_W, \hat{\rho}_W(t) \right] - \frac 1 2 \left( \hat{H}_W \, \Lambda \, \hat{\rho}_W(t) - \hat{\rho}_W \, \Lambda \, \hat{H}_W(t) \right).
\end{align}

A similar equation can also be written for evolution of operators is the Heisenberg picture, which only differs by sign.

\section{Long-time behaviour -- conservation of the Boltzmann distribution}
\label{sec:detail-long-time}

Many applications are focused on systems surrounded by a large thermal environment -- such systems will reach equilibrium as described by the canonical ensemble. Classically, the equilibrium state is described by the equilibrium phase space density
\[
    \rho_{\mathrm{eq}}(p,q) \propto \eul^{-\beta H(p,q)}
\]
and quantum-mechanically by the thermal density matrix\sidenote{Note that using the Boltzmann distribution already neglects the exchange symmetry of particles, but that is usually unimportant for atomic nuclei (except at extremely low temperatures).}
\[
    \hat{\rho}_{\mathrm{eq}} \propto \eul^{-\beta \hat{H}} .
\]

The equilibrium state ought to be stationary -- and it is easily shown that this is the case both in classical and quantum mechanics.\sidenote{In quantum mechanics, time evolution is generated by the commutator with $\hat{H}$, \eqref{eq:liouville-von-neumann}, which is clearly zero for the operator $\eul^{-\beta \hat{H}}$.} However, in approximate approaches, we need to make sure that this is still the case -- if so, the method is said to \emph{conserve the Boltzmann distribution} and therefore thermalises correctly.\sidenote{That does not ensure complete correctness -- it can still thermalise in the wrong way, for example too slowly.}

Method like RPMD exactly conserve the quantum Boltzmann operator $\eul^{-\beta \hat{H}}$, whereas current mixed quantum--classical approaches manage at most to conserve the classical Boltzmann distribution. \cite{amatiDetailedBalanceMixed2023, amatiDetailedBalanceNonadiabatic2023} This means that they thermalise systems as if they were classical -- they suffer from the problem of \emph{zero point energy leakage}. \cite{habershonZeroPointEnergy2009} For example, a quantum harmonic oscillator has a ground state of energy $\hbar \omega /2$, whereas a classical one can have an energy of exactly zero. The extra $\hbar \omega /2$ can then \enquote{leak} into the electronic subsystem and artificially heat it up. This becomes less of a problem at high temperatures, since the zero point energy is then negligible compared to $kT$.

\begin{marginfigure}[*-6]
    \centering
    \includegraphics[width=0.85\textwidth]{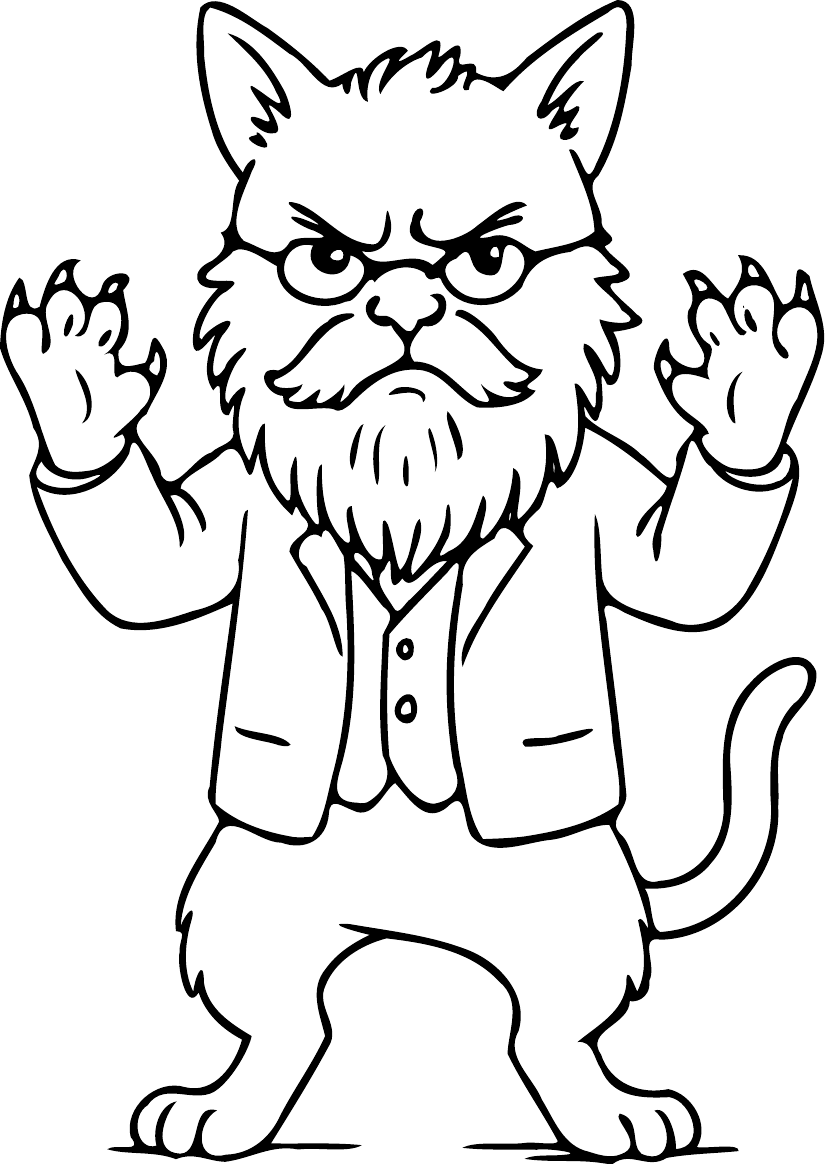}
    \caption{Ludwig Boltzmann's (1844--1906) cat avatar angry at~approaches that do not thermalise correctly. Courtesy of ChatGPT.}
\end{marginfigure}

\section{MASH estimators -- the full story}
\label{sec:detail-MASH-estimators}

We have already seen a simple example of how MASH estimates correlation functions in the CPA. How do we proceed with a general correlation function $C_{\mathcal{A}\mathcal{B}}(t) = \Tr{\hat{\mathcal{A}} \hat{\mathcal{B}}(t)}$ when both nuclear and electronic degrees of freedom might be involved? We need to take partial Wigner transforms and then decompose the electronic part into the basic operators $P_+$, $P_-$, $\sigma_x$ and $\sigma_y$, which are mapped to
\begin{align*}
    \hat{P}_+ & \mapsto h(S_z) \\
    \hat{P}_- & \mapsto h(-S_z) \\
    \hat{\sigma}_x &\mapsto S_x \\
    \hat{\sigma}_y &\mapsto S_y .
\end{align*}

The partial Wigner transform gives the exact expression\sidenote{Note that $\hat{B}(\vec{p},\vec{q};t)_W$ is the partial Wigner transform of $\hat{\mathcal{B}}(t)$.}
\[
    C_{\mathcal{A}\mathcal{B}}(t) = \frac{1}{(2\pi\hbar)^f}\int \dd\vec{p} \, \dd\vec{q} \, \, \tr{\hat{A}_W(\vec{p},\vec{q}) \hat{B}(\vec{p},\vec{q};t)_W},
\]
which is not very useful -- the Wigner transform of a time-evolved operator is not a well-behaved quantity. However, in the classical limit, we can write (see \autoref{sec:detail-classical-limit})
\[
    \hat{B}(\vec{p},\vec{q};t)_W \longrightarrow \hat{B}_W(\vec{p}_t, \vec{q}_t;t),
\]
where only the electronic part of the latter depends directly on $t$. Thus, we have transformed the dependence on $\vec{p}$ and $\vec{q}$ to something that can be evaluated for a single trajectory. However, the electronic parts still appear as quantum operators.

This is where a second crucial ingredient of MASH comes in. We shall -- just like other spin mapping approaches -- represent the electronic operator $\hat{A}$ as a function $A(\vec{S})$ of the spin variables. This is easy enough to do -- any operator is a linear combination of the basic operators $O = P_+, P_-, \sigma_x, \sigma_y$,\sidenote{We always need to decompose a general operator, because the weighting factors are only defined for the basic operators. We shall see below that this is not the case in ms-MASH.}
\[
    \hat{A}_W(\vec{p},\vec{q}) = \sum_{O} a_O(\vec{p}, \vec{q}) \, \hat{O},
\]
and then we simply apply the MASH mapping to the individual operators to yield
\[
    A(\vec{p},\vec{q},\vec{S}) = \sum_{O} a_O(\vec{p}, \vec{q}) \, O(\vec{S}),
\]
where $O(\vec{S})$ is the mapped version of $\hat{O}$ as given above.

The time-evolved version is the same, with $S_i$ replaced by $S_i(t)$. The correlation functions are then expressed as weighted averages over the Bloch sphere,
\begin{widepar}
\begin{equation}
    \label{eq:mash-estimator-full}
    C^\MASH_{\mathcal{A}\mathcal{B}}(t) = \sum_{O,O'} \frac{1}{(2\pi\hbar)^f}\int \dd\vec{p} \dd\vec{q} \, a_O(\vec{p}, \vec{q}) \int \dd \vec{S} \,
     b_{O'}(\vec{p}_t, \vec{q}_t) \, \underbrace{W_{OO'}(\vec{S}) O(\vec{S}) O'(\vec{S}_t)}_\text{basic MASH estimator},
\end{equation}    
\end{widepar}
with the proper weighting factor $W_{OO'}(\vec{S})$ for each combination of basic operators. Note that $\vec{p}_t$ and $\vec{q}_t$ depend on $\vec{S}$, so they have to be inside the inner integral. As a final note, starting from factorised initial conditions $\hat{\mathcal{A}} = \rho_\nuc \hat{A}$ (luckily) makes the partial Wigner transform trivial.

The addition of an $\vec{S}$-dependent weighting factor corresponds either to weighting the points near the poles more strongly, or to sampling points preferentially closer to the poles. The former interpretation is consistent with uniform sampling over the Bloch sphere, as is done in the MASH literature \cite{mannouchMappingApproachSurface2023,amatiDetailedBalanceMixed2023,lawrenceRecoveringMarcusTheory2024}, whereas the latter is what the ms-MASH literature uses \cite{runesonMultistateMappingApproach2023,runesonExcitonDynamicsMapping2024}.

\section{Overview of ms-MASH}
\label{sec:detail-msMASH-estimators}

The ms-MASH method is a adaptation of MASH to $N$-level systems. It is formulated in terms of wavefunction coefficients $\vec{c}$ rather than spin variables; however, the main idea is similar -- we propagate the electronic system on the classical path and we construct a surface hopping--inspired conserved energy function for the nuclei. The ms-MASH energy function is given by
\[
    E(\vec{p}, \vec{q}, \vec{c}) = \sum_i^{\text{nuc. coords}} \frac{p_i^2}{2m} + \sum_a^{\text{el. states}} V_a(\vec{q})\,\Theta_a,
\]
where $\Theta_a = 1$ if $a$ is the most populated electronic state and $0$ otherwise. Differentiating this expression gives us an expression for the ms-MASH force with hopping terms similar to what we obtained for MASH -- in fact, the dynamics for ms-MASH in the two-state case is the same as MASH dynamics.\sidenote{This is easily seen by replacing
\begin{align*}
    V_+ & = \bar{V} + V_z\\
    V_- & = \bar{V} - V_z\\
    \Theta_+ & = \frac 1 2 + \frac 1 2 \sgn(S_z) \\
    \Theta_- & = \frac 1 2 - \frac 1 2 \sgn(S_z) 
\end{align*}
}

The main difference between MASH and ms-MASH lies in the construction of estimators. ms-MASH estimators are chosen to thermalise correctly in the long-time limit and to be \emph{unitary} -- this means that the same estimator can be used for populations, coherences and even diabatic observables. This makes it much simpler compared to \eqref{eq:mash-estimator-full}. While unitarity is not an entirely natural requirement -- we have already chosen the $z$~axis as special for the dynamics -- it does make the resulting expressions much simpler.

Unlike MASH, ms-MASH does not use weighting factors -- instead, it uses different estimators\sidenote{In ms-MASH literature, the initial estimators are referred to as \enquote{initial conditions}, since they can be thought of as non-uniform sampling of the Bloch sphere.} for the initial and final operators. However, this is really the same idea in disguise, as shall become apparent in \autoref{sec:detail-MASH-pauli-matrices}.

The main problem with ms-MASH is that it is not size consistent -- adding even a fully uncoupled state will change the results. \cite{lawrenceSizeconsistentMultistateMapping2024a} This might be problematic for real-world problems, where we need to choose how many states are used for truncation. There might also be consequences for the thermalisation rate, but that has not been systematically explored. 

\section{Spin variable version of ms-MASH}
\label{sec:detail-msMASH-spin-variables}

While ms-MASH was originally formulated in terms of wavefunction coefficients, we shall find it desirable to rewrite it in terms of spin variables.\sidenote[][*0]{We could instead rewrite MASH in terms of wavefunction coefficients -- the results are outlined in \cite{runesonExcitonDynamicsMapping2024} -- but we shall find the spin-sphere formulation advantageous for describing two-level systems.} Note that we will end up with two estimators for each observable -- one for the initial operator and the other for the final one. The transformation is easily accomplished using the relationships \eqref{eq:wf-to-spin}. We shall therefore just state the results here.

The populations $P_\pm$ are initialised from the corresponding hemisphere,\sidenote[][*0]{A small note is that we have used the \enquote{cap initial conditions} from the original ms-MASH paper \cite{runesonMultistateMappingApproach2023}. Other ones have been proposed in \cite{runesonExcitonDynamicsMapping2024}, for example the \enquote{equivariant} ones, $$ P_\pm \mapsto \frac{2 + 3S_z}{4}.$$} so we obtain
\[
P_\pm \mapsto h(\pm S_z)
\]
as the initial estimator, which agrees with MASH. However, the final estimator is given by
\[
P_\pm(t) \mapsto  \frac 1 2 \pm S_z(t)  .
\]

Coherences are expressed as populations in the rotated basis. This is explained in detail in \cite{runesonMultistateMappingApproach2023}. For example, $\sigma_x$ is the population difference between the \enquote{east} and \enquote{west} hemispheres, which gives
\begin{align*}
    \sigma_x & \mapsto \sgn ( S_x) \\
    \sigma_y & \mapsto \sgn ( S_y)
\end{align*}
The final estimators are simply given by
\begin{align*}
    \sigma_x(t) & \mapsto 2 S_x(t) \\
    \sigma_y(t) & \mapsto 2 S_y(t)
\end{align*}

Note that the population--coherence correlation function, $C_{+x}(t)$ is the same in both MASH and ms-MASH (see Boxes~\ref{box:MASH-estimators} and~\ref{box:ms-MASH-estimators}) -- we shall make use of this fact in \autoref{sec:consistent-derivatives}.

\section[Bridge between MASH and ms-MASH -- the \texorpdfstring{$\sigma_z$--$\sigma_z$}{Sz--Sz} correlation function]{Bridge between MASH and ms-MASH -- the \texorpdfstring{$\vec{\sigma_z}$--$\vec{\sigma_z}$}{Sz--Sz} correlation function}
\label{sec:detail-MASH-pauli-matrices}

So far, the estimators have been defined in terms of populations and coherences. However, it seems more natural to use the basis of Pauli matrices -- i.e. to use the identity and $\sigma_z$ instead of $P_+$ and $P_-$. We can use the relationships $\hat{I} = \hat{P}_- + \hat{P}_+$ and $\hat{\sigma}_z = \hat{P}_+ - \hat{P}_-$ to write
\begin{align*}
    \hat{I} &\mapsto 1 \\
    \hat{\sigma}_z &\mapsto \sgn(S_z)
\end{align*}
so that MASH represents the identity operator exactly.\sidenote{This is known as \emph{trace preservation} and has been shown to be beneficial in non-adiabatic approaches. \cite{ullahMachineLearningMeets2025}}

For MASH, we obtain the set of correlation functions\sidenote{We have omitted the phase space integrals, but the first two identities continue to hold, since they only involve integrals over the initial spin variables.}
\begin{align*}
    C^\MASH_{II}(t) & = \int \dd\vec{S} \, 2|S_z| = 2\\
    C^\MASH_{zI}(t) & = \int \dd\vec{S} \, 2S_z = 0\\
    C^\MASH_{Iz}(t) & = \int \dd\vec{S} \, 2|S_z|\sgn(S_z(t)) \\
    C^\MASH_{zz}(t) & = \int \dd\vec{S} \, 2S_z\,\sgn(S_z(t))
\end{align*}

The ms-MASH correlation functions in this basis can be obtained in the same way to yield
\begin{align*}
    C^\MISH_{II}(t) & = \int \dd\vec{S} = 2\\
    C^\MISH_{zI}(t) & = \int \dd\vec{S} \, \sgn(S_z) = 0\\
    C^\MISH_{Iz}(t) & = \int \dd\vec{S} \, 2S_z(t) \\
    C^\MISH_{zz}(t) & = \int \dd\vec{S} \, \sgn(S_z)\, 2S_z(t)
\end{align*}

Several interesting observations can be made about these correlation functions. Firstly, when written in this form, both the MASH and ms-MASH correlation functions use a different initial and final estimator. Secondly, the ms-MASH correlations functions follow directly from the mapping
\begin{align*}
    \hat{I} &\mapsto 1 \\
    \hat{\sigma}_z &\mapsto \sgn(S_z)
\end{align*}
applied to the initial observable. (But $\hat{\sigma}_z(t)$ is represented by $2S_z(t)$.) They therefore seem to be more \enquote{natural} in some sense.

And lastly, and most importantly -- the $C_{zz}(t)$ correlation functions are just time-reversed versions of each other!\sidenote{It has also been proposed to use $2|S_z(t)|$ instead of $2|S_z|$ as the weighting factor in MASH. \cite{mannouchMappingApproachSurface2023} This would result in the same form of $C_{zz}(t)$ as ms-MASH, but the $C_{II}(t)$ and $C_{Iz}(t)$ correlation functions would fail to be reproduced exactly.} Thus, MASH and ms-MASH are not so different as they may initially have seemed; though the form of $C_{Iz}(t)$ is still quite different for both methods.

\section{MASH propagation in the Schrödinger picture}
\label{sec:detail-MASH-schrodinger-picture}

So far, we have described MASH propagation in the Heisenberg picture -- we have used the spin variables to estimate the Heisenberg time evolved operator $\hat{B}(t)$, which enters the correlation function. However, we can equivalently rewrite the expression for $C_{\rho\mathcal{B}}$ as
\[
    C_{\rho\mathcal{B}} = \Tr{ \eul^{-\imag \hat{H}t/\hbar} \rho \eul^{\imag \hat{H}t/\hbar} \hat{\mathcal{B}}}
\]
and identify
\[
    \rho_S(t) = \ket{\Psi(t)}\bra{\Psi(t)} = \eul^{-\imag \hat{H}t/\hbar} \rho \eul^{\imag \hat{H}t/\hbar}
\]
as the time-evolved density matrix in the Schrödinger picture.

\begin{figure*}[h]
    \centering
    \includegraphics[width=1.41\textwidth, angle=-90, origin=c]{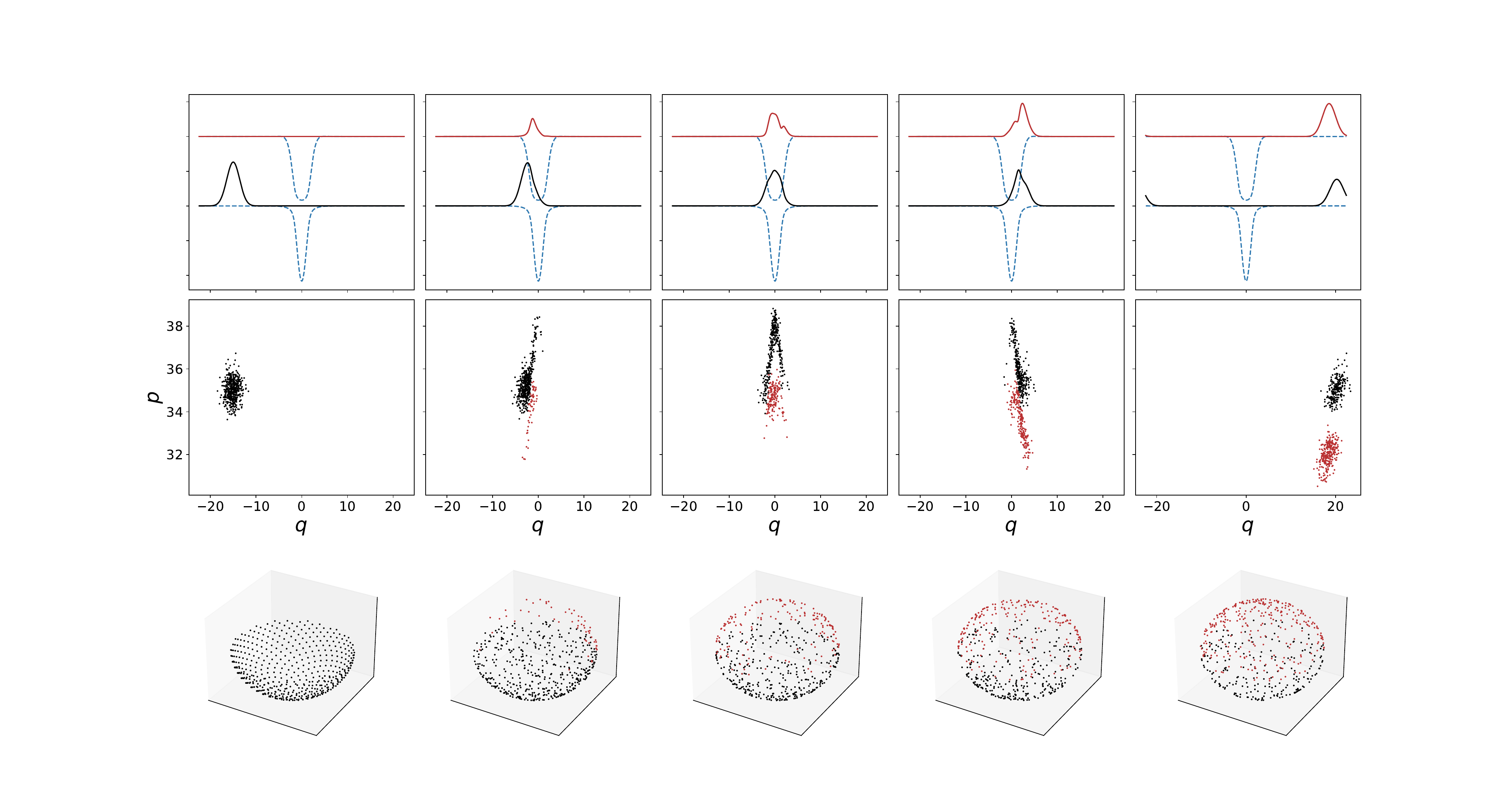}
    \caption{Time evolution of Tully's model II initialised in a wavepacket with $q_0=-15$, $p_0=35$ and $\gamma=0.5$ in the lower adiabat. The top row shows the adiabatic wavefunctions of upper and lower states. Only the absolute value $|\psi|$ is shown. The middle row shows MASH trajectories in phase space. The bottom row shows the same trajectories on the Bloch sphere. Black generally corresponds to the lower adiabat, red to the upper one.}
    \label{fig:MASH-master}
\end{figure*}

This suggests that something similar could be done in MASH -- we can propagate the initial density matrix $\rho$ to obtain a time-evolved version $\rho^\MASH_S(\vec{p}_t, \vec{q}_t,t)$. In fact, this is basically what is being represented in Figure~\ref{fig:MASH-master}. The density matrix is a distribution in the extended phase space $(\vec{p},\vec{q},\vec{S})$ and we simply follow it as we evolve the MASH dynamics.

Formally, we can obtain the expression for $\rho^\MASH_S(\vec{p}_t, \vec{q}_t,t)$ by requiring that we obtain the same result for $C_{\rho\mathcal{B}}(t)$ as in usual MASH. Thus, we need
\[
    C_{\rho\mathcal{B}}(t) = \frac{1}{(2\pi\hbar)^f} \iint \dd \vec{p}_t \dd \vec{q}_t \tr{\rho^\MASH_S(\vec{p}_t, \vec{q}_t,t) \hat{B}(\vec{p}_t, \vec{q}_t)}.
\]
Note that we have evolved the classical variables by the usual prescription; only the electronic part is treated in the Schrödinger picture. If $\rho$ is given by the product $\rho_\nuc \hat{A}$, we can make the informed guess (guided by \cite{mannouchMappingApproachSurface2023}) that\sidenote[][*-2]{The delta functions are necessary because $\vec{p}'_t$ and $\vec{q}'_t$ depend on the initial value of $\vec{S}$.}
\begin{widepar}
\begin{equation}
    \label{eq:density-matrix-MASH-evolution}
    \hat\rho^\MASH_S(\vec{p}_t, \vec{q}_t,t) = \iint \dd \vec{p}' \dd \vec{q}' \, \rho_\nuc(\vec{p}',\vec{q}') \int \dd\vec{S} \, \delta(\vec{p}_t-\vec{p}'_t) \delta(\vec{q}_t-\vec{q}'_t) \, A(\vec{S}) \, \hat{\mathcal{W}}_A(\vec{S},t)
\end{equation}
\end{widepar}
and find the form of $\hat{\mathcal{W}}_A(\vec{S},t)$. After some algebra, this reduces to the condition
\[
    \tr{\hat{\mathcal{W}}_A(\vec{S},t) \hat{B}} = W_{AB}(\vec{S})B(\vec{S}(t)),
\]
where $W_{AB}$ is the usual MASH weighting factor. Assuming $\hat{B} = \sigma_j$ and writing $\hat{\mathcal{W}}_A = \sum_i w_i\sigma_i$ for $i=0,x,y,z$ gives
\begin{equation}
    w_j = \frac 1 2 W_{AB}(\vec{S}) \, \sigma^\MASH_j [\vec{S}(t)],
\end{equation}
where $\sigma^\MASH_j[\vec{S}(t)]$ is the MASH estimator for $\sigma_j(t)$ -- i.e., $\sigma^\MASH_0 = 1$, $\sigma^\MASH_x = S_x(t)$ etc. Note that this expression can be used for any variant of MASH.

\marginnote{Another way to think about quantum jumps is similarly to \enquote{inserting the identity} in quantum mechanics. If we have the state $\sum_n c_n \ket{n}$, we can insert the identity to write it as $\sum_m \left( \sum_n c_n\braket{m}{n} \right) \ket{m}$. The overlap matrix $\braket{m}{n}$ corresponds to $\hat{\mathcal{W}}$.}

For MASH, we obtain
\begin{align*}
    w_0 & = W_{AP}/2 \\
    w_x & = W_{AC} S_x(t)/2 \\
    w_y & = W_{AC} S_y(t)/2 \\
    w_z & = W_{AP} \, \sgn(S_z(t))/2
\end{align*}
with \enquote{P} and \enquote{C} standing for population and coherence, respectively,\sidenote{This notation is used in \cite{mannouchMappingApproachSurface2023}.} and performing the sum leads to
\[
    \hat{\mathcal{W}} = \begin{pmatrix}
        W_{AP} \, h(S_z(t)) & W_{AC} \, \frac{S_x(t)-\imag S_y(t)} 2 \\
        W_{AC} \, \frac{S_x(t)+\imag S_y(t)} 2 & W_{AP} \, h(-S_z(t))
    \end{pmatrix}
\]    
as is given in \cite{mannouchMappingApproachSurface2023}. For ms-MASH, where all the weighting factors are one, we obtain
\[
    \hat{\mathcal{W}} = \begin{pmatrix}
        \frac 1 2 + S_z(t) & S_x(t)-\imag S_y(t) \\
        S_x(t)+\imag S_y(t) & \frac 1 2 - S_z(t)
    \end{pmatrix},
\]
making sure to use the final estimators rather than the initial ones.

\section{Direct method for multi-time TCFs -- limit of short hard pulses}
\label{sec:detail-time-dep-pulse}

We can think about the operator $\hat{\mathcal{U}}$ as a short hard pulse of radiation -- e.g. by adding a term like $f(t)\,\hat{\sigma}_x$ to our Hamiltonian, where $f(t)$ tends to a delta function. Considering MASH evolution under such a time-dependent Hamiltonian would lead to the direct propagation scheme proposed in \autoref{chapter:quantum-jump}.

In fact, this equivalence also tells us that we should \emph{not} rescale the momentum after the pulse, even if the active surface has changed.\sidenote{In exact quantum mechanics, this is just the Frank--Condon principle; the nuclei do not have time to adjust in such a short time. In MASH, the delta function in \eqref{eq:mash-force} makes it somewhat more subtle.} It can be shown that
\[
    d_i \propto \frac{1}{f(t)}
\]
if $f(t)$ is much larger than the separation between the energy levels. Thus, in the limit of short hard pulses, the states effectively become decoupled and no hopping occurs. The transitions happen because the pulse modifies the energy levels themselves -- or in other words, the energy for the electronic transitions comes from the radiation, not from the nuclear kinetic energy.

\pagelayout{wide} % No margins
\pagelayout{margin} % Restore margins

% !TEX root = main.tex

\chapter{Consistency of estimators with the QCLE}
\label{appendix:qcle}

In this appendix, we shall specialise the QCLE \eqref{eq:qcle} to two-level systems. We shall then require that approximate mixed quantum-classical methods should be consistent with the QCLE to first order in time, which will lead to a set of conditions on the estimators. We shall then specialise to the case of smooth MASH of \autoref{sec:all-estimators-are-cpa-consistent} and derive the conditions stated in \autoref{sec:all-estimators-are-qcle-consistent}.

For smooth MASH, the energy function is given by
\[
    E(\vec{p}, \vec{q}, \vec{S}) = \sum_i \frac{p_i^2}{2m} + \bar{V}(\vec{q}) + V_z(\vec{q})\,f (S_z),
\]
which leads to the MASH force
\[
    -\pder{\bar{V}}{q_i} - \pder{V_z}{q_i}\,f (S_z) + 2V_z(\vec{q}) d_i(\vec{q}) \, S_x f'(S_z).
\]

We can split the force into the purely nuclear component
\[
    \bar{F}_i(\vec{q}) = -\pder{\bar{V}}{q_i}
\]
and the electronic force
\begin{equation}
    \label{eq:mash-electronic-force}
    F_i(\vec{q}, \vec{S}) = - \pder{V_z}{q_i}\,f (S_z) + 2V_z(\vec{q}) d_i(\vec{q}) \, S_x f'(S_z) .
\end{equation}

\section{Derivation of the QCLE for a general operator}

For the Wigner transform of a time-evolved operator $\hat{\mathcal{B}}(t)$, we can write, similarly to \eqref{eq:qcle},
\begin{equation}
    \label{eq:qcle-operator}
    \der{\hat{B}_W(t)}{t} = \frac{\imag}{\hbar} \left[ \hat{H}_W, \hat{B}_W(t) \right] + \frac 1 2 \left( \hat{H}_W \, \Lambda \, \hat{B}_W(t) - \hat{B}_W \, \Lambda \, \hat{H}_W(t) \right).
\end{equation}    

All of this is only correct to $\mathcal{O}(\hbar^2)$, so we can expand the adiabatic Hamiltonian \eqref{eq:H-adia} to the same order, which yields\sidenote{Again, the terms with $\hbar d_i$ should properly be considered in the large mass limit rather than $\hbar\to 0$.}
\[
    \hat{H}^\adia = \sum_{i=1}^f \frac{p_i^2}{2m} + \bar{V}(\vec{q}) + V_z(\vec{q}) \, \hat{\sigma}_z^\adia + \hbar \sum_{i=1}^f \frac{p_i d_i + d_i p_i}{2m} \, \hat{\sigma}_y^\adia + \mathcal{O}(\hbar^2).
\]
Taking the Wigner transform to order $\mathcal{O}(\hbar^2)$ merely amounts to replacing $p_i d_i + d_i p_i$ by $2p_id_i$.

Henceforth, we shall drop the $W$ subscripts and the superscripts \enquote{adia}, since we shall only be working with the partially Wigner transformed operators in the adiabatic basis. Substituting into \eqref{eq:qcle-operator}, expanding the $\Lambda$ operator and only keeping terms up to $\mathcal{O}(\hbar^0)$ gives
\begin{equation}
    \label{eq:qcle-operator-final}
    \der{\hat{B}}{t} = \frac{\imag}{\hbar} \left[ \hat{H}_\el, \hat{B} \right] + \sum_i \frac{p_i}{m} \, \pder{\hat{B}}{q_i} + \sum_i \bar{F}_i \pder{\hat{B}}{p_i} + \frac{1}{2} \sum_i \left[ \hat{F}_i,\pder{\hat{B}}{p_i} \right]_+ ,
\end{equation}
where $\hat{H}_\el$ is the electronic part of $\hat{H}^\adia$ -- the rest commutes with $\hat{B}$ anyway. $\bar{F}_i$ is defined as above and the electronic force operator is
\[
    \hat{F}_i = -\pder{(V_z\hat{\sigma}_z)}{q_i} = -\pder{V_z}{q_i} \hat{\sigma}_z + 2V_z d_i \hat{\sigma}_x .
\]
This is the starting point for our derivation, which closely follows Appendix C of \cite{mannouchMappingApproachSurface2023}.

\section{Correlation functions must be consistent with the QCLE to first order in time}

The appropriate limit to consider with the QCLE is the $t\to 0$ limit. We can imagine expanding into a Taylor series in $t$ and collecting terms of different orders. In fact, the zeroth order term is correctly reproduced if our estimators are correct in the CPA. To guarantee short-time correctness, we also need to ensure that the first order terms agree -- this is what makes the quantum jump procedure work.

We shall work with correlation functions of the form
\[
    C_{AB}(t) = \frac{1}{(2\pi\hbar)^f} \int \dd \vec{p} \dd \vec{q} \, \tr{\hat{A}(\vec{p},\vec{q})\hat{B}(\vec{p}_t,\vec{q}_t;t)}
\]
and we shall assume (without loss of generality) that both $\hat{A}$ and $\hat{B}$ can be written in the factorised form $A(\vec{p},\vec{q}) \hat{\sigma}_a$ (and similarly for $\hat{B}$).

Looking at the first order in $t$ is equivalent to taking the derivative and then the limit $t\to 0$. We thus need to correctly reproduce
\[
    \dot{C}_{AB}(0) = \frac{1}{(2\pi\hbar)^f} \int \dd \vec{p} \dd \vec{q} \, \tr{\hat{A}(\vec{p},\vec{q}) \der{}{t}\hat{B}(\vec{p}_t,\vec{q}_t;t)}
\]
by the MASH derivative
\[
    \dot{C}^\MASH_{AB}(0) = \int \frac {\dd \vec{p} \dd \vec{q}} {(2\pi\hbar)^f} \int \dd \vec{S} \, W_{ab}(\vec{S}) A(\vec{p},\vec{q}, \vec{S}) \der{}{t} B(\vec{p}_t,\vec{q}_t, \vec{S}_t) \Big|_{t\to 0} .
\]

From the QCLE, we already have an expression for the exact mixed quantum-classical derivative of $\hat{B}(\vec{p}_t,\vec{q}_t;t)$. We thus need the same kind of expression for MASH. This is easy enough to obtain by partial differentiation with respect to $\vec{p}$, $\vec{q}$ and $\vec{S}$, giving
\begin{widepar}
\begin{equation}
    \label{eq:mash-operator-derivative}
    \der{}{t} B(\vec{p}_t,\vec{q}_t, \vec{S}_t) = \pder{B}{\vec{S}} \cdot \dot{\vec{S}} + \sum_i \frac{p_i}{m}\pder{B}{q_i} + \sum_i \bar{F}_i(\vec{q}) \pder{B}{p_i} + \sum_i F_i(\vec{q},\vec{S}) \pder{B}{p_i}
\end{equation}
\end{widepar}

We shall now see that the first three terms correspond exactly to the first three terms in the exact derivative \eqref{eq:qcle-operator-final}, but the final term with the electronic force leads to some conditions of $f(x)$.

The first term in \eqref{eq:qcle-operator-final} deals with the purely electronic evolution under $\hat{H}_\el$, so it results in $\dot{C}^{\mathrm{CPA}}_{AB}(0)$ for every phase space trajectory. We can use that smooth MASH is correct in the CPA,
\[
    C^{\mathrm{CPA}}_{AB}(t) = \int \dd\vec{S} \, W_{ab}(\vec{S}) \, A(\vec{S}) B(\vec{S}_t),
\]
and differentiate this expression to give\sidenote{The notation $B(\vec{S}_{0+})$ represents the $t\to 0$ limit of the final estimator for $\hat{B}$.}
\[
    \dot{C}^{\mathrm{CPA}}_{AB}(0) = \int \dd\vec{S} \, W_{ab}(\vec{S}) \, A(\vec{S}) \pder{B}{\vec{S}} \cdot \dot{\vec{S}} + \pder{W_{ab}(\vec{S})}{t}\Big|_{t\to 0} \, A(\vec{S}) B(\vec{S}_{0+}).
\]

If $W_{ab}$ is independent of time, the first term in \eqref{eq:mash-operator-derivative} is reproduced. In fact, even for MASH with weighting factor $2|S_z(t)|$, the additional term gives a vanishing contribution to the integral.

The second and third terms are easily reproduced with the factorised operators, since the nuclear parts go outside of the electronic trace in \eqref{eq:qcle-operator-final} and outside of the Bloch sphere integral in \eqref{eq:mash-operator-derivative} once we take the $t\to 0$ limit -- this is important, because in general $\vec{p}_t$ and $\vec{q}_t$ will depend on $\vec{S}$.

In order to reproduce the fourth term, we need to satisfy
\[
    \int \dd \vec{S} \, W_{ab}(\vec{S}) A(\vec{p},\vec{q}, \vec{S}) \pder{B}{p_i} F_i(\vec{q},\vec{S}) = \frac 1 2 \tr{\hat{A}(\vec{p},\vec{q}) \left[ \hat{F}_i, \pder{\hat{B}}{p_i} \right]_+}.
\]

Using the factorised forms of the operators and the expression \eqref{eq:mash-electronic-force} for $F_i(\vec{q},\vec{S})$, we can reduce this to
\begin{widepar}
\[
    \int \dd \vec{S} \, W_{ab}(\vec{S}) \sigma_a(\vec{S}) \sigma_b(\vec{S}_{0+}) \left(- \pder{V_z}{q_i}\,f (S_z) + 2V_z(\vec{q}) d_i(\vec{q}) \, S_x f'(S_z) \right) = \frac 1 2 \tr{\hat{\sigma}_a \left[ -\pder{V_z}{q_i} \hat{\sigma}_z + 2V_z d_i \hat{\sigma}_x, \hat{\sigma}_b \right]_+}.
\]    
\end{widepar}

Using the standard relationships for Pauli matrices\sidenote[][*2]{Specifically, we have
\begin{align*}
    \left[\sigma_a, \sigma_b \right]_+ = &2\delta_{ab} + 2\sigma_a \delta_{b0} \\ &+ 2\sigma_b \delta_{a0} - 4\delta_{a0}\delta_{b0}
\end{align*}
and
\[
    \tr{\sigma_a \sigma_b} = 2\delta_{ab}.
\]
} reduces the right hand side to
\[
    -2 \pder{V_z}{q_i} \left( \delta_{a0}\delta_{bz} + \delta_{az}\delta_{b0} \right) + 4V_z d_i \left( \delta_{a0}\delta_{bx} + \delta_{ax}\delta_{b0} \right).
\]

Thus, we simply need to show that
\[
    \int \dd\vec{S}\, W_{ab}\,\sigma_a(\vec{S}) \sigma_b(\vec{S}_{0_+}) \, f(S_z) = 2(\delta_{a0}\delta_{bz} + \delta_{az}\delta_{b0})
\]
and
\[
    \int \dd\vec{S}\, W_{ab}\,\sigma_a(\vec{S}) \sigma_b(\vec{S}_{0_+}) \, S_xf'(S_z) = 2(\delta_{a0}\delta_{bx} + \delta_{ax}\delta_{b0}).
\]

\section{Conditions on the mapping function}

\marginnote{
    \begin{kaocounter}{Smooth MASH estimators}
        \label{box:smooth-mash-estimators}
        \vspace*{1em}
        Smooth MASH represents the operators consistently by
        \begin{align*}
            \hat{I} &\mapsto 1 \\
            \hat{\sigma}_x &\mapsto S_x \\
            \hat{\sigma}_y &\mapsto S_y \\
            \hat{\sigma}_z &\mapsto f(S_z) \\
            \hat{\sigma}_z(t) &\mapsto WS_z(t).
        \end{align*}
        The time-evolved $\hat{\sigma}_x(t)$ is simply represented by $S_x(t)$, only $\hat{\sigma}_z$ uses different initial and final estimators.
    \end{kaocounter}
}

The left hand sides must be evaluated on a case by case basis, though some simplifications can be made -- for example, the second integral contains an $S_x$, so it will automatically evaluate to zero unless $a$ or $b$ is also $x$. Additionally, any integral involving $y$ except for $a=b=y$ also vanishes. %TODO: refer to CPA estimators in main text (or better - put into a margin box)

The cases $a=b=0$, $a=b=x$ and $a=b=y$ are trivial; we obtain
\[
    \int \dd\vec{S} \, f(S_z) = 0,
\]
which translates to
\[
    \int_{-1}^1 \dd u \, f(u) = 0
\]
using the substitution $u=\cos\theta$.

The case $a=0$, $b=z$ gives $B(\vec{S}_{0+}) = WS_z$ and 
\[
    \int \dd\vec{S} \, W S_z f(S_z) = 2,
\]
which is true by the definition of $W$.\sidenote{Remember that \[W = \frac{2}{\int_{-1}^1 \dd u \, uf(u)}.\]}

For $a=0$ and $b=x$ (or $a=x$ and $b=0$), the second integral gives
\[
    \int \dd\vec{S}\, W S_x^2 f'(S_z) = \frac{W}{2\pi} \int_{0}^{\pi} \dd\theta\, \sin\theta \int_{0}^{2\pi} \dd\phi\, \cos^2\phi \sin^2\theta f'(\cos\theta).
\]
This can be reduced to
\[
    \frac{W}{2} \int_{-1}^1 \dd u \, (1-u^2) f'(u) = W \int_{-1}^1 \dd u \, uf(u) = 2,
\]
using integration by parts and the definition of $W$.

For $a=z$ and $b=0$, we have $A(\vec{S}) = f(S_z)$ and we thus obtain the condition\sidenote{Notice that for $f(u)=u$, we need to rescale by $\sqrt{3}$. This is the how the method spin--LSC is obtained. However, this scaling also leads to the problem of inverted potentials, which MASH is able to avoid. \cite{mannouchMappingApproachSurface2023}}
\[
    \int_{-1}^1 \dd u \, f^2(u) = 2.
\]

The case $a=b=z$ gives the condition
\[
    \int_{-1}^1 \dd u \, uf^2(u) = 0.
\]

For $a=z$ and $b=x$, the integral is a bit more involved.
\[
    \int \dd\vec{S}\, Wf(S_z) S_x^2 f'(S_z) = \frac{W}{4} \int_{-1}^1 \dd u\, (1-u^2) \underbrace{2f(u)f'(u)}_{(f^2)'} = \frac{W}{2}\int_{-1}^1 \dd u\, u f^2(u)
\]
by parts, which again leads to
\[
    \int_{-1}^1 \dd u\, u f^2(u) = 0.
\]

Finally, for $a=x$ and $b=z$, we need to evaluate
\[
    \int \dd\vec{S}\, W S_x S_z S_x f'(S_z) = \frac{W}{2} \int_{-1}^1 \dd u\, (u-u^3)f'(u) = \frac{W}{2} \int_{-1}^1 \dd u\, (3u^2-1)f(u).
\]
This has to be zero and combined with $\int_{-1}^1 \dd u \, f(u) = 0$, we obtain the condition
\[
    \int_{-1}^1 \dd u\, u^2 f(u) = 0.
\]

\chapter{Integration of the MASH equations of motion}
\label{appendix:integrators}

In the MASH equations of motion, \eqref{eq:mash-force}, the term $4V_z d_i S_x \delta(S_z)$ leads to an impulsive change in momentum when the active adiabatic state changes. The precise effect is thoroughly analysed in \cite{mannouchMappingApproachSurface2023}, where the formula for momentum rescaling is derived. The main point is that we need to consider the projection of the momentum onto the non-adiabatic coupling vector,
\[
    \vec{p}_{\mathrm{proj}} = \frac{(\vec{p}\cdot\vec{d})\vec{d}}{|\vec{d}|^2}
\]
and calculate the \enquote{projected} kinetic energy $\vec{p}_{\mathrm{proj}}^2/2m$. If this is large enough to overcome the barrier, the particle will hop to the other state -- the projected component of the momentum will be rescaled to conserve energy. If not, the projected momentum will be reversed. It is shown in~\cite{mannouchMappingApproachSurface2023} that this also leads to the reversal of $\dot{S}_z$, so the spin vector will also turn backwards. Therefore, the active state in MASH can never become inconsistent with the nuclear dynamics.

However, we need to find a good way of propagating the equations of motion numerically -- we shall build up to that in several stages. The nuclear dynamics can \emph{almost} be propagated by the well-known symplectic algorithms like velocity Verlet; the only problem happens when we are crossing $S_z=0$. We shall therefore need to figure out how to propagate the hops in a consistent way. Then, we shall find a method for propagating the electronic dynamics. The discussion generally follows \cite{geutherTimeReversibleImplementationMASH2025}, but we shall use the more general language of matrix differential equations where they have used notation specific to MASH.

By generalising the results in \cite{geutherTimeReversibleImplementationMASH2025}, a 4th order \enquote{symplectic} integrator for MASH is derived. While higher-order integrators are generally not used in MD simulations, whose main goal is to sample the appropriate ensemble,\sidenote{They are still necessary to study difficult long-time behaviour -- for example, a 6th order symplectic integrator has been used in \cite{onoratoRouteThermalizationalphaFermiPastaUlam2015}.} they might find more use in calculations of low-dimensional systems like gas-phase molecules -- either because of a need for greater accuracy\sidenote{Though in all simulations we have performed, the accuracy was limited by the statistical error.} or due to the ability to take larger timesteps and the associated possibility of decreased computational cost.

\section{Classical propagation with a step potential}

For propagation of the hops, let us consider a simple one dimensional problem. A delta function force $F(x)=\lambda\delta(x)$ corresponds to a step function potential $V(x)=\lambda h(x)$. Suppose that we want to use velocity Verlet with a timestep of $\delta t$. If this were propagated naively, we would simply \enquote{skip} the point $x=0$ and continue with the same momentum.\sidenote{This is because the force is zero everywhere except at $x=0$.} This clearly violates conservation of energy and it is possible to end in a classically forbidden region.

\begin{marginfigure}
    \begin{tikzpicture}[
    %Global config
    >=latex,
    line width=1pt,
]

\draw (0,0) -- (2.2,0) -- (2.2,1) -- (4,1) node[right] {$V(x)$};

\draw[->, line width=0.7pt] (0,-0.75) -- (4.5,-0.75);
\draw[->, line width=0.7pt] (4.5,-1.0) -- (0,-1.0) ;

\node (1) at (0.5, -0.75) {};
\node (2) at (1.5, -0.75) {};
\node (3) at (2.5, -0.75) {};
\node (4) at (3.1, -0.75) {};
\node (5) at (3.7, -0.75) {};

\node  (6) at (3.7, -1.0) {};
\node  (7) at (3.1, -1.0) {};
\node  (8) at (2.5, -1.0) {};
\node  (9) at (1.9, -1.0) {};
\node (10) at (0.9, -1.0) {};

\draw[fill=white]          (1) circle(0.08);
\draw[Maroon,fill=white]   (2) circle(0.08);
\draw[Maroon,fill=white]   (3) circle(0.08);
\draw[fill=white]          (4) circle(0.08);
\draw[fill=white]          (5) circle(0.08);
\draw[fill=white]          (6) circle(0.08);
\draw[fill=white]          (7) circle(0.08);
\draw[Maroon,fill=white]   (8) circle(0.08);
\draw[Maroon,fill=white]   (9) circle(0.08);
\draw[fill=white]         (10) circle(0.08);

\draw[->]          ($ (1.north) + (1pt,2pt) $) .. controls (0.75,-0.25) and (1.25,-0.25) .. ($ (2.north)  + (-1pt,1pt) $);
\draw[->, Maroon]  ($ (2.north) + (1pt,2pt) $) .. controls (1.75,-0.25) and (2.25,-0.25) .. ($ (3.north)  + (-1pt,1pt) $);
\draw[->]          ($ (3.north) + (1pt,2pt) $) .. controls (2.65,-0.25) and (2.9, -0.25) .. ($ (4.north)  + (-1pt,1pt) $);
\draw[->]          ($ (4.north) + (1pt,2pt) $) .. controls (3.25,-0.25) and (3.55,-0.25) .. ($ (5.north)  + (-1pt,1pt) $);
\draw[->]          ($ (6.south) - (1pt,2pt) $) .. controls (3.55,-1.5) and (3.25,-1.5) ..   ($ (7.south)  - (-1pt,1pt) $);
\draw[->]          ($ (7.south) - (1pt,2pt) $) .. controls (2.9, -1.5) and (2.65,-1.5) ..   ($ (8.south)  - (-1pt,1pt) $);
\draw[->, Maroon]  ($ (8.south) - (1pt,2pt) $) .. controls (2.35,-1.5) and (2.05,-1.5) ..   ($ (9.south)  - (-1pt,1pt) $);
\draw[->]          ($ (9.south) - (1pt,2pt) $) .. controls (1.65,-1.5) and (1.15,-1.5) ..   ($ (10.south) - (-1pt,1pt) $);

\end{tikzpicture}
    \caption{Rescaling momentum at the end of a timestep is not time reversible. The timesteps that involve momentum hops are highlighted.}
    \label{fig:hysteresis}
\end{marginfigure}

Thus, we need to check whether we have crossed $x=0$ and apply momentum rescaling if necessary to ensure conservation of energy.
We could apply the momentum rescaling at the end of the timestep -- while that is exact in the $\delta t \to 0$ limit, it is only a first order method and is not time reversible. \cite{geutherTimeReversibleImplementationMASH2025} The reason is that the time reversed version would propagate on the upper surface until the end of the timestep, leading to a kind of \enquote{hysteresis} -- see Figure~\ref{fig:hysteresis}.

It turns out \cite{geutherTimeReversibleImplementationMASH2025} that we need to find the crossing time $\delta\tau$ first, propagate for $\delta \tau$, apply momentum rescaling and then propagate for the rest of the timestep, $\delta t - \delta\tau$. This yields a time-reversible method with the same order as the underlying propagation algorithm. This is true as long as we find the crossing time precisely enough, but that is always possible with enough computational resources -- in fact, since the hops constitute only a small fraction of the trajectories, this is not a significant expense.

Starting with the 4th order Yoshida algorithm \cite{yoshidaConstructionHigherOrder1990} rather than the 2nd order velocity Verlet can be used to construct a 4th order propagator for the nuclear degrees of freedom.

\marginnote{
    \begin{kaocounter}{Yoshida algorithm}
        \label{box:yoshida-algorithm}
        \vspace*{5pt}
        The variables $q$ and $v$ are propagated in a leapfrog way,
        \begin{align*}
            q_1 & \gets q(t) + v(t) \, c_1 \delta t \\
            v_1 & \gets v(t) + a(q_1) \, d_1 \delta t \\
            q_2 & \gets q_1 + v_1 \, c_2 \delta t \\
            v_2 & \gets v_1 + a(q_2) \, d_2 \delta t \\
            q_3 & \gets q_2 + v_2 \, c_3 \delta t \\
            v_3 & \gets v_2 + a(q_3) \, d_3 \delta t \\
            q(t + \delta t) & \gets q_3 + v_3 \, c_4 \delta t
        \end{align*}
        with coefficients
        \begin{align*}
            c_1 = c_4 & = x_1/2 \\
            c_2 = c_3 & = (x_0+x_1)/2 \\
            d_1 = d_3 & = x_1 \\
            d_2 & = x_0 \\
            x_0 & = -\frac{2^{1/3}}{2-2^{1/3}} \\
            x_1 & = \frac{1}{2-2^{1/3}}
        \end{align*}
    \end{kaocounter}
}

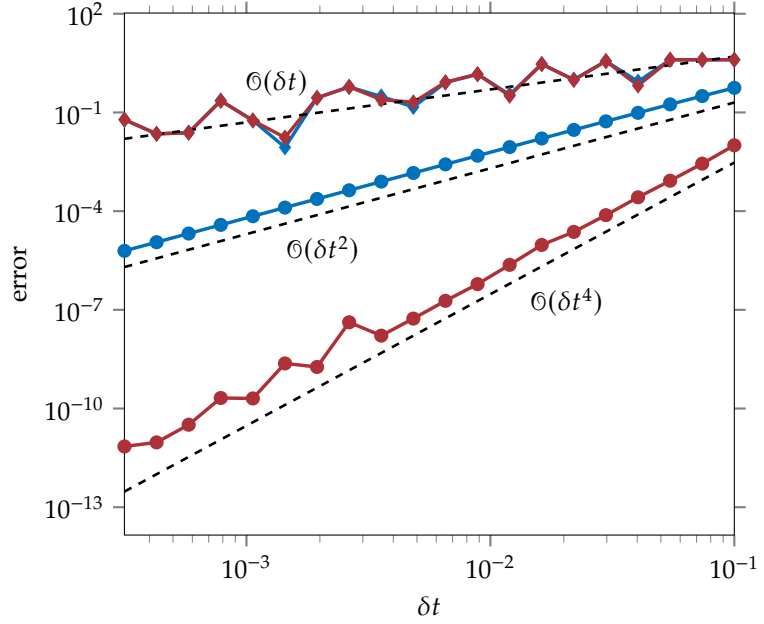
\begin{figure}
    \begin{tikzpicture}
    \pgfplotsset{compat=1.18}
    \begin{loglogaxis}[enlarge x limits=false,
                 tick align=outside,
                 xlabel=$\delta t$,
                 ylabel=error,
                 width=0.75\textwidth,
                 scale only axis
                ]

        \addplot[mark=diamond*, line width=1.3pt, RoyalBlue]  table[x index = 0, y index = 1] {data/int-err-hop.dat};
        \addplot[mark=*, line width=1.3pt, RoyalBlue]   table[x index = 0, y index = 2] {data/int-err-hop.dat};
        \addplot[mark=diamond*, line width=1.3pt, Maroon]   table[x index = 0, y index = 3] {data/int-err-hop.dat};
        \addplot[mark=*, line width=1.3pt, Maroon]   table[x index = 0, y index = 4] {data/int-err-hop.dat};

        \addplot[mark=none, line width = 1.0pt, dashed, domain=0.000316:0.1] {30.0*x^4}
            node[pos=0.65,below right] {$\mathcal{O}(\delta t^4)$};

        \addplot[mark=none, line width = 1.0pt, dashed, domain=0.000316:0.1] {20.0*x^2}
            node[pos=0.25,below right] {$\mathcal{O}(\delta t^2)$};

        \addplot[mark=none, line width = 1.0pt, dashed, domain=0.000316:0.1] {50.0*x}
            node[pos=0.25,above=5pt] {$\mathcal{O}(\delta t)$};

    \end{loglogaxis}

\end{tikzpicture}
    \caption{Error when propagating the potential \eqref{eq:harmonic-step-potential} with the velocity Verlet (blue) and Yoshida (red) algorithms. The naive method with momentum rescaling at the end (diamonds) is first order, whereas the corrected method (circles) has the same order as the underlying algorithm.}
    \label{fig:plot-int-err-hop}
\end{figure}

We can show these properties numerically as well. Consider the model system
\begin{equation}
    \label{eq:harmonic-step-potential}
    V(x) = \frac 1 2 k x^2 + \lambda h(x),
\end{equation}
a harmonic oscillator with a step at the minimum. This can be solved both exactly and approximately and the different methods compared. The results are shown in Figure~\ref{fig:plot-int-err-hop}. It is easy to see that the naive algorithm is indeed 1st order, but we recover the 2nd order error of velocity Verlet with the corrected method. The results are the same for the Yoshida algorithm -- the naive version reduces to a 1st order method, but the corrected one regains its 4th order charm.

\section{Two hops in one timestep -- a phantom menace}

So far, we have seen that the finding the hopping point correctly regains the error of the underlying algorithm. In principle, we have solved the problem; the algorithm will be $\mathcal{O}(\delta t^n)$ as $\delta t\to 0$. However, in practice, we may not want to use infinitesimal timesteps -- in fact, we want to use as large a timestep as possible. A new problem then appears -- trajectories could hop twice (or even multiple times) during one timestep.\sidenote{While this has not been observed in two-state MASH, ms-MASH simulations with multiple states have been observed to do so.~\cite{runesonMultistateMappingApproach2023}}

\begin{figure}
    \begin{tikzpicture}
    \pgfplotsset{compat=1.18}
    \begin{loglogaxis}[enlarge x limits=false,
                 ymin = 1e-13,
                 ymax = 100,
                 tick align=outside,
                 xlabel=$\delta t$,
                 ylabel=error,
                 width=0.75\textwidth,
                 scale only axis
                ]

        \addplot[mark=*, line width=1.3pt, RoyalBlue]  table[x index = 0, y index = 1] {data/int-err-double-hop.dat};
        \addplot[mark=*, line width=1.3pt, Maroon]   table[x index = 0, y index = 2] {data/int-err-double-hop.dat};

        \addplot[mark=none, line width = 1.0pt, dashed, domain=0.000316:0.02] {100.0*x^4}
            node[pos=0.25,below right] {$\mathcal{O}(\delta t^4)$};

        \addplot[mark=none, line width = 1.0pt, dashed, domain=0.000316:0.02] {70.0*x^2}
            node[pos=0.25,below right] {$\mathcal{O}(\delta t^2)$};

        \draw[line width = 1.0pt, dotted] ({axis cs:0.007,0}|-{rel axis cs:0,0}) -- ({axis cs:0.007,0}|-{rel axis cs:0,1})
            node[pos=0.05,right] {$\delta t=\delta t^*$};
        
    \end{loglogaxis}

\end{tikzpicture}
    \caption{Error when propagating the potential \eqref{eq:harmonic-double-step-potential} with the velocity Verlet (blue) and Yoshida (red) algorithms. The jump at the point $\delta t > \delta t^*$ when \enquote{hop skipping} start happening can be clearly seen. Note that for very small $\delta t$, the Yoshida algorithm is limited by floating point precision.}
    \label{fig:plot-int-err-double-hop}
\end{figure}
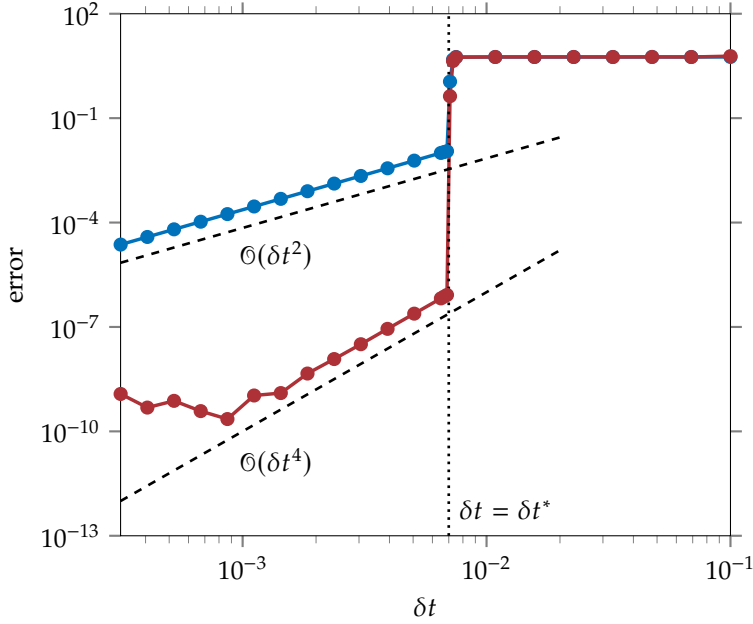

\begin{marginfigure}
    \centering
    \begin{tikzpicture}[yscale=2]
    \draw[->] (0,0) -- (0,1.0) node[above] {$y$};
    \draw[->] (-1.4,0) -- (1.4,0) node[right] {$x$};

    \draw[dashed] (-1.4,0.3) -- (-0.1,0.3);
    \draw[<->] (-1,0) -- (-1,0.3) node[pos=0.5,left] {$\lambda$};

    \draw (-0.1,0) -- (-0.1,-0.2);
    \draw ( 0.1,0) -- ( 0.1,-0.2);

    \draw[->] (-0.35,-0.13) -- (-0.1,-0.13);
    \draw[->] ( 0.35,-0.13) -- ( 0.1,-0.13);

    \draw (0, -0.2) node[below] {$2a$};

    \draw[line width=1, domain=-1:-0.1, smooth, variable=\x, Maroon] plot ({\x}, {(\x + 0.1)^2});
    \draw[line width=1, Maroon] (-0.1,0) -- (-0.1,0.3) -- (0.1,0.3) -- (0.1, 0);
    \draw[line width=1, domain=0.1:1, smooth, variable=\x, Maroon] plot ({\x}, {(\x - 0.1)^2});

\end{tikzpicture}
    \caption{The double step potential defined by~\eqref{eq:harmonic-double-step-potential}.}
\end{marginfigure}

Of course, if we estimate \textit{both} hopping times correctly, we can easily deal with the situation. However, hops are usually detected by comparing the initial and final points -- a hop from $\ket{+}$ to $\ket{-}$ and then back to $\ket{+}$ would thus not be detectable. We shall simulate such a situation with the model potential
\begin{equation}
    \label{eq:harmonic-double-step-potential}
    V(x) = \begin{cases}
        \frac 1 2 k (x-a)^2 &      x > a \\
        \lambda             & -a < x < a \\
        \frac 1 2 k (x+a)^2 &      x < -a \\
    \end{cases}
\end{equation}

If the timestep is large enough, we can go directly from the left parabolic well to the right one in one timestep -- this corresponds to an undetectable hop. In fact, if $E$ is the total energy, we can easily predict that the minimum timestep for \enquote{hop skipping} is given by
\[
    \delta t^* = \frac{2a}{v_\lmax} = \frac{2a}{\sqrt{2E/m}} .
\]

That is exactly what we observe in Figure~\ref{fig:plot-int-err-double-hop} -- for $\delta t < \delta t^*$, the integrators regain their nice behaviour, but once we make the timestep too large, the error degrades massively. For real-world problems, the consequences of \enquote{hop skipping} might not be so severe, but this example shows that it is something to be wary of.

\section{Propagation of the electronic variables}

The equation of motion for the electronic degrees of freedom (i.e. the vector wavefunction $\vec{\psi}$) is
\[
    \dot{\vec{\psi}} = -\frac{\imag}{\hbar} \vec{V}[\vec{q}(t)] \vec{\psi},
\]
and we shall define $\vec{F}[\vec{q}(t)] \equiv -\frac{\imag}{\hbar} \vec{V}[\vec{q}(t)] $ for convenience.

If $\vec{\psi}$ were just a number, $\psi$, we would obtain a simple first-order differential equation
\[
    \dot{\psi} = f[q(t)] \,\psi,
\]
whose solution is easily shown to be
\[
    \psi(t) = \eul^{\int_0^t \dd t' f[q(t')]} \, \psi(0).
\]
However, this is not true in the matrix case because the $\vec{V}[\vec{q}(t)]$ at different times do not commute with each other.
All is not lost -- we can get very close with the \emph{Magnus expansion}.\sidenote{Another possibility is the \emph{Fer expansion} \cite{blanesMagnusFerExpansions1998a}, which uses \[\vec{\psi}(t) = e^{\vec{F}_1(t)} e^{\vec{F}_2(t)}\cdots\vec{\psi}(0).\] Both of these are superior to the usual Dyson series in that they preserve unitarity at every step of the expansion. \cite{tannorIntroductionQuantumMechanics2007}} \cite{tannorIntroductionQuantumMechanics2007,blanesMagnusFerExpansions1998a} If we represent the solution to the original equation as 
\[
    \vec{\psi}(t) = e^{\vec{\Omega}(t)} \vec{\psi}(0),
\]
we can expand $\vec{\Omega}(t)$ as a series
\[
    \vec{\Omega}(t) = \vec{\Omega}_1(t) + \vec{\Omega}_2(t) + \cdots
\]
where the first two terms are given by
\[
    \vec{\Omega}_1(t) = \int_0^t \dd t' \, \vec{F}[\vec{q}(t')],
\]
which coincides with the scalar expression, and
\[
    \vec{\Omega}_2(t) = \frac{1}{2} \int_0^t \dd t' \int_0^{t'} \dd t'' \, \Big[\vec{F}[\vec{q}(t')], \vec{F}[\vec{q}(t'')]\Big],
\]
which takes the non-commutativity into account \textit{via} a commutator at two different times. Similarly, the higher order terms contain nested commutators of $\vec{F}$.

In order to devise a numerical propagation algorithm, let us consider the scalar case first. We can rewrite the equation as
\[
    \der{}{t} \ln \psi = f[q(t)] .
\]
This is the same as the equation for velocity propagation,
\[
    \der{v}{t} = a[q(t)],
\]
after the identification
\begin{align*}
    v & \mapsto \ln\psi \\
    a & \mapsto f .
\end{align*}
We can therefore directly use the classical algorithms like velocity Verlet or Yoshida to propagate $\ln \psi$, just like we would propagate $v$. For velocity Verlet, this gives
\[
    \psi(t + \delta t) = \eul^{f[q(t+\delta t)] \delta t /2} \eul^{f[q(t)]\delta t /2} \, \psi(t), 
\]
where $q(t+\delta t)$ is calculated by the standard propagation.

We can also write this in the form
\[
    \psi(t + \delta t) = \eul^{F(t)} \, \psi(t)
\]
with
\[
    F(t) = \Big\{ f[q(t)] + f[q(t+\delta t)] \Big\} \, \frac {\delta t} 2 .
\]

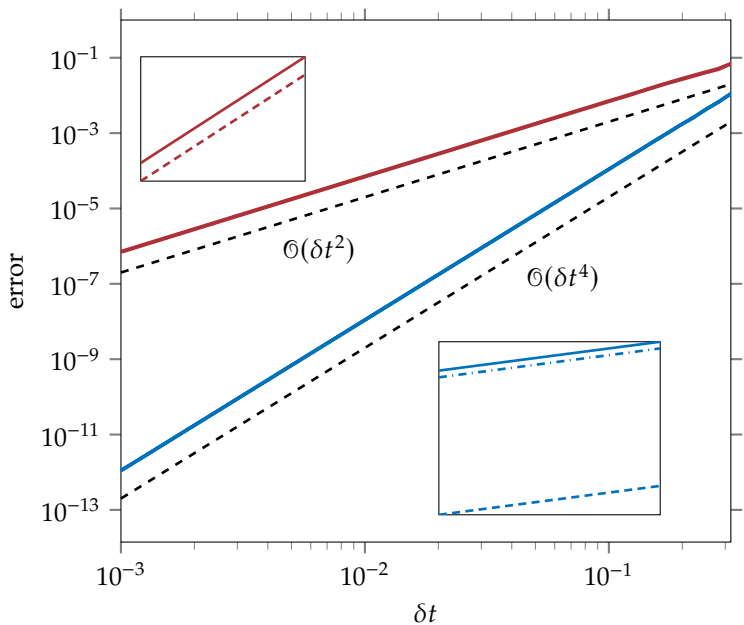
\begin{figure}
    \begin{tikzpicture}
    \pgfplotsset{compat=1.18}
    \begin{loglogaxis}[enlarge x limits=false,
                 tick align=outside,
                 xlabel=$\delta t$,
                 ylabel=error,
                 width=0.75\textwidth,
                 scale only axis
                ]

        \addplot[mark=none, line width=1.3pt, Maroon] table {data/int-err-elec-Verlet-F.dat};
        \addplot[mark=none, line width=1.3pt, Maroon] table {data/int-err-elec-Verlet-M.dat};

        \addplot[mark=none, line width=1.3pt, RoyalBlue] table {data/int-err-elec-Yoshida-F.dat};
        \addplot[mark=none, line width=1.3pt, RoyalBlue] table {data/int-err-elec-Yoshida-M.dat};
        \addplot[mark=none, line width=1.3pt, RoyalBlue] table {data/int-err-elec-Yoshida-M3.dat};

        \addplot[mark=none, line width = 1.0pt, dashed, domain=0.001:0.316] {0.2*x^4}
            node[pos=0.65,below right] {$\mathcal{O}(\delta t^4)$};

        \addplot[mark=none, line width = 1.0pt, dashed, domain=0.001:0.316] {0.2*x^2}
            node[pos=0.25,below right] {$\mathcal{O}(\delta t^2)$};

        \coordinate (inset1) at (axis cs:0.0012, 0.11);
        \coordinate (inset2) at (axis cs:0.02,3.0e-9);
        
    \end{loglogaxis}

    \begin{loglogaxis}[at={(inset1)},enlarge x limits=false,enlarge y limits=false,anchor={outer north west},ticks=none,width=0.35\textwidth]
        \addplot[mark=none, line width=1pt, Maroon] coordinates {
        (0.018858632787726495, 0.00025630262092633804)
        (0.021209508879201904, 0.0003241154057594259)};

        \addplot[mark=none, densely dashed, line width=1pt, Maroon] coordinates {
            (0.018858632787726495, 0.0002462625229847498)
            (0.021209508879201904, 0.0003115786188572084)};
    \end{loglogaxis}

    \begin{axis}[at={(inset2)},
                       enlarge x limits=false,
                       enlarge y limits=false,
                       anchor={outer north west},
                       ticks=none,
                       width=0.42\textwidth]
        \addplot[mark=none, line width=1pt, RoyalBlue, domain=0:0.0003] 
            {-9.91599767140578 + 3.999797527504998 * x};

        \addplot[mark=none, dashdotted, line width=1pt, RoyalBlue, domain=0:0.0003]
            {-9.916276613525868 + 3.9998623130653845 * x};

        \addplot[mark=none, densely dashed, line width=1pt, RoyalBlue, domain=0:0.0003]
            {-9.92198485823717 + 3.9996935846570465 * x };
        
    \end{axis}

\end{tikzpicture}
    \caption{Error when propagating the electronic Hamiltonian \eqref{eq:model-elec-hamiltonian} with a harmonic nuclear potential. The various velocity Verlet (blue) and Yoshida (red) methods give almost identical results. The insets show the Fer (solid) and Magnus (dashed) version of each method. For Yoshida, the three-term Magnus (dashdotted) version is also shown. Note that the differences between Magnus and Fer versions are only around 1\%.}
    \label{fig:plot-int-err-elec}
\end{figure}

While these two form are equivalent for scalars, they are different for matrices. In fact, \cite{runesonMultistateMappingApproach2023} suggests using the former prescription, whereas \cite{geutherTimeReversibleImplementationMASH2025} uses the latter. Both of them give second-order methods, since the commutator term $\vec{\Omega}_2(t)$ only enters the equations in third order\sidenote{The argument here is a bit more subtle -- $\mathcal{O}(\delta t^3)$ is the \emph{local} error, but that still leads to an $\mathcal{O}(\delta t^2)$ global error for a time-reversible algorithm. \cite{blanesImprovedHighOrder2000}} in $\delta t$. Both methods only require one evaluation of a matrix exponential per step\sidenote{Not two, since $f[q(t + \delta t)]$ becomes $f[q(t)]$ of the next step, so we do not have to calculate it again.} and seem to give very similar results (see Figure~\ref{fig:plot-int-err-elec}).

The Yoshida version can be obtained by the same argument, giving the somewhat scary
\[
    \psi(t + \delta t) = \eul^{f[q_3]d_3\delta t}\eul^{f[q_2]d_2\delta t}\eul^{f[q_1]d_1\delta t} \psi(t).
\]
It turns out that this is exact to 4th order even after taking non-commutativity into account. We shall denote this the \enquote{Fer version} of the propagator, since it involves a product of exponentials.

The naive single-exponential \enquote{Magnus version}, $\psi(t + \delta t) = \eul^{F(t)} \, \psi(t)$ with
\[
    F(t) = \Big\{ d_1f[q_1] + d_2f[q_2] + d_3f[q_3] \Big\} \, \delta t
\]
is not correct to fourth order for the non-commutative case -- while it correctly approximates $\vec{\Omega}_1(t)$, is does not make any effort with $\vec{\Omega}_2(t)$.\sidenote{It is enough to approximate $\vec{\Omega}_2(t)$ for a 4th order method -- the $\vec{\Omega}_3(t)$ term is $\mathcal{O}(\delta t^5)$. See \cite{buddSolutionLinearDifferential1999,blanesImprovedHighOrder2000}.}

With some algebraic manipulation,\sidenote{This entails expressing all quantities as Taylor series and equating terms. The Python module SymPy was extremely helpful in this task. \cite{meurerSymPySymbolicComputing2017}} $\vec{\Omega}_2(t)$ can be approximated by
\[
    -\frac{\delta t^2}{12} \, \Big[f[q(t)], f[q(t + \delta t)]\Big].
\]
However, that would require four force evaluations per step and it would be preferable if we could find an expression using only the already calculated forces. In fact, this is possible with
\[
    -\frac{\delta t^2}{24c_2} \, \Big[f[q_1], f[q_3]\Big],
\]
where $c_2$ is the fractional timestep from the Yoshida algorithm (see Box~\ref{box:yoshida-algorithm}).

These propagators have been tested numerically with the nuclear potential $V(q)=\frac 1 2 kq^2$ and the electronic \enquote{Hamiltonian}
\begin{equation}
    \label{eq:model-elec-hamiltonian}
    \vec{F}[q] = \begin{pmatrix}
        0 & q^3 & 0 \\
        0 & 0   & q \\
        0 & 0   & 0
    \end{pmatrix} .
\end{equation}
    
The elegant feature of this matrix is that all third and higher order commutators vanish (a courtesy of the Heisenberg algebra\sidenote{
    The Heisenberg algebra is a Lie algebra generated by $X$, $Y$ and $Z$ with the commutator relations $[X,Y] = Z$ and $[X,Z]=[Y,Z]=0$. The matrix $\vec{F}[q]$ can be written as $q^3 X + q Y$ using its three-dimensional representation.
}), so the Magnus series only has two terms and we can obtain the exact results by simple integration. The results are shown in Figure~\ref{fig:plot-int-err-elec}. It is apparent that each method regains its respective global error. The differences between the Magnus and Fer versions are inconsequential and they depend on the exact Hamiltonian used.\sidenote{For example, for the $4\times 4$ nilpotent Hamiltonian \[ \begin{pmatrix}
    0 & q & 0 & 0 \\
    0 & 0 & q^3 & 0 \\
    0 & 0 & 0 & -q \\
    0 & 0 & 0 & 0
\end{pmatrix},\] the Fer versions work slightly better then the Magnus ones. (Data not shown.)} From a computational perspective, the Magnus versions would be preferable due to the cost of matrix exponentiation -- this is especially true for multi-level problems.

\section{A 4th order \enquote{symplectic} integrator}

We can combine what we derived in the previous two sections into a 4th order propagator for MASH. Note that MASH dynamics is not generated by a Hamiltonian, so the algorithm -- despite being time reversible and energy conserving -- should not properly be called symplectic.

In fact, all the pieces are properly in place. We use the corrected Yoshida algorithm for the propagation of the nuclear degrees of freedom and a corresponding 4th order electronic algorithm to propagate the wavefunction.

The final ingredient is the determination of the hopping time. Two methods were proposed -- bisection \cite{runesonMultistateMappingApproach2023} or spline interpolation \cite{geutherTimeReversibleImplementationMASH2025}. The former is easy to implement, but the latter is more efficient -- in fact, since the timestep is generally quite small, linear interpolation will usually already be good enough. Currently, we have no general solution to the double hop problem.

In this work, we have used the Fer version of the electronic propagation together with the bisection method. This allowed us to use an order of magnitude larger timesteps, which was crucial for obtaining the well-converged results for the derivatives or for calculating the multi-time correlation functions in a reasonable time.

If this algorithm were used for propagation of real-world electronic problems, it would have to be adapted into a locally diabatic version.~\cite{geutherTimeReversibleImplementationMASH2025} We do not see any major obstacles in doing that, but it is beyond the scope of the current work.

% !TEX root = main.tex

\chapter{Variance of MASH estimators}
\label{appendix:variance}

It has been observed that in many simulations (of both the Tully models and the spin-boson model -- see e.g. Figure~\ref{fig:plot-derivatives-p10}), the population--population correlation function generally converges faster with ms-MASH than with MASH. Put differently, ms-MASH has a lower statistical error than MASH for the same number of trajectories. We think that this is something worth exploring further.

A first thought could be that this is due to the two Heaviside functions in the MASH estimator. As a simple argument in favour of this, consider sampling a variable $X$ uniformly from the interval $(-1/2, 1/2)$. The random variables
$h(X)$ and $X+1/2$ both have average value $1/2$. However, the latter \enquote{smooth} version has 3$\times$ lower variance. This argument is too naive, however -- the MASH estimator has an additional factor of $2|S_z|$, which makes it smoother than a pure step function.

We shall thus derive the expressions for the variance of the MASH and ms-MASH estimators for $C_{++}(t)$ in the CPA to see whether we can formalise the observation. We shall use the formula
\[
    \mathrm{Var}[X] = E[X^2] - E[X]^2,
\]
and since we already know the expectation value $E[X]$ -- it is simply given by the correlation function $C_{++}(t)$ -- it is sufficient to calculate $E[X^2]$. However, we need to be careful to take a true spherical average with a factor of $1/4\pi$ rather than the Bloch sphere average. Effectively, the estimators will get an additional factor of two.

For MASH, we therefore have
\[
    C_{++}(t) = \avg{4|S_z|h(S_z)h(S_z(t))}_\mathrm{sph},
\]
so we need to calculate
\[
    \avg{16|S_z|^2h(S_z)^2h(S_z(t))^2}_\mathrm{sph} = \avg{16S_z^2h(S_z)h(S_z(t))}_\mathrm{sph},
\]
using $h(x)^2=h(x)$.\sidenote{Astute readers might notice that this is not true at $x=0$ with the usual definition $h(0)=1/2$. However, this does not matter for the integration.}

This is easily achieved using the method from Appendix B in \cite{mannouchMappingApproachSurface2023}, which leads to the integral
\[
    \frac{1}{4\pi}\int_0^\pi \dd\theta' \sin\theta' \int_{\gamma(t)}^\pi \dd \phi' 16\sin^2\theta' \sin^2\phi'
\]
with $\gamma(t) = \cos^{-1}(C_{zz}(t)/2)$. This equals
\[
    \frac{8}{3\pi}\Big(\pi - \gamma(t) + \sin\gamma(t) \cos\gamma(t)\Big)
\]
or
\[
    \frac{8}{3\pi}\left(\pi - \cos^{-1}\left(\frac{C_{zz}(t)}{2}\right) + \frac{C_{zz}(t)}{2} \sqrt{1 - \frac{C_{zz}(t)^2}{4}} \right).
\]

For ms-MASH, the population--population correlation function is
\[
    C_{++}(t) = \avg{2h(S_z)\left(S_z(t) + \frac 1 2 \right)}_\mathrm{sph},
\]
so we need to evaluate
\[
    \avg{4h(S_z) \left[ S_z(t)^2 + \left(S_z(t) + \frac 1 2 \right) - \frac 1 4 \right]}_\mathrm{sph}.
\]

We have
\[
    \avg{h(S_z)}_\mathrm{sph} = \frac 1 2 \int_0^{\pi/2} \dd\theta \sin\theta = \frac 1 2
\]
and
\[
    \avg{4h(S_z)\left(S_z(t) + \frac 1 2 \right)}_\mathrm{sph} = 2C_{++}(t).
\]
The only remaining term is thus
\[
    \avg{4h(S_z)S_z(t)^2}_\mathrm{sph}.
\]
We can expand $S_z(t)$ as (see \cite{mannouchMappingApproachSurface2023})
\[
    S_z(t) = \frac{1}{2} \Big( C_{xz}(t)S_x + C_{yz}(t)S_y + C_{zz}(t)S_z \Big).
\]
After taking the square and integrating, the cross terms will evaluate to zero, so we are left with
\begin{align*}
    \frac{1}{4\pi}\int_0^{\pi/2}\dd\theta \sin\theta \int_0^{2\pi} &\dd\phi \, C_{xz}(t)^2 \sin^2\theta \cos^2\phi \, + \\
    & + C_{yz}(t)^2 \sin^2\theta \sin^2\phi + C_{zz}(t)^2 \cos^2\theta ,
\end{align*}
    
which yields
\[
    \frac{1}{2} \times \frac{1}{3} \Big( C_{xz}(t)^2 + C_{yz}(t)^2 + C_{zz}(t)^2 \Big) = \frac{2}{3}.
\]
\begin{marginfigure}
    \begin{tikzpicture}
    \pgfplotsset{compat=1.18}
    \begin{axis}[enlarge x limits=false,
                 ymin = 0,
                 tick align=outside,
                 xlabel=$C_{++}(t)$,
                 ylabel=$ \mathrm{Var} \, C^\mathrm{est.}_{++}(t) $,
                 width=0.75\textwidth,
                 scale only axis,
                 xlabel style={font=\footnotesize},
                 ylabel style={font=\footnotesize},
                 yticklabel style = {font=\footnotesize,xshift=0.5ex},
                 xticklabel style = {font=\footnotesize,yshift=0.5ex}
                ]

        \addplot[mark=none, line width=1.3pt, RoyalBlue] table[x index = 1, y index = 2] {data/variance.dat};
        \addplot[mark=none, line width=1.3pt, Maroon]    table[x index = 1, y index = 3] {data/variance.dat};
        
    \end{axis}

\end{tikzpicture}
    \caption{Variance of MASH (blue) and ms-MASH (red) estimators as a function of $C_{++}(t)$ in the classical path approximation.}
    \label{fig:plot-variance}
\end{marginfigure}
We thus obtain
\[
    \frac{2}{3} + 2C_{++}(t) - \frac{1}{2} = \frac{1}{6} + 2C_{++}(t)
\]
for the expectation value of the square.

\begin{figure}[h]
    \begin{tikzpicture}
    \begin{axis}[enlarge x limits=false,
                 ymin=0,
                 tick align=outside,
                 xlabel=$C^\mathrm{est.}_{++}(t)$,
                 ylabel=$ \mathrm{Var} \, C^\mathrm{est.}_{++}(t) $,
                 width=0.85\textwidth,
                 scale only axis,
                 set layers
                ]

        \addplot[only marks, mark=*, RoyalBlue, mark size=0.8pt] table[x index = 0, y index = 1] {data/spin-boson-variance-1.dat};
        \addplot[only marks, mark=*, Maroon   , mark size=0.8pt] table[x index = 2, y index = 3] {data/spin-boson-variance-1.dat};
        
        \addplot[only marks, mark=*, RoyalBlue, mark size=0.8pt] table[x index = 0, y index = 1] {data/spin-boson-variance-2.dat};
        \addplot[only marks, mark=*, Maroon   , mark size=0.8pt] table[x index = 2, y index = 3] {data/spin-boson-variance-2.dat};
        
        \begin{pgfonlayer}{axis foreground}
            \addplot[mark=none, line width=1.5pt] table[x index = 1, y index = 2] {data/variance.dat}
                node[pos=0.9,above left,RoyalBlue] {\textbf{MASH}};
            \addplot[mark=none, line width=1.5pt] table[x index = 1, y index = 3] {data/variance.dat}
                node[pos=0.85,below right,Maroon] {\textbf{ms-MASH}};
        \end{pgfonlayer}

    \end{axis}

\end{tikzpicture}
    \caption{Variance of MASH (blue) and ms-MASH (red) estimators for $C_{++}(t)$ for the four cases of the spin--boson model considered in \cite{runesonMultistateMappingApproach2023}, compared with the CPA expressions (black). Note that \textit{adiabatic} populations are used here.}
    \label{fig:spin-boson-variance}
\end{figure}
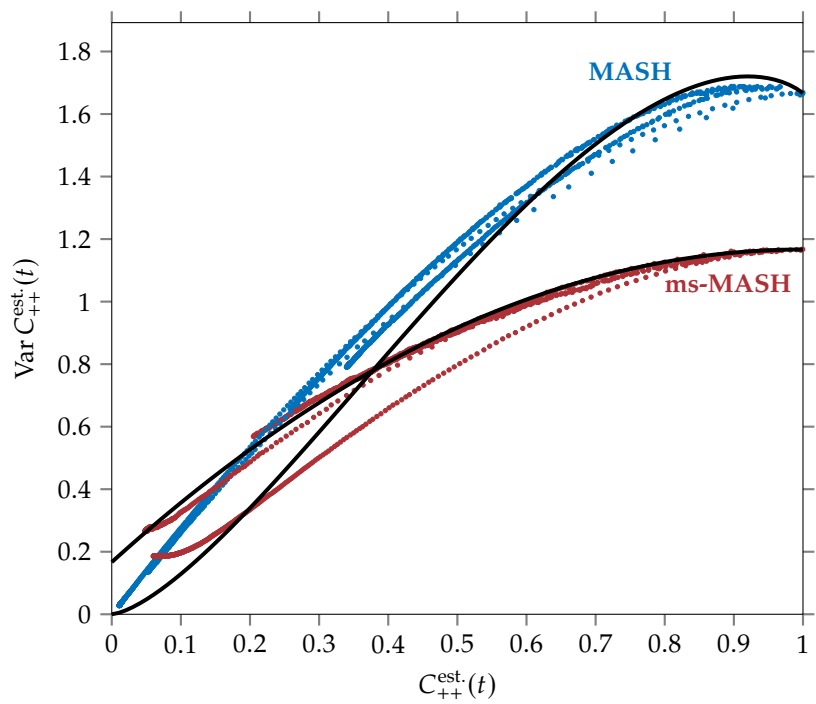

We can see that the variance will generally depend on the value of $C_{++}(t)$, or equivalently $C_{zz}(t)$. The results are shown in Figure~\ref{fig:plot-variance}. We can see that ms-MASH does generally have lower variance, but not universally -- if $C_{++}(t)\lessapprox 0.37$, the statistical error of MASH is lower. We assume that -- since the dynamics is the same -- this result continues to hold at least qualitatively even when nuclear dynamics is taken into account. This is confirmed for the spin--boson model in Figure~\ref{fig:spin-boson-variance}, where the averaging over different $\vec{p}$ and $\vec{q}$ is shown to only have a small effect on the general trend.

% !TEX root = main.tex

\chapter{Implementation details}
\label{appendix:impl-details}

% TODO: and summarised in \autoref{alg:MASH-propagation}
Propagation of the MASH equations of motion has already been covered in \autoref{appendix:integrators}. However, several other implementation details are worth mentioning and will be collected here.

\section{Stable analytical eigendecomposition of a \texorpdfstring{2$\times$2}{2×2} matrix}
\label{sec:analytical-2x2-matrix}

For the problems treated in this work, we shall be working in the diabatic basis, since that generally proves to be the more convenient choice.
However, for hopping between surfaces and for some correlation function calculations, we need to transform to the adiabatic basis instead.
The general problem is therefore to diagonalise the matrix
\[
\begin{pmatrix}
    \kappa & \Delta \\
    \Delta & -\kappa
\end{pmatrix}.
\]
While we could use a general purpose LAPACK routine like \texttt{dsyev}; for a 2$\times$2 matrix, an analytical solution will be much more efficient. Additionally, this will allow us to easily deal with the eigenvector sign ambiguity.\sidenote{With \texttt{dsyev}, we need to make sure that the eigenvectors point in the same direction as the ones at the neighbouring grid points and reverse them otherwise.}

The eigenvalues of the matrix are easily found to be $\pm \sqrt{\kappa^2 + \Delta^2}$. The corresponding eigenvectors can be written in many equivalent ways, one possibility\sidenote{These eigenvectors are not normalised to avoid clutter.} is
\begin{equation}
    \label{eq:evecs-k-gt-0}
    \begin{pmatrix}
        \kappa + \sqrt{\kappa^2 + \Delta^2} \\
        \Delta
    \end{pmatrix}
    \quad
    \text{and}
    \quad
    \begin{pmatrix}
        \Delta \\
        -\kappa - \sqrt{\kappa^2 + \Delta^2}
    \end{pmatrix}.
\end{equation}

While these are perfectly exact expressions, using them for propagation of the Tully models would result in a painful sea of NaNs. The problem is that these expressions are numerically unstable as $\Delta\to 0$, even though the eigenvectors are well defined. More specifically, when $\kappa<0$, the expression $-\kappa - \sqrt{\kappa^2 + \Delta^2}$ leads to catastrophic cancellation and when normalising the vector, we will obtain $0/0$.

However, we can rewrite the eigenvectors differently, since
\[
\frac{1}{\kappa + \sqrt{\kappa^2 + \Delta^2}} = \frac{-\kappa + \sqrt{\kappa^2 + \Delta^2}}{\Delta^2}.
\]
This leads to the expressions
\begin{equation}
    \label{eq:evecs-k-lt-0}
    \begin{pmatrix}
        \Delta \\
        -\kappa + \sqrt{\kappa^2 + \Delta^2}
    \end{pmatrix}
    \quad
    \text{and}
    \quad
    \begin{pmatrix}
        -\kappa + \sqrt{\kappa^2 + \Delta^2} \\
        -\Delta
    \end{pmatrix},
\end{equation}
which are easily seen to be stable for $\kappa<0$, but not for $\kappa>0$, as $\Delta\to 0$.

Also note that we have reversed the sign of the second vector in \eqref{eq:evecs-k-lt-0}. This is done so that the two expressions coincide in that $\kappa\to 0$ limit -- i.e. to eliminate the aforementioned sign problem.

Thus, for a stable decomposition of a 2$\times$2 matrix, we need to use \eqref{eq:evecs-k-gt-0} if $\kappa>0$ and \eqref{eq:evecs-k-lt-0} if $\kappa<0$ to ensure numerical stability when the coupling term is small.\marginnote{This is significant mainly for scattering problems, where $\Delta$ decays quickly outside of a small coupling region.}
We can note that this is very similar to obtaining a numerically stable solution of a quadratic equation. %TODO: reference

\section{Pseudocode for the propagation of MASH equations of motion}

We mentioned in \autoref{appendix:integrators} that one could find the hopping point by bisection or by spline interpolation. We chose the former method and show one way of implementing it in Algorithm~\ref{alg:mash-bisect}.

With the bisection algorithm in place, MASH propagation (Algorithm~\ref{alg:mash-propagate}) is quite straightforward. We assume $N_\mathrm{jumps}$ quantum jumps are to be performed, with $N_\mathrm{steps}[n_j]$ steps after jump number $n_j$. The algorithm generates a series of data points $n_s=0,1,\dots,N_\mathrm{steps}[n_j]$ for each jump, which can then be used to estimate the appropriate CFs -- we shall discuss that later. It would be possible to calculate the CFs immediately after every timestep using the same method, but we have chosen to implement a two-pass algorithm for simplicity.

Note that because MASH estimators are integrals over $\vec{S}$ with $\int \dd\vec{S} = 2$, but the Monte Carlo average over trajectories gives the usual spherical average, we need to multiply by two at the end and \textit{after every quantum jump}.

\begin{algorithm}
    \caption{Propagating a hop with bisection search}
    \label{alg:mash-bisect}
    $\delta\delta\tau \gets \delta\tau / 2$ \Comment*[r]{bisection timestep (halved every time)}
    $\delta\tau_\mathrm{hop} \gets \delta\delta\tau$ \Comment*[r]{hopping time}
    \For{ $n = 1,\dots,N_\mathrm{bisect}$ } {
        propagate by $\delta\delta\tau$ \;
        $\delta\delta\tau \gets |\delta\delta\tau| / 2$ \Comment*[r]{bisection}
        \If{ new surface active } {
            $\delta\delta\tau \gets -\delta\delta\tau$ \Comment*[r]{Go back.}
        }
        $\delta\tau_\mathrm{hop} \gets \delta\tau_\mathrm{hop} + \delta\delta\tau$ \;
    }
    propagate by $\delta\delta\tau$ \Comment*[r]{This gets us to the hopping point.}
    \;
    $\vec{d}_\mathrm{norm} \gets \vec{d} / |\vec{d}|$ \Comment*[r]{normalised NACV}
    $p_i \gets \vec{p}\cdot \vec{d}_\mathrm{norm}$ \Comment*[r]{initial momentum}
    $\vec{p}_\mathrm{proj} \gets p_i \vec{d}_\mathrm{norm}$ \;
    $p_f^2 \gets p_i^2 \pm 4mV_z$ \Comment*[r]{$\pm$ depends on active surface}
    \eIf {$p_f^2 < 0$} {
        $\vec{p} \gets \vec{p} - 2\vec{p}_\mathrm{proj}$ \Comment*[r]{Frustrated hop.}
    }{
        $\vec{p} \gets \vec{p} - \vec{p}_\mathrm{proj} + \sgn(p_i) \sqrt{p_f^2} \, \vec{d}_\mathrm{norm}$ \Comment*[r]{Successful hop.}
    }

    propagate by $\delta\tau - \delta\tau_\mathrm{hop}$ \Comment*[r]{Do rest of hop}
    \Comment*[r]{on new active surface.}

\end{algorithm}

\begin{algorithm}
    \caption{Propagation of MASH.}
    \label{alg:mash-propagate}
    $\mathrm{CF}_\mathrm{acc}[\cdots] \gets 0$ \Comment*[r]{initialise CF accumulator}
    \For{ $n_t = 1,\dots,N_\mathrm{traj}$ } {
        initialise $\vec{p}$ and $\vec{q}$ \Comment*[r]{e.g. Wigner distribution}
        \For{ $n_j = 0,\dots,N_\mathrm{jumps}$ } {
            initialise $\vec{S}$ \Comment*[r]{e.g. full-sphere sampling}
            save data for $n_s=0$ \;
            \For{ $n_s = 1,\dots,N_\mathrm{steps}[n_j]$ } {
                propagate by $\delta t$ \;
                \If{ active surface changed } {
                    propagate by $-\delta t$ \Comment*[r]{Go back.}
                    bisect and hop \Comment*[r]{Algorithm~\ref{alg:mash-bisect}}
                }
                save data \;
            }
        }
        calculate CFs \Comment*[r]{factor of 2 after every jump}
        $\mathrm{CF}_\mathrm{acc} \gets \mathrm{CF}_\mathrm{acc} + \mathrm{CF}$ \;
    }
    $\mathrm{CF} \gets \mathrm{CF}_\mathrm{acc} /N_\mathrm{traj}$ \;
\end{algorithm}

\section{Calculation of TCFs in Liouville space}

We have introduced Schrödinger picture propagation in \autoref{chapter:quantum-jump} in connection with the quantum jump procedure. However, it also allows for an efficient calculation of correlation functions, especially when many different ones are sought from a single simulation.

We shall need to upgrade our description from Hilbert space to Liouville space -- now, we can express all operators and density matrices as vectors in the basis $\big\{\hat{P}_+, \hat{P}_-, \hat{\sigma}_x, \hat{\sigma}_y\big\}$.\sidenote{This is quite a naughty basis, since it is orthogonal, but not orthonormal -- $\braket{\hat{P}_+}{\hat{P}_+}=1$, but $\braket{\hat{\sigma}_x}{\hat{\sigma}_x}=2$. (Remember that $\braket{\hat{A}}{\hat{B}} = \tr{\hat{A}\hat{B}}$ in Liouville space.) This will come back to bite us later.}

Suppose that we want to estimate the $P_+$--$P_-$ correlation function. We shall write the initial density matrix as a row vector
\[
P_+ = \begin{pmatrix}
    1 & 0 & 0 & 0
\end{pmatrix} .
\]
We first need to turn this into its the MASH version, which is
\[
P_+(\vec{S}) = \begin{pmatrix}
    h(S_z) & 0 & 0 & 0
\end{pmatrix} .
\]
For a general operator, this is achieved through multiplication by the initial estimator matrix
\[
    \vec{E} = \begin{pmatrix}
        h(S_z) & 0 & 0 & 0 \\
        0 & h(-S_z) & 0 & 0 \\
        0 & 0 & S_x & 0 \\
        0 & 0 & 0 & S_y \\
    \end{pmatrix}
\]
or its equivalent for other methods like ms-MASH.

This is then evolved in time to give $\rho^\MASH_S(t)$. This is accomplished by multiplication by the Liouville space version of $\hat{\mathcal{W}}$, which becomes a 4$\times$4 matrix $\vec{W}$.

Since the Liouville space inner product corresponds to the trace, we might assume that multiplying by the final observable expressed as a column vector,
\[
    P_- = \begin{pmatrix}
        0 \\ 1 \\ 0 \\ 0
    \end{pmatrix} ,
\]
would give us the final estimator.\sidenote{Notice that the final estimator is not expressed in terms of spin variables -- all time evolution is contained in $\vec{W}$.} However, the inner product of $\sum_O a_O \hat{O}$ and $\sum_O b_O \hat{O}$ is
\[
    a_+b_+ + a_-b_- + 2a_xb_x + 2a_yb_y,
\]
so we need to introduce the factors of two into our description. This is easily achieved by pre-multiplying the final operator by the overlap matrix
\[
    \vec{\Sigma} = \begin{pmatrix}
    1 & 0 & 0 & 0 \\
    0 & 1 & 0 & 0 \\
    0 & 0 & 2 & 0 \\
    0 & 0 & 0 & 2 \\
\end{pmatrix}.
\] 

The $\vec{W}$ matrix is best written as $\vec{W} = \vec{W'}(t) \vec{\Sigma}^{-1}$ with
\[
    \vec{\Sigma}^{-1} = \begin{pmatrix}
        1 & 0 & 0 & 0 \\
        0 & 1 & 0 & 0 \\
        0 & 0 & 1/2 & 0 \\
        0 & 0 & 0 & 1/2 \\
    \end{pmatrix}.
\]
In MASH, $\vec{W'}(t)$ can then be written as the outer product
\[
    \begin{pmatrix}
        W_{PP}  & W_{PC} \\
        W_{PP}  & W_{PC} \\
        W_{CP}  & W_{CC} \\
        W_{CP}  & W_{CC}
    \end{pmatrix}
    \begin{pmatrix}
        h(S_z(t)) & h(-S_z(t)) & 0 & 0 \\
        0 & 0 & S_x(t) & S_y(t)
    \end{pmatrix}.
\]
For ms-MASH, where all weighting factors are one, we have the even simpler
\[
    \begin{pmatrix}
        1 \\
        1 \\
        1 \\
        1
    \end{pmatrix}
    \begin{pmatrix}
        \frac 1 2 + S_z(t) & \frac 1 2 - S_z(t) & 2S_x(t) & 2S_y(t)
    \end{pmatrix}.
\]

For the $P_+$--$P_-$ correlation function, we thus need to evaluate (while the factors of $\vec{\Sigma}$ cancel here, they will be important for the quantum jumps)
\begin{widepar}
\[
\underbrace{
\begin{pmatrix}
    1 & 0 & 0 & 0
\end{pmatrix}
}_{\text{initial operator}}
\underbrace{
\begin{pmatrix}
    h(S_z) & 0 & 0 & 0 \\
    0 & h(-S_z) & 0 & 0 \\
    0 & 0 & S_x & 0 \\
    0 & 0 & 0 & S_y \\
\end{pmatrix}
}_{\text{initial estimators}}
\underbrace{
\begin{pmatrix}
    2|S_z|  & 2 \\
    2|S_z|  & 2 \\
    2  & 3 \\
    2  & 3
\end{pmatrix}
}_{\text{weighting factors}}
\underbrace{
\begin{pmatrix}
    h(S_z(t)) & h(-S_z(t)) & 0 & 0 \\
    0 & 0 & S_x(t) & S_y(t)
\end{pmatrix}
}_{\text{final estimators}}
\vec{\Sigma}^{-1}
\vec{\Sigma}
\underbrace{
\begin{pmatrix}
    0 \\ 1 \\ 0 \\ 0
\end{pmatrix}
}_{\text{final operator}} .
\]
\end{widepar}
This might seem like an absurd amount of work to write down
\[
2|S_z|h(S_z)h(-S_z(t)).
\]
However, this method can be generalised to \emph{any} observable (even diabatic ones -- the final state vector would then be a time-dependent quantity). Also, many of the matrix products can be evaluated at the beginning of the simulation. And in fact, it gets even better -- if we use
\[
\begin{pmatrix}
    1 & 0 & 0 & 0 \\
    0 & 1 & 0 & 0
\end{pmatrix}
\]
as the initial operator and
\[
    \begin{pmatrix}
        1  & 0 \\
        0  & 1 \\
        0  & 0 \\
        0  & 0
    \end{pmatrix}
\]
as the final one, the end result will be a 2$\times$2 matrix containing all four correlation functions between $P_+$ and $P_-$! This is only possible because we expanded our description from Hilbert space to Liouville space, which allowed us to separate all the different components.

The final formula to estimate the correlation functions of several initial operators $\vec{\rho}(\vec{p},\vec{q})$ and final operators $\vec{\hat{\mathcal{B}}}(\vec{p},\vec{q})$ is therefore
\begin{equation}
    \label{eq:MASH-Liouville-estimators}
  \underbrace{
    \vec{\rho}(\vec{p},\vec{q})
    \vec{E}
    \vec{W'}(t)
    \vec{\Sigma}^{-1}
}_{\displaystyle \vec{\rho}(\vec{p},\vec{q};t)}
    \;\cdot\;
    \vec{\Sigma}
    \vec{\hat{B}}(\vec{p}_t,\vec{q}_t) ,  
\end{equation}
where the notation $\vec{\rho}(\vec{p},\vec{q};t)$ suggests that we have evolved the electronic variables, but the nuclear ones are handled separately.

Note that this would be much easier if we used the identity and the Pauli matrices as a basis. Then, $\vec{\Sigma}$ would simply be equal to twice the identity matrix. In fact, that would be beneficial, since it would easily take care of converting between the factor of $1/2\pi$ in $\int\dd\vec{S}$ and $1/4\pi$ when averaging over a sphere. While we wanted to stay consistent with the current literature in this chapter, we believe that using this basis instead is a better way of implementing this algorithm.

\section{Multi-time TCFs with quantum jumps}
\label{sec:impl-multi-time-tcfs}

The formula \eqref{eq:MASH-Liouville-estimators} can be generalised to the case of quantum jumps by evolving the density operator during each period as suggested, by multiplying the evolved density matrix by additional factors of~$\hat{\mathcal{W}}$. If we jump at~$t_0$ and then evolve for a further time~$t_1$, we would therefore obtain
\[
    \vec{\rho}(\vec{p},\vec{q})
    \vec{E}
    \vec{W'}(t_0)
    \vec{\Sigma}^{-1}
    \cdot
    \vec{E}
    \vec{W'}(t_1)
    \vec{\Sigma}^{-1}
    \cdot
    \vec{\Sigma}
    \vec{\hat{B}}(\vec{p}_{t_0+t_1},\vec{q}_{t_0+t_1}) , 
\]

For $n$ jumps, we can slightly abuse notation to write 
\[
    \vec{\rho}(\vec{p},\vec{q})
    \left(
    \vec{E}
    \vec{W'}
    \vec{\Sigma}^{-1}\right)^n
    \vec{\Sigma}
    \vec{\hat{B}}(\vec{p}_t,\vec{q}_t)
\]
for the final estimator. This implicitly includes rescaling by a factor $2$ every jump, since we are resampling the spin coordinates and introducing another factor of $1/2\pi$.

For multi-time TCFs, we need to change the Hilbert space density matrix to
\[
    \rho \to \hat{\mathcal{U}}\rho \hat{\mathcal{U}}^\dagger .
\]
In the Liouville space description, this can be achieved by a single matrix multiplication
\[
    \rho \to \rho \, \hat{\hat{\mathcal{U}}},
\]
where $\hat{\hat{\mathcal{U}}}$ is a 4$\times$4 matrix in Liouville space. There does not seem to be an elegant way to find $\hat{\hat{\mathcal{U}}}$ in terms of $\hat{\mathcal{U}}$ -- one simply has to expand everything in terms of the basis operators and use the product relationships between them. In any case, this is an easy enough task for a computer.

In the case of $\hat{\mathcal{U}} = \hat{\sigma}_x$, $\hat{\hat{\mathcal{U}}}$ can be obtained by inspection -- we simply swap the two populations and reverse the sign of $\sigma_y$, giving
\[
    \hat{\hat{\mathcal{U}}} = \begin{pmatrix}
        0 & 1 & 0 & 0 \\
        1 & 0 & 0 & 0 \\
        0 & 0 & 1 & 0 \\
        0 & 0 & 0 & -1 \\
    \end{pmatrix}.
\]

% !TEX root = main.tex

\chapter{Chebyshev propagation}
\label{appendix:chebyshev}

Exact description of quantum dynamics entails solving the time-dependent Schrödinger equation
\[
    \imag \hbar \pder{\Psi}{t} = \hat{H}\Psi .
\]

The form of the equation already suggests two steps that need to be overcome -- we need to apply $\hat{H}$ to the wavefunction and we need to devise a scheme for time propagation.

Time propagation could be achieved naively by any method for solving first-order ODEs -- however, such methods are not commonly used. More numerically stable methods exist, which leverage the special properties of the quantum-mechanical problem; e.g. the split-operator method, the short iterative Lanczos propagator or Chebyshev propagation. All three of these work very well and can be converged to any desired precision. \cite{leforestierComparisonDifferentPropagation1991,tannorIntroductionQuantumMechanics2007}

Applying $\hat{H}$ usually relies on the specific form of the Hamiltonian,\sidenote{Hamiltonians with a more complicated form -- e.g. a charged particle in a magnetic field -- are more difficult to propagate. However, that is also true in classical mechanics.} $\hat{H} = \hat{T}(p) + \hat{V}(x)$. We need to choose a specific representation for the wavefunction, but it usually turns out that if the $\hat{V}$ operator is easy to apply in the given representation, applying the $\hat{T}$ operator is more difficult and \textit{vice versa}.

\section[Calculating \texorpdfstring{$\hat{H}\psi$}{Hψ}]{Calculating \texorpdfstring{$\bm{\hat{H}\psi}$}{Hψ}}

For numerical work, we need to represent the wavefunction as a sum of finitely many basis functions. The traditional way of doing this is to use an orthogonal basis set -- e.g. the harmonic oscillator wavefunctions -- which is known as the \emph{spectral} basis. However, we could also think about representing the wavefunction by its values on a set of grid points. This is equivalent to considering a basis set of highly localised functions, which is known as the \emph{pseudospectral} basis.\sidenote{There is a duality between spectral and pseudospectral bases. The theory is related to the subject of Gaussian quadrature and can be formalised with the use of projection operators on the space spanned by the finite basis set. See \cite{tannorIntroductionQuantumMechanics2007} for a nice discussion.}

By the above discussion, applying the Hamiltonian to a wavefunction can be decomposed into applying $\hat{T}$ and applying $\hat{V}$ separately. It is generally the case that in the spectral basis, applying the kinetic energy operator is easy (since the derivatives of the basis functions are known), whereas the potential energy is difficult -- we would need to express the product $V(x)\psi(x)$ in terms of the basis functions, which is not impossible, but will usually require the evaluation of complicated integrals. On the other hand, in the pseudospectral basis, it is easy to calculate the potential energy. Since the basis functions are highly localised, we can approximate them by delta functions, and the calculation of potential energy reduces to simple multiplication.\sidenote{This approximation is not variational -- if we use this method to solve the time-independent Schrödinger equation, we will not generally obtain an upper bound to the energies. \cite{colbertNovelDiscreteVariable1992}}

However, the spectral and pseudospectral bases are related by a unitary transformation -- our strategy will therefore be to represent the wavefunction in a pseudospectral basis, but transform to a spectral basis for the kinetic energy calculation. In our case, the spectral basis\sidenote{The corresponding pseudospectral basis consists of $\mathrm{sinc}$ functions, but we shall not need to know that explicitly.} will be given by the plane wave basis functions $e^{ikx}$, so that the transformation can be efficiently represented by the fast Fourier transform.

We shall use a grid of points $(x_1, x_2, \dots , x_N)$ to discretise our wavefunction into a vector $ \vec{\psi} = (\psi(x_1), \psi(x_2), \dots , \psi(x_N))$. Applying the potential energy is then as simple as element-wise multiplication of $\psi$ with the vector $\vec{V} = (V(x_1), V(x_2), \dots , V(x_N))$.\sidenote{The element-wise (or Hadamard) product will be denoted using the operator $\odot$.}

Note that for multi-state wavefunctions, the general algorithm is the same, but the wavefunction is a vector at each $x$ and the potential energy is a matrix. The element-wise multiplication now becomes \enquote{element-wise} matrix--vector multiplication instead.

For the kinetic energy operator, we first transform to Fourier space and then use the fact that $\der{}{x}{2}$ is equivalent to multiplication by $-k^2 = -p^2 / \hbar^2$. Then, we transform the wavefunction back into position space using the inverse FFT. In most FFT implementations, the $k$ values are given by
\[
k_i = \begin{cases}
    \frac{2\pi (i-1)}{N\Delta x}    & \text{for} \quad i \leq N/2\\
    \frac{-2\pi (N+1-i)}{N\Delta x} & \text{for} \quad i > N/2\\
\end{cases}
\]
for $N$ even\sidenote{$k_i$ can also be defined for odd $N$, but we generally want a power of two for efficient FFT.} and $i=1,\dots,N$.

\begin{algorithm}
    \caption{Calculating $\hat{H}\psi$.}
    \label{alg:H-psi}
    \For{ $i = 1, \dots, N$ } {
        $V_i \gets V(x_i)$ \Comment*[r]{Construct $\vec{V}$.}
    }
    \For{ $i = 1, \dots, N$ } {
        $T_i \gets k_i^2 / 2m$ \Comment*[r]{Construct $\vec{T}$ in Fourier space.}
    }
    $\vec{\phi}\gets \mathrm{FFT} ( \vec{\psi} )$ \Comment*[r]{Calculate $\vec{T\psi}$.}
    $\vec{\phi} \gets \vec{T} \odot \vec{\phi}$ \;
    $\vec{T\psi} \gets \mathrm{IFFT} ( \vec{\phi} )$ \;
    $\vec{V\psi} \gets \vec{V} \odot \vec{\psi}$ \Comment*[r]{Calculate $\vec{V\psi}$.}
    $\vec{H\psi} \gets \vec{T\psi} + \vec{V\psi}$ \Comment*[r]{$\hat{H} = \hat{T}+\hat{V}$.}
\end{algorithm}

\section{Chebyshev polynomials}

The Chebyshev polynomials (of the first kind), $T_n(x)$, are a sequence of orthogonal polynomials for the weight function\sidenote{This means that $$\int_{-1}^1 \frac{T_n(x)T_m(x)}{\sqrt{1-x^2}} = 0$$ if $n\neq m$. } $(1-x^2)^{-1/2}$. They can be used to represent any function on the interval $[-1, 1]$ as a series
\[
    f(x) = \sum_{n=0}^\infty a_n T_n(x) .
\]

\marginnote{The Chebyshev series has very nice convergence properties, which makes it useful in all kinds of numerical work. This short section does not do the Chebyshev polynomials enough justice.}

For our purposes, we shall need the recurrence relation
\begin{equation}
    \label{eq:cheb-recurrence}
    T_{n+1}(x) = 2xT_n(x) - T_{n-1}(x)
\end{equation}
together with the starting points $T_0(x)=1$ and $T_1(x)=x$.

We shall also need the Chebyshev series for the complex exponential, which is expressed in terms of the Bessel functions $J_n$,
\begin{equation}
    \eul^{-\imag tx} = \sum_{n=0}^\infty (2-\delta_{n0}) J_n(t) T_n(-\imag x),
\end{equation}
and converges for $x$ in the interval $[-1, 1]$. \cite{tannorIntroductionQuantumMechanics2007}

\begin{algorithm}
    \caption{Propagating $\psi_0$ through a time $t$ to give $\psi_t$. This is easily generalised to the case when multiple $t$s are needed by accumulating arrays of wavefunctions.}
    \label{alg:cheb-propagate}
    $\alpha \gets \Delta E t / 2 \hbar$ \Comment*[r]{For convenience.}
    $n \gets 1$ \;
    $\phi_{n-1} \gets \psi_0$ \;
    $\phi_{n} \gets -\imag \hnorm \psi_0$ \Comment*[r]{Using Algorithm~\ref{alg:H-psi}.}
    $\psi_t \gets J_0(\alpha)\phi_{n-1}$ \Comment*[r]{Start with $n=0$ term.}
    \While{ $n \leq n_\lmax$ } {
        $\phi_{n+1} \gets -2\imag \hnorm \phi_n - \phi_{n-1}$ \Comment*[r]{Chebyshev recurrence.}
        $\psi_t \gets \psi_t + 2J_n(\alpha)\phi_{n}$ \Comment*[r]{Accumulate sum.}
        $n \gets n + 1$ \Comment*[r]{Shift everything by one.}
        $\phi_{n-1} \gets \phi_n$ \;
        $\phi_n \gets \phi_{n+1}$ \;
    }
    $\psi_t \gets \eul^{-\imag \bar{E} t/\hbar} \psi_t$ \Comment*[r]{Final phase factor.}
\end{algorithm}

\section{Propagating the time-dependent Schrödinger equation}

The idea behind Chebyshev propagation is to represent the exact propagator $\eul^{-\imag\hat{H}t/\hbar}$ as a Chebyshev series. \cite{tal-ezerAccurateEfficientScheme1984,chenChebyshevPropagatorQuantum1999} However, in order for the expansion to converge, we need to ensure that the Hamiltonian only has eigenvalues in the interval $[-1, 1]$.\sidenote{Thus, we should not use the Chebyshev propagator for non-Hermitian Hamiltonians, e.g. those with absorbing potentials. In practice, Chebyshev propagation can tolerate a small degree of non-Hermiticity and there are modifications to deal with absorbing boundary conditions. \cite{mandelshtamSimpleRecursionPolynomial1995}} We therefore need to normalise the Hamiltonian to
\[
    \hnorm = \frac{\hat{H} - \bar{E}}{\Delta E / 2},
\]
where $\Delta E = E_\lmax - E_\lmin$ and $\bar{E} = (E_\lmax + E_\lmin)/2$. Since we are working on a grid, the maximum and minimum energies are readily estimated as
\[
    E_\lmin = V_\lmin
\]
and
\[
    E_\lmax = V_\lmax + T_\lmax,
\]
where $V_\lmin$ and $V_\lmax$ are the minimum and maximum potential energy in our chosen range of $x$ and $T_\lmax$ is the maximum kinetic energy supported by the grid spacing $\Delta x$,\sidenote{For multi-dimensional problems, we sum the kinetic energy contributions of each dimension separately.}
\[
    T_\lmax = \frac{\hbar^2 \pi^2}{2m\Delta x^2}.
\]
\begin{marginfigure}
    \centering
    \includegraphics[width=0.65\textwidth]{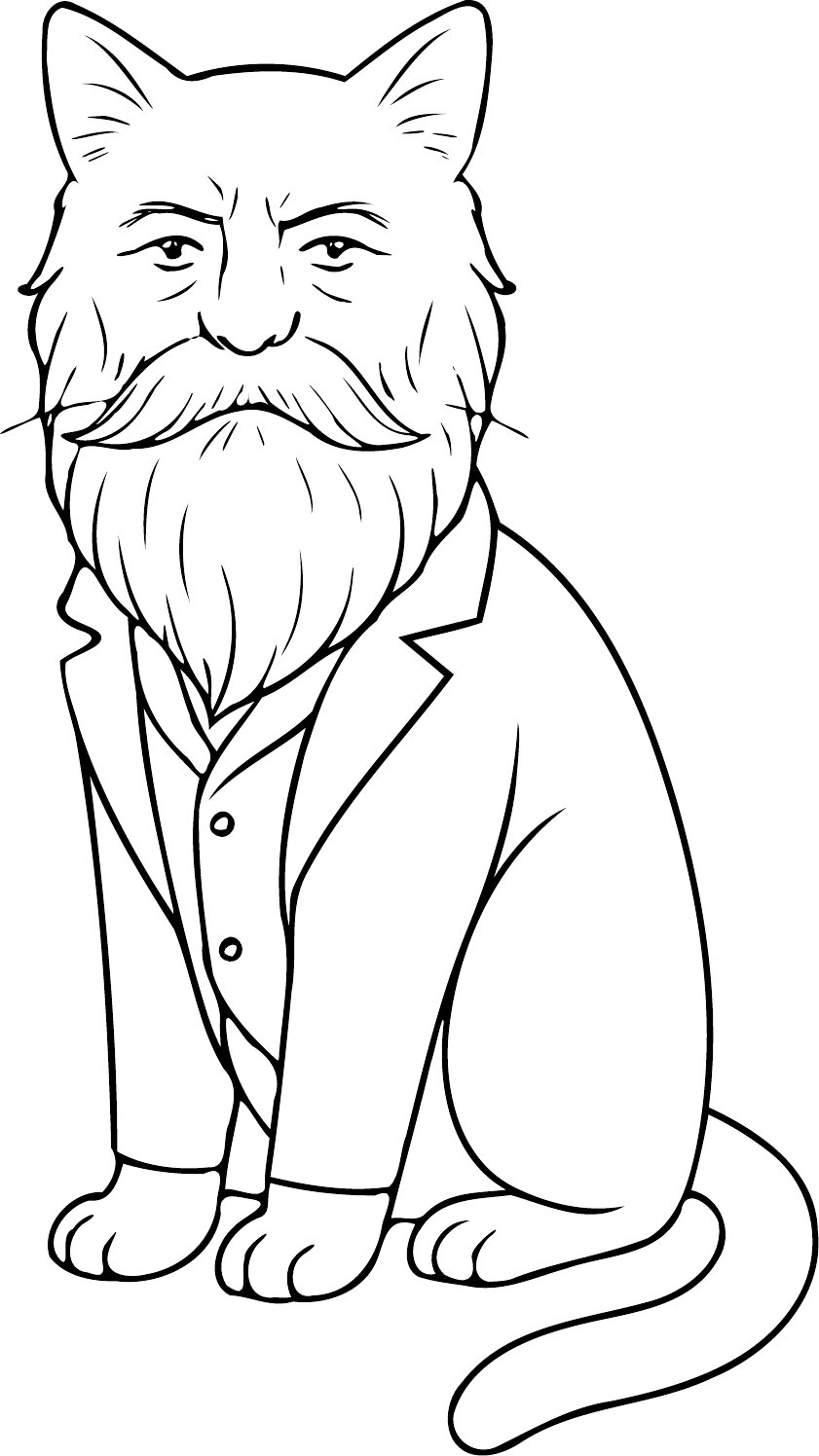}
    \caption{Pafnuty L. Chebyshev (1821--1894) in his feline form. Courtesy of ChatGPT.}
\end{marginfigure}

The time-evolved wavefunction is then represented as
\[
    \psi(t) = \eul^{-\imag\hat{H}t/\hbar} \psi(0) = \eul^{-\imag \bar{E} t/\hbar} \eul^{-\imag\hnorm \Delta E t/2\hbar} \psi(0)
\]
and the evolution under the normalised Hamiltonian is expanded as a Chebyshev series,

\[
    \psi(t) = \eul^{-\imag \bar{E} t/\hbar} \sum_{n=0}^\infty (2-\delta_{n0}) \, J_n \left(\frac{\Delta E t}{2\hbar} \right) \, T_n(-\imag \hnorm) \psi(0).
\]

The calculation of $\phi_n \equiv T_n(-\imag \hnorm) \psi(0)$ is easily achieved through the use of the recurrence relation \eqref{eq:cheb-recurrence} and Algorithm~\ref{alg:H-psi},
\[
    \phi_{n+1} = -2\imag \hnorm \phi_n - \phi_{n-1}.
\]

Of course, we cannot sum the infinite amount of terms in the Cheybshev expansion. However, the Bessel function $J_n(x)$ as a function of $n$ decays very quickly once $n>x$. \cite{tal-ezerAccurateEfficientScheme1984} Thus, we can choose
\[
    n_\lmax = \left\lfloor \frac{\Delta E t}{2\hbar} \right\rfloor + C ,
\]
e.g. with $C=100$, or we could iterate until the terms are smaller than a given tolerance.

For multi-time correlation functions, we first propagate the initial $\psi$ by the full range of $t_0$. We store this array, apply the operator $\hat{\mathcal{U}}$ to each wavefunction and use these as starting points for the propagation through $t_1$. The propagation through $t_0$ can thus be performed only once.

% !TEX root = main.tex

\chapter{All estimators give similar results}
\label{appendix:all-estimators}

This appendix presents data to support the statement that when using MASH dynamics, all estimators give similar results. We shall use the spin--boson model and Tully's model~II as examples, but the results presumably generalise to other situations. Aside from the MASH and ms-MASH estimators, we shall look at the modified MASH estimators from \autoref{sec:other-cpa-estimators}, at time-reversed ms-MASH estimators and at the QCLE-consistent mixed estimator from \autoref{sec:all-estimators-are-cpa-consistent}.

\begin{figure*}[h]
    \begin{tikzpicture}[
    subcaption/.style={anchor=north, font=\small}
    ]

    \begin{groupplot}[
            group style={
                group name=plotGroup,
                group size=2 by 2,
                horizontal sep=2.0cm,
                vertical sep=2.0cm
            },
            width=0.45\textwidth,
            height=0.45\textwidth,
            enlargelimits=false,
            tick align=outside,
            scale only axis,
            axis on top,
            groupplot xlabel=$t$,
            groupplot ylabel=$\avg{P_1 P_1(t)}$
            ]
            
        \nextgroupplot[ymax=1.0, xmin=0]

        \addplot[mark=none, line width=1.3pt, Gray] table[x index = 0, y index = 2] {data/spin-boson-a-all-est.dat};
        \addplot[mark=none, line width=1.3pt, Gray] table[x index = 0, y index = 3] {data/spin-boson-a-all-est.dat};
        \addplot[mark=none, line width=1.3pt, Gray] table[x index = 0, y index = 4] {data/spin-boson-a-all-est.dat};
        \addplot[mark=none, line width=1.3pt, Gray] table[x index = 0, y index = 6] {data/spin-boson-a-all-est.dat};
        \addplot[mark=none, line width=1.3pt, Gray] table[x index = 0, y index = 7] {data/spin-boson-a-all-est.dat};
        % draw these last to overlap any gray lines
        \addplot[mark=none, line width=1.3pt, RoyalBlue] table[x index = 0, y index = 1] {data/spin-boson-a-all-est.dat};
        \addplot[mark=none, line width=1.3pt, Maroon] table[x index = 0, y index = 5] {data/spin-boson-a-all-est.dat};

        \addplot[mark=none, line width=1.3pt, densely dashed] table {data/spin-boson-a-exact.dat};

        \nextgroupplot[ymin=0.2, ymax=1.0, xmin=0]

        \addplot[mark=none, line width=1.3pt, Gray] table[x index = 0, y index = 2] {data/spin-boson-b-all-est.dat};
        \addplot[mark=none, line width=1.3pt, Gray] table[x index = 0, y index = 3] {data/spin-boson-b-all-est.dat};
        \addplot[mark=none, line width=1.3pt, Gray] table[x index = 0, y index = 4] {data/spin-boson-b-all-est.dat};
        \addplot[mark=none, line width=1.3pt, Gray] table[x index = 0, y index = 6] {data/spin-boson-b-all-est.dat};
        \addplot[mark=none, line width=1.3pt, Gray] table[x index = 0, y index = 7] {data/spin-boson-b-all-est.dat};
        % % draw these last to overlap any gray lines
        \addplot[mark=none, line width=1.3pt, RoyalBlue] table[x index = 0, y index = 1] {data/spin-boson-b-all-est.dat};
        \addplot[mark=none, line width=1.3pt, Maroon] table[x index = 0, y index = 5] {data/spin-boson-b-all-est.dat};

        \addplot[mark=none, line width=1.3pt, densely dashed] table {data/spin-boson-b-exact.dat};

        \nextgroupplot[ymin=0, ymax=1.0, xmin=0]

        \addplot[mark=none, line width=1.3pt, Gray] table[x index = 0, y index = 2] {data/spin-boson-c-all-est.dat};
        \addplot[mark=none, line width=1.3pt, Gray] table[x index = 0, y index = 3] {data/spin-boson-c-all-est.dat};
        \addplot[mark=none, line width=1.3pt, Gray] table[x index = 0, y index = 4] {data/spin-boson-c-all-est.dat};
        \addplot[mark=none, line width=1.3pt, Gray] table[x index = 0, y index = 6] {data/spin-boson-c-all-est.dat};
        \addplot[mark=none, line width=1.3pt, Gray] table[x index = 0, y index = 7] {data/spin-boson-c-all-est.dat};
        % % draw these last to overlap any gray lines
        \addplot[mark=none, line width=1.3pt, RoyalBlue] table[x index = 0, y index = 1] {data/spin-boson-c-all-est.dat};
        \addplot[mark=none, line width=1.3pt, Maroon] table[x index = 0, y index = 5] {data/spin-boson-c-all-est.dat};

        \addplot[mark=none, line width=1.3pt, densely dashed] table {data/spin-boson-c-exact.dat};

        \nextgroupplot[ymin=0, ymax=1.0, xmin=0]

        \addplot[mark=none, line width=1.3pt, Gray] table[x index = 0, y index = 2] {data/spin-boson-d-all-est.dat};
        \addplot[mark=none, line width=1.3pt, Gray] table[x index = 0, y index = 3] {data/spin-boson-d-all-est.dat};
        \addplot[mark=none, line width=1.3pt, Gray] table[x index = 0, y index = 4] {data/spin-boson-d-all-est.dat};
        \addplot[mark=none, line width=1.3pt, Gray] table[x index = 0, y index = 6] {data/spin-boson-d-all-est.dat};
        \addplot[mark=none, line width=1.3pt, Gray] table[x index = 0, y index = 7] {data/spin-boson-d-all-est.dat};
        % % draw these last to overlap any gray lines
        \addplot[mark=none, line width=1.3pt, RoyalBlue] table[x index = 0, y index = 1] {data/spin-boson-d-all-est.dat};
        \addplot[mark=none, line width=1.3pt, Maroon] table[x index = 0, y index = 5] {data/spin-boson-d-all-est.dat};

        \addplot[mark=none, line width=1.3pt, densely dashed] table {data/spin-boson-d-exact.dat};

    \end{groupplot}

    \node[subcaption] at (plotGroup c1r1.below south) {(a) high temperature};
    \node[subcaption] at (plotGroup c2r1.below south) {(b) low temperature};
    \node[subcaption] at (plotGroup c1r2.below south) {(c) activationless};
    \node[subcaption] at (plotGroup c2r2.below south) {(d) inverted regime};

\end{tikzpicture}
    \caption{The $\avg{P_1 P_1(t)}$ correlation function for the spin--boson model. The MASH (blue) and ms-MASH (red) estimators are shown, as well as the other estimators discussed in this chapter (grey). The four cases correspond exactly to Figure~8 in \cite{runesonMultistateMappingApproach2023}. Exact HEOM results (black, dashed) were generously provided by Cole Hunt.}
    \label{fig:all-est-spin-boson}
\end{figure*}
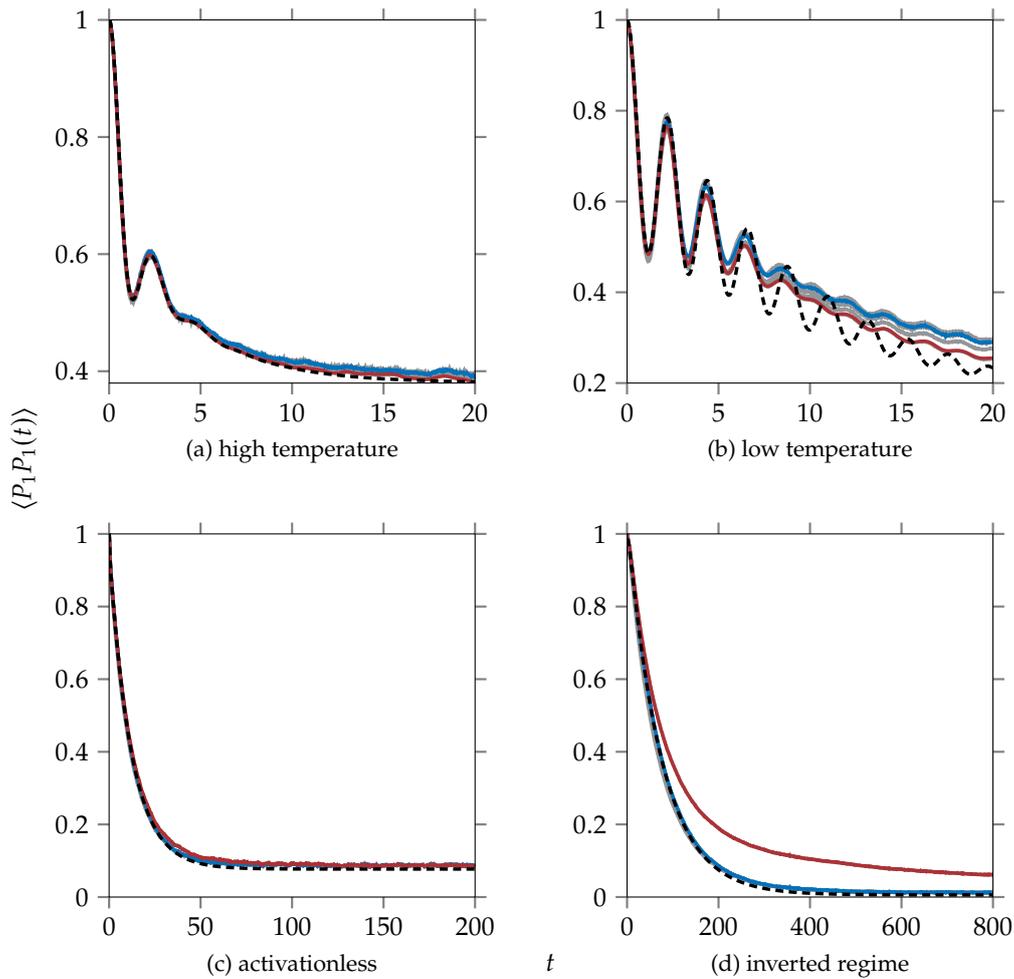

For the spin--boson model (Figure~\ref{fig:all-est-spin-boson}), we see that all other estimators (grey) overlap with either MASH or ms-MASH. The only visible difference is in the low temperature case (b), but it is quite small. There is a difference between MASH and ms-MASH in the inverted regime case (d), which was already observed in~\cite{runesonMultistateMappingApproach2023}. However, the fact that MASH does better than ms-MASH in this case should not be taken too seriously -- for other situations (like Tully's model~II with $p_0=25$), ms-MASH does better than the original MASH, so their relative performace is likely highly dependent on the particular model and parameters used.

\begin{figure*}[h]
    \begin{tikzpicture}[
    subcaption/.style={anchor=north, font=\small}
    ]

    \begin{groupplot}[
            group style={
                group name=plotGroup,
                group size=2 by 2,
                horizontal sep=2.0cm,
                vertical sep=2.0cm
            },
            width=0.45\textwidth,
            height=0.45\textwidth,
            enlarge x limits=false,
            tick align=outside,
            scale only axis,
            axis on top,
            groupplot xlabel=$t$,
            xtick={0, 600, 1200, 1800, 2400},
            xticklabels={$0$, $600$, $1200$, $1800$, $2400$}
            ]
            
        \nextgroupplot[ylabel=$C_{-+}(t)$, ymin=0]

        \addplot[mark=none, line width=1.3pt, Gray] table[x index = 0, y index = 1] {data/tully-ii-all-est-MASH-8Sz.dat};
        \addplot[mark=none, line width=1.3pt, Gray] table[x index = 0, y index = 1] {data/tully-ii-all-est-MASH-3Sz-3Szt.dat};
        \addplot[mark=none, line width=1.3pt, Gray] table[x index = 0, y index = 1] {data/tully-ii-all-est-MASH-Szt.dat};
        \addplot[mark=none, line width=1.3pt, Gray] table[x index = 0, y index = 1] {data/tully-ii-all-est-MISH-rev.dat};
        \addplot[mark=none, line width=1.3pt, Gray] table[x index = 0, y index = 1] {data/tully-ii-all-est-MISH-QCLE.dat};
        % draw these last to overlap any gray lines
        \addplot[mark=none, line width=1.3pt, RoyalBlue] table[x index = 0, y index = 1] {data/tully-ii-all-est-MASH.dat};
        \addplot[mark=none, line width=1.3pt, Maroon] table[x index = 0, y index = 1] {data/tully-ii-all-est-MISH.dat};

        \addplot[mark=none, line width=1.0pt, densely dashed] table[x index = 0, y index = 1] {data/tully-ii-p25-cfs-exact.dat};

        \nextgroupplot[ylabel=$C_{+x}(t)$]

        \addplot[mark=none, line width=1.3pt, Gray] table[x index = 0, y index = 2] {data/tully-ii-all-est-MASH-8Sz.dat};
        \addplot[mark=none, line width=1.3pt, Gray] table[x index = 0, y index = 2] {data/tully-ii-all-est-MASH-3Sz-3Szt.dat};
        \addplot[mark=none, line width=1.3pt, Gray] table[x index = 0, y index = 2] {data/tully-ii-all-est-MASH-Szt.dat};
        \addplot[mark=none, line width=1.3pt, Gray] table[x index = 0, y index = 2] {data/tully-ii-all-est-MISH-rev.dat};
        \addplot[mark=none, line width=1.3pt, Gray] table[x index = 0, y index = 2] {data/tully-ii-all-est-MISH-QCLE.dat};
        % draw these last to overlap any gray lines
        \addplot[mark=none, line width=1.3pt, RoyalBlue] table[x index = 0, y index = 2] {data/tully-ii-all-est-MASH.dat};
        \addplot[mark=none, line width=1.3pt, Maroon] table[x index = 0, y index = 2] {data/tully-ii-all-est-MISH.dat};

        \addplot[mark=none, line width=1.0pt, densely dashed] table[x index = 0, y index = 2] {data/tully-ii-p25-cfs-exact.dat};

        \nextgroupplot[ylabel=$C_{x+}(t)$]

        \addplot[mark=none, line width=1.3pt, Gray] table[x index = 0, y index = 3] {data/tully-ii-all-est-MASH-8Sz.dat};
        \addplot[mark=none, line width=1.3pt, Gray] table[x index = 0, y index = 3] {data/tully-ii-all-est-MASH-3Sz-3Szt.dat};
        \addplot[mark=none, line width=1.3pt, Gray] table[x index = 0, y index = 3] {data/tully-ii-all-est-MASH-Szt.dat};
        \addplot[mark=none, line width=1.3pt, Gray] table[x index = 0, y index = 3] {data/tully-ii-all-est-MISH-rev.dat};
        \addplot[mark=none, line width=1.3pt, Gray] table[x index = 0, y index = 3] {data/tully-ii-all-est-MISH-QCLE.dat};
        % draw these last to overlap any gray lines
        \addplot[mark=none, line width=1.3pt, RoyalBlue] table[x index = 0, y index = 3] {data/tully-ii-all-est-MASH.dat};
        \addplot[mark=none, line width=1.3pt, Maroon] table[x index = 0, y index = 3] {data/tully-ii-all-est-MISH.dat};

        \addplot[mark=none, line width=1.0pt, densely dashed] table[x index = 0, y index = 3] {data/tully-ii-p25-cfs-exact.dat};

        \nextgroupplot[ylabel=$C_{xx}(t)$]

        \addplot[mark=none, line width=1.3pt, Gray] table[x index = 0, y index = 4] {data/tully-ii-all-est-MASH-8Sz.dat};
        \addplot[mark=none, line width=1.3pt, Gray] table[x index = 0, y index = 4] {data/tully-ii-all-est-MASH-3Sz-3Szt.dat};
        \addplot[mark=none, line width=1.3pt, Gray] table[x index = 0, y index = 4] {data/tully-ii-all-est-MASH-Szt.dat};
        \addplot[mark=none, line width=1.3pt, Gray] table[x index = 0, y index = 4] {data/tully-ii-all-est-MISH-rev.dat};
        \addplot[mark=none, line width=1.3pt, Gray] table[x index = 0, y index = 4] {data/tully-ii-all-est-MISH-QCLE.dat};
        % draw these last to overlap any gray lines
        \addplot[mark=none, line width=1.3pt, RoyalBlue] table[x index = 0, y index = 4] {data/tully-ii-all-est-MASH.dat};
        \addplot[mark=none, line width=1.3pt, Maroon] table[x index = 0, y index = 4] {data/tully-ii-all-est-MISH.dat};

        \addplot[mark=none, line width=1.0pt, densely dashed] table[x index = 0, y index = 4] {data/tully-ii-p25-cfs-exact.dat};

    \end{groupplot}

    \node[subcaption] at (plotGroup c1r1.below south) {(a) $C_{-+}(t)$};
    \node[subcaption] at (plotGroup c2r1.below south) {(b) $C_{+x}(t)$};
    \node[subcaption] at (plotGroup c1r2.below south) {(c) $C_{x+}(t)$};
    \node[subcaption] at (plotGroup c2r2.below south) {(d) $C_{xx}(t)$};

\end{tikzpicture}
    \caption{Various correlation functions for Tully's model~II initialised with $q_0 = -15$, $p_0=25$ and $\gamma = 0.5$. The MASH (blue) and ms-MASH (red) estimators are shown, as well as the other estimators discussed in this chapter (grey). Exact results (black, dashed) were obtained using Chebyshev propagation.}
    \label{fig:all-est-tully-ii}
\end{figure*}
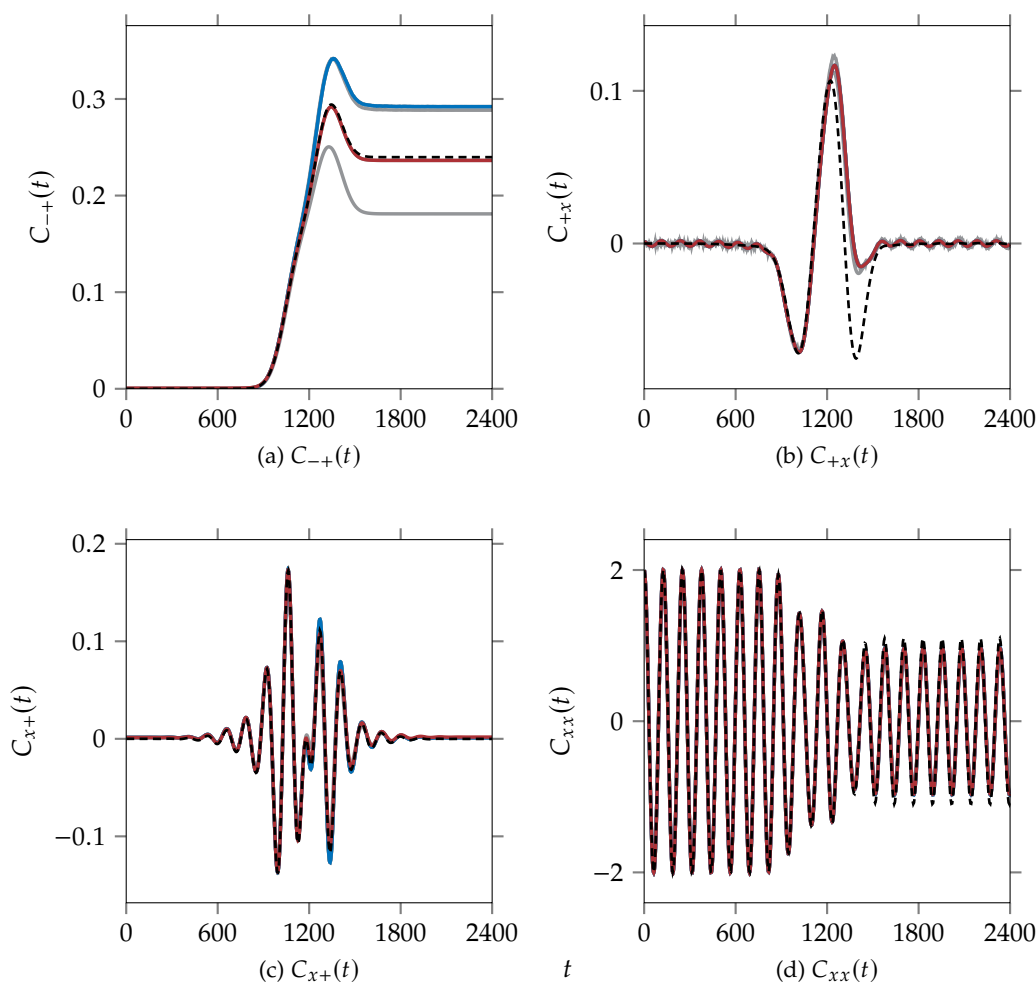

For Tully's model~II (Figure~\ref{fig:all-est-tully-ii}), we have can see that all estimators generally give the same result for all correlation functions, except for the population-population one in (a). There, the three lines we see are due to MASH (blue), ms-MASH (red) and MASH with $2|S_z(t)|$ as the weighting factor (grey). The latter shows that MASH gives a significant microscopic reversibility error in this case, which can be remedied by a quantum jump (see \autoref{appendix:single-jump}). Note that the \enquote{reversed ms-MASH} estimator overlaps with the MASH estimator here -- this seems to generally be the case.\sidenote{This is presumably due to the fact that the $C_{zz}(t)$ correlation function in MASH is just the time-reversed version of ms-MASH, as we have seen in \autoref{sec:detail-MASH-pauli-matrices}.}

Note that all even the CFs involving coherences are estimated very well -- the only problematic one is $C_{+x}(t)$, where the latter peak is much less pronounced than it should be. This is likely because this case involves significant interference between the two states, which MASH is not able to precisely capture. However, all estimators do equally badly here. Surprisingly, the $C_{x+}(t)$ and $C_{xx}(t)$ are reproduced well, despite also involving similar interference effects.

% !TEX root = main.tex

% figure on facing page
% https://tex.stackexchange.com/questions/248826/add-a-figure-on-even-page-before-chapter-starts
% https://tex.stackexchange.com/questions/11707/how-to-force-output-to-a-left-or-right-page

\cleardoubleevenemptypage

\thispagestyle{pagenumhead.scrheadings}

\stepcounter{chapter}
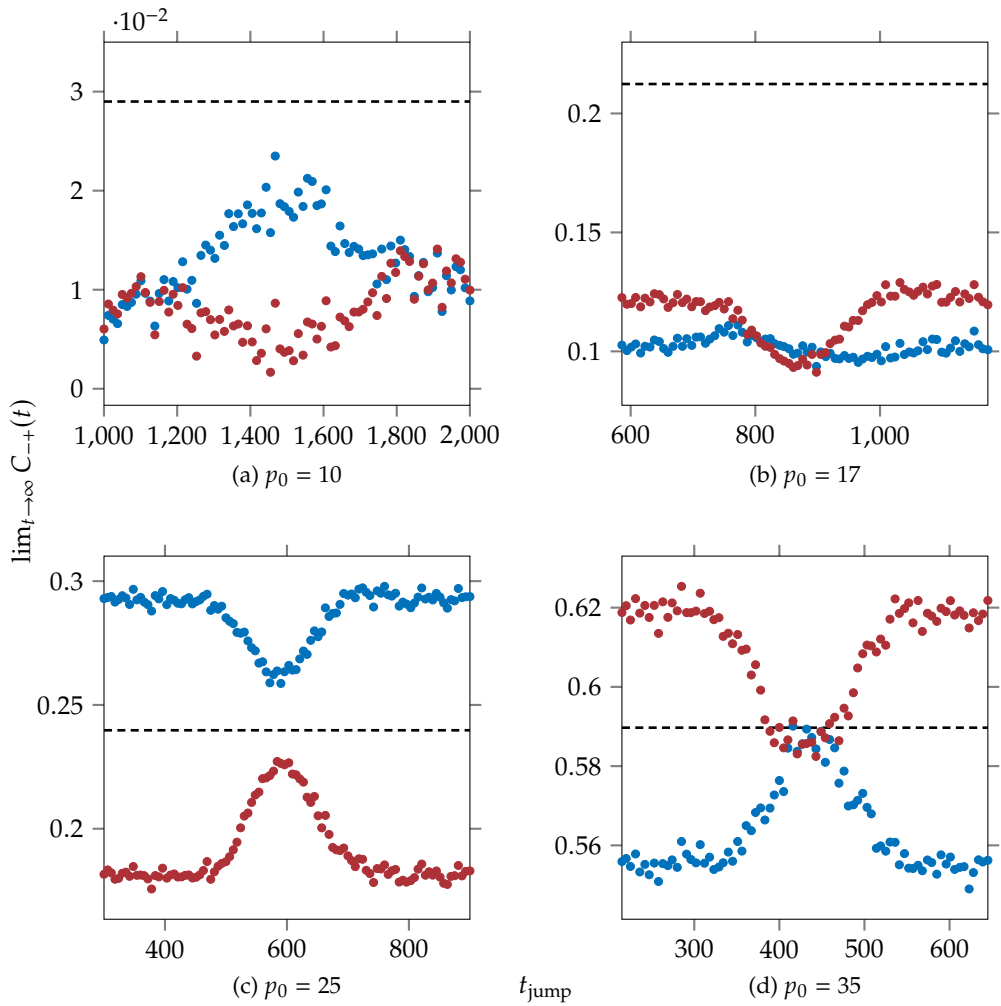
\begin{figure*}[h]
    \begin{tikzpicture}[
    subcaption/.style={anchor=north, font=\small}
    ]

    \begin{groupplot}[
            group style={
                group name=plotGroup,
                group size=2 by 2,
                horizontal sep=2.0cm,
                vertical sep=2.0cm
            },
            width=0.45\textwidth,
            height=0.45\textwidth,
            enlarge x limits=false,
            tick align=outside,
            scale only axis,
            axis on top,
            groupplot xlabel=$t_\mathrm{jump}$,
            groupplot ylabel=$\lim_{t\to\infty} C_{-+}(t)$
            ]
            
        \nextgroupplot[ymax = 0.035]

        \addplot[only marks, mark=*, mark size=1.5pt, RoyalBlue] table[x index = 0, y index = 1] {data/tully-ii-1-jump-test-p10.dat};
        \addplot[only marks, mark=*, mark size=1.5pt, Maroon   ] table[x index = 0, y index = 2] {data/tully-ii-1-jump-test-p10.dat};

        \draw[line width = 1.0pt, densely dashed] ({rel axis cs:0,0}|-{axis cs:0,0.0290}) -- ({rel axis cs:1,0}|-{axis cs:0,0.0290});

        \nextgroupplot[ymax = 0.23]

        \addplot[only marks, mark=*, mark size=1.5pt, RoyalBlue] table[x index = 0, y index = 1] {data/tully-ii-1-jump-test-p17.dat};
        \addplot[only marks, mark=*, mark size=1.5pt, Maroon   ] table[x index = 0, y index = 2] {data/tully-ii-1-jump-test-p17.dat};

        \draw[line width = 1.0pt, densely dashed] ({rel axis cs:0,0}|-{axis cs:0,0.21238}) -- ({rel axis cs:1,0}|-{axis cs:0,0.21238});

        \nextgroupplot

        \addplot[only marks, mark=*, mark size=1.5pt, RoyalBlue] table[x index = 0, y index = 1] {data/tully-ii-1-jump-test-p25.dat};
        \addplot[only marks, mark=*, mark size=1.5pt, Maroon   ] table[x index = 0, y index = 2] {data/tully-ii-1-jump-test-p25.dat};

        \draw[line width = 1.0pt, densely dashed] ({rel axis cs:0,0}|-{axis cs:0,0.23977}) -- ({rel axis cs:1,0}|-{axis cs:0,0.23977});

        \nextgroupplot
                       
        \addplot[only marks, mark=*, mark size=1.5pt, RoyalBlue] table[x index = 0, y index = 1] {data/tully-ii-1-jump-test-p35.dat};
        \addplot[only marks, mark=*, mark size=1.5pt, Maroon   ] table[x index = 0, y index = 2] {data/tully-ii-1-jump-test-p35.dat};

        \draw[line width = 1.0pt, densely dashed] ({rel axis cs:0,0}|-{axis cs:0,0.5897}) -- ({rel axis cs:1,0}|-{axis cs:0,0.5897});

    \end{groupplot}

    \node[subcaption] at (plotGroup c1r1.below south) {(a) $p_0 = 10$};
    \node[subcaption] at (plotGroup c2r1.below south) {(b) $p_0 = 17$};
    \node[subcaption] at (plotGroup c1r2.below south) {(c) $p_0 = 25$};
    \node[subcaption] at (plotGroup c2r2.below south) {(d) $p_0 = 35$};

\end{tikzpicture}
    \caption{The final population of Tully's model~II initialised in the $\ket{-}$ state in a wavepacket with $q_0=-15$, $\gamma=0.5$ and various momenta $p_0$. One quantum jump has been performed at $t_\mathrm{jump}$ for MASH (blue) and MASH with $2|S_z(t)|$ as the weighting factor (red). The exact final population (black) has been calculated by Chebyshev propagation. The results are an average over $10^6$ trajectories.}
    \label{fig:1-jump-test}
\end{figure*}
\addtocounter{chapter}{-1}

\chapter{A single jump is all it takes}
\label{appendix:single-jump}

While MASH does not work very well for Tully's model II, the quantum jump procedure can lead to significant improvements. However, the choice of when to jump is critical -- we explore its impact here and show that a judiciously applied single jump is very good even for the low momentum case ($p_0=25$) discussed in \cite{mannouchMappingApproachSurface2023}, where convergence could not be obtained even with four quantum jumps.

The results for various initial momenta are shown in Figure~\ref{fig:1-jump-test}. We show both the result of MASH and of MASH with $2|S_z(t)|$ as a weighting factor -- the difference between these two gives the \emph{microscopic reversibility error}, which can be used to quantify how well MASH performs.

It is apparent that the case $p_0=35$, which was shown to converge with two jumps in \cite{mannouchMappingApproachSurface2023}, can easily be converged with a single jump at around $t_\mathrm{jump}\approx 420$. Even the difficult $p_0 = 25$ case gets significant improvement from a single jump.

For the very low momentum cases,\sidenote{Note that for $p_0=10$, the correct $t\to\infty$ limit is of course zero. The particle does not have enough energy to leave in the upper state, since $p_0^2/2m < \varepsilon$. Figure~\ref{fig:1-jump-test} instead shows the population at $t=2|q_0|m/p_0$ of the long-lived resonance that is set up in the upper state.} $p_0=10$ and $p_0=17$, MASH fails dramatically. The final microscopic reversibility error does not give a~good indication of MASH performance, though it would be apparent at~intermediate times. However, MASH still ought to work if enough jumps are applied -- it would just be very difficult to converge.

The results of this appendix show that for a quantum jump, the choice of time is absolutely crucial. In fact, we usually cannot afford to do more than a few jumps, even in simple one-dimensional models, so a good method or heuristic for finding the optimal time would be a helpful development for MASH in general. Of course, another avenue is the development of approximate schemes like decoherence corrections -- these have in successfully used in \cite{mannouchMappingApproachSurface2023} and \cite{lawrenceRecoveringMarcusTheory2024} -- but that discussion is beyond the scope of this work.

% End of the main document content
\backmatter
\setchapterstyle{plain} % Output plain chapters from this point onwards

\printbibliography[heading=bibintoc, title=Bibliography] % Add the bibliography heading to the ToC, set the title of the bibliography and output the bibliography note

\end{document}